\documentclass[a4paper,11pt]{article}
\pdfoutput=1 

\usepackage{jheppub} 

\usepackage[T1]{fontenc} 

\title{\boldmath Aspects of Witten Diagrams for Holographic Defects}

 \author[a,b,c]{Dean Carmi,}
 \author[c]{Sudip Ghosh}
 \author[c]{and Trakshu Sharma}
 \affiliation[a]{Theoretical Physics Department, CERN, 1211 Geneva 23, Switzerland}
 \affiliation[b]{Department of Mathematics and Physics University of Haifa at Oranim, Kiryat Tivon 36006, Israel.}
 \affiliation[c]{Department of Physics and Haifa Center for Physics and Astrophysics, University of Haifa, Haifa 3498838, Israel}
 \emailAdd{deancarmi1@gmail.com, sudip112phys@gmail.com, trakshusharma@gmail.com}

\abstract{In this paper, we study the conformal block decomposition of Witten diagrams for $d$-dimensional holographic CFTs in the presence of a $p$-dimensional conformal defect. The holographic dual in this case contains a probe AdS$_{p+1}$ brane embedded inside AdS$_{d+1}$. In particular, we focus on contact, tree-level exchanges and some one-loop two-point Witten diagrams, which contribute to the two-point function of CFT bulk scalar operators. We also consider a tree-level exchange diagram for a three-point function involving one CFT bulk scalar operator and two scalar operators localized on the defect. Employing the split representation of AdS propagators, adapted to the probe brane setup, we perform the direct channel conformal block decompositions of these diagrams. In the case of tree-level diagrams, we obtain explicit expressions for the OPE coefficients in the direct channel decompositions. For two-point tree-level exchange diagrams, we derive recursion relations for the coefficients in the crossed-channel block expansions and compute the seed coefficients which serve as inputs for these relations. Our explicit results for the block decomposition coefficients for tree-level Witten diagrams are potentially useful for further developing the analytic functional approach to bootstrapping two-point functions of bulk operators in general defect CFTs. We also study the crossing kernel, which encodes the bulk channel partial wave expansion of a defect channel partial wave. Using the bulk channel Lorentzian inversion formula for defect CFTs, we derive closed form expressions for this defect-to-bulk channel crossing kernel for zero-dimensional defects in $d=2,4$ dimensions and surface defects in $d=4,6$ dimensions.}

\begin{document} 
\maketitle
\flushbottom

\section{Introduction}

The AdS/CFT correspondence \cite{Maldacena:1997re} posits a remarkable duality between quantum gravity in asymptotically anti-de Sitter (AdS) spacetimes and conformal field theories (CFTs) living on their boundary. This equivalence provides a powerful framework for studying a class of strongly coupled CFTs which admit a dual weakly coupled holographic description in AdS. In a CFT, correlation functions of local operators are the most basic observables. The holographic correspondence enables their computation through a perturbative expansion in terms of Witten diagrams in AdS \cite{Witten:1998qj,Gubser:1998bc}. In order to extract the CFT data encoded in holographic correlators, it is crucial to understand the conformal block decomposition of Witten diagrams.

In this paper, we are interested in extensions of the usual AdS/CFT setup to CFTs in the presence of defects, also known as defect conformal field theories (DCFTs). A $p$-dimensional conformal defect in a $d$-dimensional CFT breaks the global conformal symmetry $SO(d+1,1)$ to the subgroup $SO(p+1,1)\times SO(d-p)$. DCFTs are of interest more broadly due to their applications in a variety of physical contexts. These include, for example, the study of surface critical phenomena, impurities in condensed matter systems, line and surface operators in gauge theories, and branes in string theory. The introduction of defects enriches the set of CFT observables, which can now involve correlators of bulk local operators, defect-localized operators, and mixed correlators involving both bulk and defect operators. These correlators can be expanded in conformal blocks in different channels, leading to seemingly distinct representations of the same observable. The equivalence of these representations imposes nontrivial crossing relations, whose analysis lies at the heart of the bootstrap program for defect CFTs \cite{Liendo:2012hy, Gliozzi:2015qsa, Lemos:2017vnx, Liendo:2019jpu, Bissi:2018mcq, Mazac:2018biw, Kaviraj:2018tfd, Barrat:2022psm, Bianchi:2022ppi}. 

\begin{figure}[htp]
    \centering
    \includegraphics[width=8.cm]{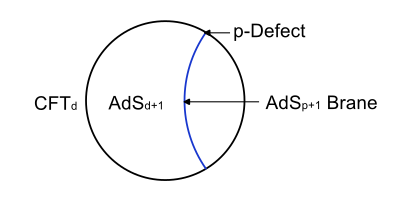}
    \caption{AdS$_{d+1}$ with a $p+1$ dimensional AdS brane extending in it. The $d$-dimensional CFT lies on the boundary of AdS$_{d+1}$ while the boundary of AdS$_{p+1}$ brane corresponds to a $p$-dimensional defect in the CFT$_d$. Figure adapted from reference \cite{Gimenez-Grau:2023fcy}.}
    \label{fig:AdS_CFT_Defect}
\end{figure}

Holographic realizations of DCFTs typically involve a probe AdS$_{p+1}$ brane embedded in the dual AdS$_{d+1}$ geometry, as illustrated schematically in Fig.\ref{fig:AdS_CFT_Defect} \cite{Karch:2000gx, Karch:2000ct, Aharony:2003qf, DeWolfe:2001pq,Erdmenger:2002ex}. The conformal boundary of AdS$_{p+1}$ is identified with the $p$-dimensional defect in the CFT. Witten diagrams in this setup generally involve interaction vertices in the ambient AdS$_{d+1}$ bulk as well as those localized on the AdS$_{p+1}$ brane. In recent years, the computation of holographic DCFT correlation functions has garnered significant attention \cite{Drukker:2020swu, Barrat:2021yvp, Chen:2023yvw, Zhou:2024ekb, Chen:2024orp, Chen:2025yxg, Chen:2026ium, Pufu:2023vwo, Gimenez-Grau:2023fcy, Giombi:2018qox, Giombi:2018hsx,Komatsu:2020sup, Barrat:2020vch}, spurred in part by advances in the analytic bootstrap of holographic correlators in certain supersymmetric theories in the absence of defects \cite{Rastelli:2016nze, Rastelli:2017udc}. For a review of these developments in the defect-free case and further references, we refer the reader to \cite{Bissi:2022mrs}.

Beyond their fundamental importance for holographic correlators, Witten diagrams have emerged as the essential ingredients in the analytic functional approach for bootstrapping generic CFTs\footnote{The analytic functional approach is intimately related to CFT dispersion relations \cite{Carmi:2019cub,Caron-Huot:2020adz, Gopakumar:2021dvg, Paulos:2020zxx}.}. In this formalism, also referred to as the Polyakov bootstrap, crossing symmetric combinations of tree-level exchange Witten diagrams, up to additional contact diagrams, furnish an alternative basis for expanding CFT correlators \cite{Gopakumar:2016wkt,Gopakumar:2016cpb, Gopakumar:2018xqi}. The conformal block decomposition of these Witten diagrams encodes the action of a complete basis of analytic functionals \cite{Mazac:2018mdx,Mazac:2018ycv,Mazac:2019shk}. The application of such functionals to the crossing equation yields sum rules that constrain the CFT data. In the case of defects, the details of this framework have been worked out for two-point functions of bulk operators in boundary CFTs (BCFTs) \cite{Mazac:2018biw, Kaviraj:2018tfd}, which can be regarded as CFTs with co-dimension one defects. In these papers, the conformal block expansion of two-point tree-level Witten diagrams for BCFTs was thoroughly studied\footnote{An equivalent route to the conformal block decomposition of Witten diagrams is through geodesic Witten diagrams \cite{Hijano:2015zsa}. Their analogue for BCFTs has been explored in \cite{Karch:2017wgy, Sato:2017gla, Rastelli:2017ecj}. }. The functional approach has not yet been developed in detail for higher codimension defects. A systematic understanding of the conformal block decomposition of tree-level Witten diagrams in this context would constitute a crucial step towards extending the Polyakov bootstrap program for general codimension defects. 

Motivated by the above mentioned developments, in this paper, we study the conformal block expansions of Witten diagrams for CFTs containing a flat $p$-dimensional conformal defect. We do not restrict attention to the holographic duals of any specific DCFT. Instead, we study individual diagrams within an effective field theory description in AdS. We focus primarily on two-point Witten diagrams where the external insertions correspond to scalar bulk primary operators in the DCFT. The two-point function in the presence of a defect is not fixed by symmetry and depends non-trivially on two cross-ratios. The diagrams that we consider include a contact diagram, tree-level exchange diagrams and certain one-loop diagrams. The tree-level exchanges are of two types. One of these involves exchanged fields that propagate in the ambient AdS$_{d+1}$ spacetime. The other is a defect channel exchange, which arises due to fields propagating on the probe AdS$_{p+1}$ brane. A summary of our main results is as follows. 

For tree-level scalar exchanges in both the bulk and defect channels, and a spin-$2$ tree-level exchange in the bulk channel, we obtain explicit results for the coefficients in the direct channel block decompositions of these diagrams. In addition, we discuss how to expand a class of one-loop diagrams, featuring exchanges in the bulk channel, in terms of bulk channel blocks. We also derive the block decomposition of a tree-level exchange diagram for a three-point function that involves two scalar operators localized on the defect and a scalar bulk operator. A central role in these analyses is played by the spectral and split representation of AdS propagators \cite{Penedones:2010ue}, but now adapted to the situation where we have a probe brane in AdS. This representation provides an efficient method to obtain the partial wave expansion in the direct channel\footnote{For CFTs in the absence of defects, the split representation has been extensively used to obtain the conformal partial wave expansion of Witten diagrams. See for example \cite{Costa:2014kfa,Sleight:2017fpc,Giombi:2017hpr, Chen:2017yia,Zhou:2018sfz,Carmi:2019ocp, Carmi:2021dsn, Carmi:2024tzj}.}. In the context of defects, the application of the split representation to study Witten diagrams in BCFTs has appeared in \cite{Rastelli:2017ecj, Mazac:2018biw, Kaviraj:2018tfd} and for DCFTs in  \cite{Goncalves:2018fwx,Gimenez-Grau:2023fcy, Alday:2024srr}. 

We further study the expansion of two-point tree-level exchange diagrams in terms of blocks in the crossed-channel. By exploiting the equations of motion satisfied by the AdS propagators, we derive recursion relations for the coefficients in the crossed channel block decompositions. These recursion relations can be solved if a set of seed coefficients is provided as input. The seeds are OPE coefficients of operators lying on the leading twist trajectory in the crossed channel. We obtain closed-form expressions for these coefficients by employing the Mellin representation of bulk and defect channel exchange Witten diagrams. 

Finally, we study the crossing kernel for DCFTs. In particular, we consider the defect-to-bulk channel crossing kernel, which encodes the decomposition of a defect channel partial wave in terms of the bulk channel partial waves. Using the Lorentzian inversion formula derived in \cite{Liendo:2019jpu}, we compute this crossing kernel explicitly for zero-dimensional defects in $d=2,4$ dimensions, and surface defects in $d=4,6$ dimensions.

This paper is organized as follows. In section \ref{sec:prelim}, we set up some notations and conventions, define the various propagators that appear throughout the paper and review the basic kinematics of two-point functions in DCFTs. In section \ref{2pt_tree_direct}, we derive the direct channel block expansions for $2$-point contact diagrams and tree-level scalar exchange diagrams, including both defect and bulk channels. Section \ref{2pt_spin_exch} features the bulk channel decomposition of a spin-$2$ bulk channel exchange diagram. In section \ref{2pt_exch_cross}, we provide a recursive method to compute the coefficients in the crossed channel decompositions of tree-level $2$-point exchange diagrams. Here, we also compute the seed coefficients for the recursion relations using Mellin space. Section \ref{2pt_loop} contains the analysis of bulk channel block expansions of a set of two-point $1$-loop diagrams. In section \ref{3pt_corr}, we perform the block decomposition of a $3$-point Witten diagram for the form factor. In section \ref{sec:def6j}, we present results for the defect-to-bulk channel crossing kernel. We conclude in section \ref{sec:concl} with a discussion of future directions. The Appendices \ref{app:blocks}--\ref{app:1ptloop} contain details of intermediate results that appear in the main part of the paper.

\section{Preliminaries}
\label{sec:prelim}

\subsection{Notations and conventions}
\label{subsec:notconv}

In this section, we introduce the notations that we will use to distinguish points in different regions of the setup shown in Fig.~\ref{fig:AdS_CFT_Defect}, and define the various propagators that appear throughout the paper. We work in Euclidean AdS$_{d+1}$ in Poincaré coordinates, and denote a generic bulk point as
\begin{equation}
z = (z_0, z_{\perp}^i, z^a), \quad \text{with} \quad a=1,\ldots, p, \quad i=p+1,\ldots,d ,
\end{equation}

\noindent where $z_{\perp}^i$ and $z^a$ denote directions orthogonal and parallel to the AdS$_{p+1}$ brane. The points that lie on the brane will be indicated by the hatted variables $\hat{z}= (\hat{z}, \hat{z}^{a})$. We label points on the boundary of AdS$_{d+1}$ as $x$ and use $\hat{x}$ to mark points on the boundary of AdS$_{p+1}$. 

\vskip 4pt
Let us now set up the notation for the propagators that we will encounter in the computation of Witten diagrams. The usual bulk-to-boundary propagator for a scalar field of mass $m$ in AdS$_{d+1}$, dual to a CFT operator with dimension $\Delta$, will be denoted as\footnote{In the literature, bulk-to-boundary propagators sometimes include an overall normalization factor. In our conventions, these factors will not appear.}
\begin{equation}
\label{bulk2bdy}
K_\Delta(x,z) = \left( \frac{z_0}{z_0^2 + |z_{\perp}^i - x_{\perp}^i|^2 + |z^a - x^a|^2} \right)^\Delta , 
\end{equation}

\noindent where $m^{2}=\Delta(\Delta-d)$. If the bulk point $z$ lies on the AdS$_{p+1}$ brane, we will set $z^{i}_{\perp}=0$, replace $(z_{0},z^{a})\rightarrow (\hat{z}_{0},\hat{z}^{a})$ in Eq.~\eqref{bulk2bdy} and refer to $K_{\Delta}(x,\hat{z})$ as the brane-to-boundary propagator. We will also have brane-to-defect propagators where one endpoint lies on the AdS$_{p+1}$ brane and the other endpoint lies on the boundary of AdS$_{p+1}$, i.e., on the defect in the CFT. For a scalar field of mass $\widehat{m}$ living on the AdS$_{p+1}$ brane and dual to an operator with dimension $\hat{\Delta}$ localized on the defect, the brane-to-defect propagator is 
\begin{equation}
\label{brane2defect}
\widehat{K}_{\hat\Delta}(\hat z,\hat x) = \left( \frac{\hat{z}_0}{\hat{z}_0^2 + |\hat{z}^a - \hat{x}^a|^2} \right)^{\hat\Delta}, 
\end{equation}

\noindent where $\widehat{m}^{2}=\hat{\Delta}(\hat{\Delta}-d)$. Besides the above propagators, we have bulk-to-bulk, bulk-to-brane and brane-to-brane propagators, where the terminology depends on where the end-points are localized. In this paper, we will frequently use the spectral and split representations of these propagators. For a scalar exchange bulk-to-bulk propagator in AdS$_{d+1}$, this is given by
\begin{equation}
\label{bulk2bulk}
G_\Delta(z_1,z_2) = \int_{-\infty}^{\infty} d\nu \ P(\nu,\Delta) \int \limits_{\partial AdS_{d+1}} d^d x \; K_{\frac{d}{2}+i\nu}(x,z_1)\, K_{\frac{d}{2}-i\nu}(x,z_2) ,
\end{equation}

\noindent where the measure factor $P(\nu,\Delta)$ is
\begin{align}
\label{specmeasblk2blk}
    P(\nu,\Delta)=\frac{\nu^{2}}{\pi}\frac{C_{\frac{d}{2}+i\nu}C_{\frac{d}{2}-i\nu}}{[\nu^{2}+ (\Delta-\frac{d}{2})^{2}]}, \ C_{\Delta}=\frac{\Gamma(\Delta)}{2\pi^{\frac{d}{2}} \Gamma\left(\Delta+1-\frac{d}{2}\right)}.
\end{align}

\noindent The split representation of a bulk-to-brane propagator $G_\Delta(z_{1},\hat z_{2})$ is of the same form as in Eq.~\eqref{bulk2bulk} but with $z_{2}$ replaced by $\hat{z}_{2}$ signifying the restriction of this point to the $AdS_{p+1}$ brane. Finally, the brane-to-brane propagator in the split representation is,
\begin{equation}
\label{brane2brane}
\widehat{G}_{\hat \Delta}(\hat z_1,\hat z_2) = \int_{-\infty}^{\infty} d\nu \ \widehat{P}(\nu,\hat \Delta) \int \limits_{\partial AdS_{p+1}} d^p \hat x \; \widehat{K}_{\frac{p}{2}+i\nu}(\hat z_1, \hat x)\, \widehat{K}_{\frac{p}{2}-i\nu}(\hat z_2,\hat x), 
\end{equation}

\noindent where $\widehat{P}(\nu, \hat \Delta)$ 
\begin{align}
\label{specmeasbrane2brane}
    \widehat{P}(\nu,\hat \Delta)=\frac{\nu^{2}}{\pi}\frac{\widehat{C}_{\frac{p}{2}+i\nu}\widehat{C}_{\frac{p}{2}-i\nu}}{[\nu^{2}+(\hat{\Delta}-\frac{p}{2})^{2}]}, \ \widehat{C}_{\hat \Delta}=\frac{\Gamma(\hat \Delta)}{2\pi^{\frac{p}{2}} \Gamma\left(\hat \Delta+1-\frac{p}{2}\right)}.
\end{align}

\noindent The following notation will be employed for integrals over points in AdS$_{d+1}$ and AdS$_{p+1}$
\begin{align}
    \int\limits_{AdS_{d+1}} d^{d+1}z = \int \frac{d^{d+1}z}{z_{0}^{d+1}}, \quad  \int\limits_{AdS_{p+1}} d^{p+1}\hat{z} = \int \frac{d^{d+1}\hat{z}}{\hat{z}_{0}^{p+1}} .
\end{align} 

\noindent We also adopt the convention of \cite{Meltzer:2019nbs} to denote operators that belong to the principal series representation of the $SO(d+1,1)$ conformal group as $\underline{\mathcal{O}}$. Their scaling dimensions will be denoted as  $\underline{\Delta}=\frac{d}{2}+i\nu$. Similarly, for operators in the principal series representation of $SO(p+1,1)$ will be indicated as $\underline{\widehat{\mathcal{O}}}$ and their scaling dimensions as $\underline{\hat{\Delta}}=\frac{p}{2}+i\nu$.

\subsection{Two-point function in defect CFT}

In this section, we briefly review some general kinematical properties of correlation functions in defect CFTs that will be relevant to us. Further details can be found in \cite{Gadde:2016fbj, Billo:2016cpy}.

We consider a $d$-dimensional CFT in the presence of a flat $p$-dimensional defect. The main observable of interest to us is the $2$-point function of scalar operators that live in the ambient $d$-dimensional spacetime. Such operators will be referred to as bulk scalar primary operators throughout this paper. We denote their $2$-point function as\footnote{We have used the double bracket notation in Eq.~\eqref{2ptfunction} to emphasise that the correlator is being evaluated in the presence of the defect.}
\begin{align}
    \label{2ptfunction}
    \langle\hspace{-0.09cm}\langle O_{\Delta_{1}}(x_{1})O_{\Delta_{2}}(x_{2})\rangle\hspace{-0.09cm}\rangle = \frac{\mathcal{G}(\xi,\eta)}{|x_{1,\perp}|^{\Delta_{1}}|x_{2,\perp}|^{\Delta_{2}}} , 
\end{align}

\noindent where $(\xi,\eta)$ are $SO(p+1,1)\times SO(d-p)$ invariant cross-ratios and are defined as
\begin{align}
\label{crossratios}
    & \xi = \frac{x_{12}^{2}}{|x_{1,\perp}||x_{2,\perp}|}, \quad \eta= \frac{x_{1,\perp}.x_{2,\perp}}{|x_{1,\perp}||x_{2,\perp}|} .
\end{align}

The function $\mathcal{G}(\xi,\eta)$ can be expanded in conformal blocks in two different channels called bulk and defect channels. The defect channel expansion arises from using the bulk to defect operator expansion, which allows us to express a bulk primary that approaches close to the defect in terms of operators localised on the defect. The block expansion in this channel takes the form
\begin{align}
    \label{defchblock}
    \mathcal{G}(\xi,\eta)= \sum_{\widehat{\mathcal{O}}} b_{1\widehat{\mathcal{O}}} b_{2\widehat{\mathcal{O}}} \  \widehat{f}_{\hat{\Delta},s}(\xi,\eta), 
\end{align}

\noindent where $\widehat{f}_{\hat{\Delta},s}(\xi,\eta)$ is the defect channel block for the exchange of a defect primary operator $\widehat{O}$ having dimension $\hat{\Delta}$ and transverse spin $s$. The expression for $\widehat{f}_{\hat{\Delta},s}(\xi,\eta)$ is given in Appendix \ref{app:defectblock}.  $b_{1\widehat{\mathcal{O}}},b_{2\widehat{\mathcal{O}}}$ denote coefficients in the $2$-point function of the external bulk primary and the exchanged defect primary.

In the bulk-channel, using the OPE between the two bulk operators $\mathcal{O}_{1},\mathcal{O}_{2}$ leads to a decomposition of the following form
\begin{align}
    \label{bulkchblock}
    \mathcal{G}(\xi,\eta)= \xi^{-\frac{\Delta_{1}+\Delta_{2}}{2}}\sum_{\mathcal{O}} c_{12\mathcal{O}} a_{\mathcal{O}} f_{\Delta,J}(\xi,\eta)
\end{align}

\noindent where $f_{\Delta,J}(\xi,\eta)$ is the bulk channel block for the exchange of a bulk primary operator with dimension $\Delta$ and spin $J$. We refer the reader to Appendix \ref{app:bulkblock} for further details about the bulk blocks. $c_{12\mathcal{O}}$ is the $3$-point function coefficient involving $\mathcal{O}_{1},\mathcal{O}_{2}$ and the exchanged bulk primary $\mathcal{O}$, $a_{\mathcal{O}}$ is the coefficient appearing in the $1$-point function of $\mathcal{O}$.

\section{Tree-level $2$-point Witten diagrams}
\label{2pt_tree_direct}

In this section, we derive the conformal block expansions of tree-level Witten diagrams which contribute to the $2$-point function of CFT bulk scalar operators.

\subsection{Contact diagram}
\label{sec:contactdiag}

\begin{figure}[htp]
    \centering
    \includegraphics[width=4.5cm]{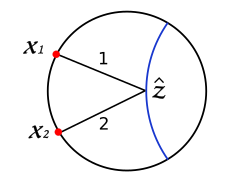}
    \caption{Contact Witten diagram.}
    \label{fig:contact_diag}
\end{figure}

The Witten diagram for a contact interaction of the form $\int_{\mathrm{AdS}_{p+1}}\phi_{1}\phi_{2}$, involving two bulk scalars on the brane, in Fig \ref{fig:contact_diag} is given by
\begin{align}
\label{contact}
W_{c}(x_{1},x_{2}) = \int\limits_{AdS_{p+1}} d^{p+1}\hat{z} \ K_{\Delta_{1}}(x_{1},\hat{z}) K_{\Delta_{2}}(x_{2},\hat{z}) 
\end{align}

\noindent where $K_{\Delta}(x,\hat{z})$ is the bulk-to-brane propagator defined in  section \ref{subsec:notconv}. The above integral can be computed in closed form \cite{Gimenez-Grau:2023fcy}. However, we will not need it here since our interest lies in deriving the conformal partial wave expansion in the defect channel for this diagram. For this purpose, we first insert a delta function $\delta(\hat{w}-\hat{z})$ within the integral in Eq.~\eqref{contact}. Then we use the following completeness property satisfied by the harmonic function $\widehat{\Omega}_{\nu}(\hat{w},\hat{z})$ in $AdS_{p+1}$
\begin{align}
\label{harmonicfuncid}
\int_{-\infty}^{\infty} d\nu  \ \widehat{\Omega}_{\nu}(\hat{w},\hat{z}) =\delta(\hat{w}-\hat{z}) .
\end{align}

\noindent The harmonic function $\widehat{\Omega}_{\nu}(\hat{w},\hat{z})$ admits the the split representation
\begin{align}
    \label{splitrepAdSp}
    \widehat{\Omega}_{\nu}(\hat{w},\hat{z})= \frac{\nu^{2}}{\pi}\int\limits_{\partial AdS_{p+1}} d^{p}\hat{x} \ \widehat{C}_{\frac{p}{2}+i\nu} \widehat{C}_{\frac{p}{2}-i\nu} \ \widehat{K}_{\frac{p}{2}+i\nu}(\hat{x},\hat{w}) \widehat{K}_{\frac{p}{2}-i\nu}(\hat{x},\hat{z})
\end{align}

\noindent where the brane-to-defect propgator $\widehat{K}_{\hat{\Delta}}(\hat{x},\hat{z})$ and the coefficient $\widehat{C}_{\hat{\Delta}}$ are defined in section \ref{sec:prelim}. Using this split representation, we can then express the contact diagram as
\begin{align}
    \label{contact3}
    W_{c}(x_{1},x_{2}) & = \int_{-\infty}^{\infty} d\nu \ \frac{\nu^{2}}{\pi} \int\limits_{\partial AdS_{p+1}} d^{p}\hat{x} \ \bigg[\int\limits_{AdS_{p+1}} d^{p+1}\hat{w}\ K_{\Delta_{1}}(x_{1},\hat{w}) \widehat{K}_{\frac{p}{2}+i\nu}(\hat{x},\hat{w}) \bigg] \times \nonumber\\
   &\hspace{4.5cm}\bigg[\int\limits_{AdS_{p+1}} d^{p+1}\hat{z} \ K_{\Delta_{2}}(x_{2},\hat{z}) \widehat{K}_{\frac{p}{2}-i\nu}(\hat{x},\hat{z}) \bigg] .
\end{align}

\noindent Now each of the $AdS_{p+1}$ integrals appearing in the above expression yields a $2$-point function $\langle\hspace{-0.09cm}\langle\mathcal{O}\ \widehat{\mathcal{O}} \rangle\hspace{-0.09cm}\rangle$ involving a CFT bulk operator $\mathcal{O}$ and a defect operator $\widehat{\mathcal{O}}$. More explicitly, we have 
\begin{align}
    \label{bulkdef2ptfuncs}
    & \int\limits_{AdS_{p+1}} d^{p+1}\hat{w} \ K_{\Delta_{1}}(x_{1},\hat{w}) \widehat{K}_{\frac{p}{2}+i\nu}(\hat{x},\hat{w}) = b_{1\underline{\hat{\Delta}}}\langle \hspace{-0.09cm}\langle\mathcal{O}_{\Delta_{1}}(x_{1})\widehat{\mathcal{O}}_{\frac{p}{2}+i\nu}(\hat{x})\rangle \hspace{-0.09cm}\rangle\nonumber\\
    & \int\limits_{AdS_{p+1}} d^{p+1}\hat{z} \ K_{\Delta_{2}}(x_{2},\hat{z}) \widehat{K}_{\frac{p}{2}-i\nu}(\hat{x},\hat{z}) = b_{2\underline{\widetilde{\hat{\Delta}}}}\langle \hspace{-0.09cm}\langle\mathcal{O}_{\Delta_{2}}(x_{2})\widehat{\mathcal{O}}_{\frac{p}{2}-i\nu}(\hat{x})\rangle\hspace{-0.09cm}\rangle
\end{align}

\noindent where $\langle \hspace{-0.09cm}\langle\mathcal{O}_{\Delta}(x)\widehat{\mathcal{O}}_{\hat{\Delta}}(\hat{x})\rangle\hspace{-0.09cm}\rangle$ is fixed by the $SO(p+1,1)\times SO(d-p)$ symmetry and takes the following form. 
\begin{align}
    \label{bulkdef2ptfunc}
    \langle \hspace{-0.09cm}\langle \mathcal{O}_{\Delta}(x)\widehat{\mathcal{O}}_{\hat{\Delta}}(\hat{x})\rangle \hspace{-0.09cm}\rangle = \frac{|x^{i}_{\perp}|^{\hat{\Delta}-\Delta}}{(|x^{i}_{\perp}|^{2}+(x^{a}-\hat{x}^{a})^{2})^{\hat{\Delta}}}
\end{align}

\noindent and the coefficients $b_{1\underline{\hat{\Delta}}}, b_{2\underline{\widetilde{\hat{\Delta}}}}$ are given by
\begin{align}
    \label{bulkdefopecoeff}
    b_{1\underline{\hat{\Delta}}}= \pi^{p/2}\frac{\Gamma\left(\frac{\Delta_{1}-\frac{p}{2}-i\nu}{2}\right)\Gamma\left(\frac{\Delta_{1}-\frac{p}{2}+i\nu}{2}\right)}{2\Gamma(\Delta_{1})}, \quad b_{2\underline{\widetilde{\hat{\Delta}}}}= \pi^{p/2}\frac{\Gamma\left(\frac{\Delta_{2}-\frac{p}{2}+i\nu}{2}\right)\Gamma\left(\frac{\Delta_{2}-\frac{p}{2}-i\nu}{2}\right)}{2\Gamma(\Delta_{2})} . 
\end{align}

\noindent The integral over the boundary of $AdS_{p+1}$ now involves a product of the $2$-point function structures in Eq.~\eqref{bulkdef2ptfuncs}. This gives the integral representation of the conformal partial wave $ \widehat{\mathcal{F}}_{\frac{p}{2}+i\nu,0}(x_{1},x_{2})$, for scalar exchange in the defect channel 
\begin{align}
    \label{defpwintrep}
    \widehat{\mathcal{F}}_{\frac{p}{2}+i\nu,0}(x_{1},x_{2})= \int d^{p}\widehat{x} \ \langle \hspace{-0.09cm}\langle\mathcal{O}_{\Delta_{1}}(x_{1})\widehat{\mathcal{O}}_{\frac{p}{2}+i\nu}(\widehat{x})\rangle \hspace{-0.09cm}\rangle \langle \hspace{-0.09cm}\langle \widehat{\mathcal{O}}_{\frac{p}{2}-i\nu}(\widehat{x})\mathcal{O}_{\Delta_{2}}(x_{2})\rangle \hspace{-0.09cm}\rangle .
\end{align}

\noindent The partial wave is related to the defect channel conformal blocks as  
\begin{align}
\label{defparwave}
\widehat{\mathcal{F}}_{\hat{\Delta},0}(x_{1},x_{2}) = \frac{1}{|x_{1,\perp}|^{\Delta_{1}}|x_{2,\perp}|^{\Delta_{2}}} \left[\widehat{\mathcal{K}}_{\hat{\Delta}} \widehat{f}_{\hat{\Delta},0}(\chi)+ \widehat{\mathcal{K}}_{p-\hat{\Delta}} \widehat{f}_{p-\hat{\Delta},0}(\chi)\right]
\end{align}

\noindent where $\widehat{f}_{\hat{\Delta},0}(\chi)$ is the defect channel scalar block. The definition of the cross-ratio $\chi$ and the expressions for the block and the normalization factor $\widehat{\mathcal{K}}_{\hat{\Delta}}$ are given in Appendix \ref{app:defectblock}. We thus obtain the defect channel partial wave expansion of the contact diagram
\begin{align}
    \label{contactpwexp}
   W_{c}(x_{1},x_{2}) 
   &=\int_{-\infty}^{\infty} d\nu \ \frac{\nu^{2}}{\pi} \ \widehat{C}_{\frac{p}{2}+i\nu} \widehat{C}_{\frac{p}{2}-i\nu} \ b_{1\underline{\hat{\Delta}}} b_{2\underline{\widetilde{\hat{\Delta}}}} \ \widehat{\mathcal{F}}_{\frac{p}{2}+i\nu,0}(x_{1},x_{2})  .
\end{align}

\vskip 4pt
\noindent In the above equation, the contribution from the conformal block and its shadow are identical due to the $\nu$ going to $-\nu$ symmetry of the integral. Using this fact, we can rewrite Eq.\eqref{contactpwexp} as 
\begin{align}
    \label{contactblockinteg}
     & W_{c}(x_{1},x_{2})  =\frac{4}{|x_{1,\perp}|^{\Delta_{1}}|x_{2,\perp}|^{\Delta_{2}}}\int_{-\infty}^{\infty} \frac{d\nu}{2\pi} \ \nu^{2} \ \widehat{C}_{\frac{p}{2}+i\nu} \widehat{C}_{\frac{p}{2}-i\nu} \ b_{1\underline{\hat{\Delta}}} b_{2\underline{\widetilde{\hat{\Delta}}}} \widehat{\mathcal{K}}_{\frac{p}{2}+i\nu} \ \widehat{f}_{\frac{p}{2}+i\nu,0}(\chi) .
\end{align}

\vskip 4pt
\noindent We can now write this as a sum over defect channel conformal blocks by deforming the $\nu$-integration contour. Note that the integral has simple poles at
\begin{align}
    \label{polescontact}
    i\nu= \Delta_{1}+2n-\frac{p}{2}, \ i\nu= \Delta_{2}+2n-\frac{p}{2}, \quad n\in \mathbb{Z}_{\ge 0}. 
\end{align}

\noindent These poles come from the coefficients $b_{1\underline{\hat{\Delta}}},  b_{2\underline{\widetilde{\hat{\Delta}}}}$. They correspond to the defect channel exchange of operators of the form $(\partial_{i}\partial^{i})^{n}\mathcal{O}_{1}$ and $(\partial_{i}\partial^{i})^{n}\mathcal{O}_{2}$, with scaling dimensions $\Delta_{1}+2n$ and $\Delta_{2}+2n$ respectively. Here $\mathcal{O}_{1}, \mathcal{O}_{2}$ are the CFT operators in the $2$-point function, and the derivatives are taken along directions which are transverse to the defect. Then, deforming the $\nu$-contour to pick up the contribution from the above set of poles, we get
\begin{align}
\label{contactdefblockexp}
W_{c}(x_{1},x_{2}) & =\frac{\mathcal{W}_{c}(\chi)}{|x_{1,\perp}|^{\Delta_{1}}|x_{2,\perp}|^{\Delta_{2}}}
\end{align}

\noindent where $\mathcal{W}_{c}(\chi)$ has the following expansion in the defect channel
\begin{align}
\label{contactdefblockexp1}
\mathcal{W}_{c}(\chi) & =\sum_{n=0}^{\infty} \left[\widehat{a}^{(1)}_{n}\widehat{f}_{\Delta_{1}+2n,0}(\chi)+ \widehat{a}^{(2)}_{n}\widehat{f}_{\Delta_{2}+2n,0}(\chi)\right].
\end{align}

\noindent The coefficient $\widehat{a}^{(1)}_{n}$ is given by
\begin{align}
\label{contactdefchnlcoeff}
\widehat{a}^{(1)}_{n}= \frac{(-1)^{n}}{n!}\frac{\pi^{\frac{p}{2}}}{2} \ \frac{\Gamma\left(\Delta_{1}+2n\right)\Gamma\left(\Delta_{1}+n-\frac{p}{2}\right)\Gamma\left(\frac{\Delta_{2}-\Delta_{1}-2n}{2}\right)\Gamma\left(\frac{\Delta_{1}+\Delta_{2}-p}{2}+n\right)}{\Gamma(\Delta_{1})\Gamma(\Delta_{2})\Gamma\left(\Delta_{1}+2n-\frac{p}{2}\right)}
\end{align}

\noindent and $\widehat{a}^{(2)}_{n}$ has the same expression but with $\Delta_{1}$ replaced by $\Delta_{2}$. It is easy to verify the above result for the decomposition coefficients by comparing with the closed-form expression of the contact diagram given in \cite{Gimenez-Grau:2023fcy}.




\subsection{Tree-level scalar exchange: Defect channel}
\label{sec:defchexch}

\begin{figure}[htp]
    \centering
    \includegraphics[width=4.5cm]{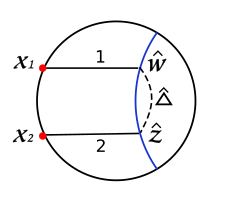}
    \caption{Defect channel Witten exchange diagram.}
    \label{fig:defect_channel_diag}
\end{figure}

We will now consider the diagram shown in Fig \ref{fig:defect_channel_diag}. We take the interaction vertices to be $\int_{\mathrm{AdS}_{p+1}}\phi_{i}\widehat{\phi}, \ i=1,2$, where $\widehat{\phi}$ is a scalar field living on the AdS$_{p+1}$ brane and $\phi_{1},\phi_{2}$ are AdS$_{d+1}$ bulk scalar fields localised on the brane. 

The expression for this diagram is 
\begin{align}
\label{defchspin0exch}
    \widehat{W}_{\hat{\Delta},0}(x_{1},x_{2}) = \int\limits_{AdS_{p+1}}d^{p+1}\hat{w} \ d^{p+1}\hat{z}\ K_{\Delta_{1}}(x_{1},\hat{w}) \widehat{G}_{\hat{\Delta}}(\hat{w},\hat{z})K_{\Delta_{2}}(x_{2},\hat{z})
\end{align}

\noindent where $\widehat{G}_{\hat{\Delta}} (\hat{w},\hat{z})$ is the brane-to-brane propagator for scalar exchange. We will now obtain the conformal block decomposition of this diagram in the defect channel. We employ the spectral and split representation of $\widehat{G}_{\hat{\Delta}} (\hat{w},\hat{z})$ given in Eq.~\eqref{brane2brane}. This allows us to write the Witten diagram as
\begin{align}
\label{defchspin0exch1}
    & \widehat{W}_{\hat{\Delta},0}(x_{1},x_{2}) = \int_{-\infty}^{\infty} d\nu \ \widehat{P}(\nu,\hat{\Delta})  \int\limits_{\partial AdS_{p+1}} d^{p}\hat{x} \bigg[\int\limits_{AdS_{p+1}}d^{p+1}\hat{w}\ K_{\Delta_{1}}(x_{1},\hat{w}) \widehat{K}_{\frac{p}{2}+i\nu}(\hat{x},\hat{w}) \bigg] \nonumber\\
   &\hspace{2.2cm}\times \bigg[\int\limits_{AdS_{p+1}} d^{p+1}\hat{z}\ K_{\Delta_{2}}(x_{2},\hat{z}) \widehat{K}_{\frac{p}{2}-i\nu}(\hat{x},\hat{z}) \bigg].
\end{align}

\noindent Now, we have already noted in the previous section that each of the two $AdS_{p+1}$ integrals appearing in the above expression yields a CFT $2$-point function with one bulk and one defect CFT operator. The integral over the point $\hat{x}$ on the boundary of $AdS_{p+1}$ then gives the defect channel partial wave. We thus get the following partial wave expansion of the defect exchange diagram.
\begin{align}
    \label{defchspin0exch1}
   \widehat{W}_{\hat{\Delta},0}(x_{1},x_{2})
   &=\int_{-\infty}^{\infty} d\nu \ \widehat{P}(\nu,\hat{\Delta})  \ b_{1\underline{\hat{\Delta}}} b_{2\underline{\widetilde{\hat{\Delta}}}} \ \widehat{\mathcal{F}}_{\frac{p}{2}+i\nu,0}(x_{1},x_{2}) \nonumber\\
   &= \frac{2}{|x_{1,\perp}|^{\Delta_{1}}|x_{2,\perp}|^{\Delta_{2}}}\int_{-\infty}^{\infty} d\nu \ \widehat{P}(\nu,\hat{\Delta})  \ b_{1\underline{\hat{\Delta}}} b_{2\underline{\widetilde{\hat{\Delta}}}} \ \widehat{\mathcal{K}}_{\frac{p}{2}+i\nu}\ \widehat{f}_{\frac{p}{2}+i\nu,0}(\chi).
\end{align}

\noindent where in the second line of the above equation we have used the $\nu\leftrightarrow -\nu$ symmetry of the integrand to replace the partial wave with the defect channel block. 

\vskip 4pt
\noindent Let us now express Eq.~\eqref{defchspin0exch1} as a sum over defect channel blocks by deforming the $\nu$ contour and picking up the poles at 
\begin{align}
    \label{nupolesdefch}
    i\nu= \hat{\Delta}-\frac{p}{2}, \ i\nu= \Delta_{1}-\frac{p}{2}+2n, \ i\nu= \Delta_{2}-\frac{p}{2}+2n, \ n\in \mathbb{Z}_{\ge 0}.
\end{align}

\noindent Here, the first pole comes from the measure factor $\widehat{P}(\nu,\hat{\Delta})$. In the boundary CFT, this corresponds to the exchange of a defect scalar primary operator with scaling dimension $\hat{\Delta}$. The second and third set of poles are due to the factors $b_{1\underline{\hat{\Delta}}}$ and $ b_{2\underline{\widetilde{\hat{\Delta}}}}$ respectively. These imply the exchange of transverse derivative operators of the type encountered in the defect channel block decomposition of the contact diagram in the previous section. Evaluating the residues at these poles, we then get,
\begin{align}
    \label{defchspin0exch2}
   \widehat{W}_{\hat{\Delta},0}(x_{1},x_{2})
   &= \frac{\widehat{\mathcal{W}}_{\hat{\Delta},0}(\chi)}{|x_{1,\perp}|^{\Delta_{1}}|x_{2,\perp}|^{\Delta_{2}}} 
\end{align}

\noindent where $\widehat{\mathcal{W}}_{\hat{\Delta},0}(\chi)$ has the following defect channel block decomposition
\begin{align}
    \label{defchspin0exch3}
    \widehat{\mathcal{W}}_{\hat{\Delta},0}(\chi)= \widehat{A}_{\hat{\Delta}} \widehat{f}_{\hat{\Delta},0}(\chi) + \sum_{n=0}^{\infty} \left[\widehat{B}^{(1)}_{n}\widehat{f}_{\Delta_{1}+2n,0}(\chi)+ \widehat{B}^{(2)}_{n}\widehat{f}_{\Delta_{2}+2n,0}(\chi)\right] .
\end{align}

\noindent The coefficients appearing in the above expansion are given by
\begin{align}
    \label{defchcoeffsspin0}
    & \widehat{A}_{\hat{\Delta}} = \frac{\pi^{\frac{p}{2}}\Gamma(\hat{\Delta})\Gamma\left(\frac{\Delta_{1}-\hat{\Delta}}{2}\right)\Gamma\left(\frac{\Delta_{2}-\hat{\Delta}}{2}\right)\Gamma\left(\frac{\Delta_{1}-p+\hat{\Delta}}{2}\right)\Gamma\left(\frac{\Delta_{2}-p+\hat{\Delta}}{2}\right)}{8\Gamma(\Delta_{1})\Gamma(\Delta_{2})\Gamma\left(\hat{\Delta}+1-\frac{p}{2}\right)}, \nonumber\\
    & \widehat{B}^{(1)}_{n} = \frac{(-1)^{n}}{n!} \ \frac{\pi^{\frac{p}{2}}\left(\Delta_{1}\right)_{2n}\Gamma\left(\Delta_{1}+n-\frac{p}{2}\right)\Gamma\left(\frac{\Delta_{2}-\Delta_{1}-2n}{2}\right)\Gamma\left(\frac{\Delta_{1}+\Delta_{2}-p}{2}+n\right)}{2(\hat{\Delta}-\Delta_{1}-2n)(\hat{\Delta}+\Delta_{1}+2n-p)\Gamma(\Delta_{2})\Gamma\left(\Delta_{1}+2n-\frac{p}{2}\right)}
\end{align}

\noindent and $\widehat{B}^{(2)}_{n}$ is the same as $\widehat{B}^{(1)}_{n}$ but with $\Delta_{1}$ replaced by $\Delta_{2}$. Note that in obtaining the above result, we have assumed $\Delta_{1}\ne \Delta_{2}$. For $\Delta_{1} = \Delta_{2}=\Delta$, we have double poles in the integrand of Eq.~\eqref{defchspin0exch1} at $i\nu=\Delta+2n-\frac{p}{2}$. Taking this into account, the resulting block expansion becomes
\begin{align}
\widehat{\mathcal{W}}_{\hat{\Delta},0}(\chi)  = \frac{1}{|x_1^i|^{\Delta_1} |x_2^i|^{\Delta_2}} \bigg[ \hat{\alpha}_{\hat{\Delta}}  \widehat{f}_{\hat{\Delta},0}(\chi) + \sum_{n =0}^{\infty} & \left( \frac{2(\gamma-H_n) \ \hat{\beta}_n+\partial_n \hat{\beta}_n}{\Gamma(n+1)^2} \  \widehat{f}_{\Delta+2n,0}(\chi) \right.  \notag \\ 
&\left. + \frac{\hat{\beta}_n}{\Gamma(n+1)^2}  \ \partial_n \widehat{f}_{\Delta+2n,0}(\chi)\right)\bigg] 
\end{align}

\noindent where, $\gamma$ and $H_n$ are Euler's constant and the Harmonic number, respectively, and
\begin{align}
\hat{\alpha}_{\hat{\Delta}}  &= \frac{\pi^{\frac{p}{2}}\Gamma(\hat{\Delta})\Gamma\left(\frac{\Delta-\hat{\Delta}}{2}\right)^2\Gamma\left(\frac{\Delta-p+\hat{\Delta}}{2}\right)^2}{16\Gamma(\Delta)^2\Gamma\left(\hat{\Delta}+1-\frac{p}{2}\right)}, \nonumber \\
\hat{\beta}_n &= -\frac{\pi^{\frac{p}{2}}\left(\Delta\right)_{2n}\Gamma\left(\Delta+n-\frac{p}{2}\right)\Gamma\left(\frac{2\Delta-p}{2}+n\right)}{2(\hat{\Delta}-\Delta-2n)(\hat{\Delta}+\Delta+2n-p)\Gamma(\Delta)\Gamma\left(\Delta+2n-\frac{p}{2}\right)} .
\end{align} 

\subsection{Tree level scalar exchange: bulk channel}
\label{sec:bulk_exch_spin0}

\begin{figure}[htp]
    \centering
    \includegraphics[width=4.0cm]{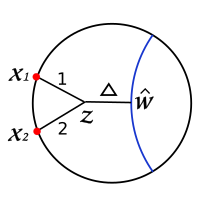}
    \caption{Bulk channel Witten exchange diagram.}
    \label{fig:bulk_channel_diag}
\end{figure}

In this section, we consider a $2$-point Witten diagram, displayed in Fig \ref{fig:bulk_channel_diag}, which arises from a cubic coupling $\int_{\mathrm{AdS}_{d+1}} \phi_{1}\phi_{2}\phi$ involving three bulk scalar fields and a one-point vertex $\int_{\mathrm{AdS}_{p+1}}\phi$ on the brane. The integral expression for this diagram is

\begin{align}
\label{bulkchspin0exch}
    W_{\Delta,0}(x_{1},x_{2}) = \int\limits_{AdS_{d+1}}d^{d+1}z\int\limits_{AdS_{p+1}}d^{p+1}\hat{w}\ K_{\Delta_{1}}(x_{1},z)K_{\Delta_{2}}(x_{2},z) G_{\Delta}(z,\hat{w}) ,
\end{align}

\noindent where $G_{\Delta}(z,\hat{w})$ is the scalar bulk-to-brane propagator. We will now derive the bulk channel decomposition of this diagram. Employing the split representation of the bulk-to-brane propagator, we get
\begin{align}
\label{bulkchspin0exch1}
 W_{\Delta,0}(x_{1},x_{2}) &= \int_{-\infty}^{\infty} d\nu \ P(\nu,\Delta)  \int\limits_{\partial AdS_{d+1}} d^{d}x \bigg[\int\limits_{AdS_{p+1}}d^{p+1}\hat{w}\ K_{\frac{d}{2}-i\nu}(x,\hat{w})\bigg] \times\nonumber\\
 &\hspace{1.0cm}\bigg[\int\limits_{AdS_{d+1}}d^{d+1}z \ K_{\Delta_{1}}(x_{1},z)K_{\Delta_{2}}(x_{2},z)K_{\frac{d}{2}+i\nu}(x,z)\bigg]. 
\end{align}

\noindent Now, in the above equation, the $AdS_{d+1}$ integral gives
\begin{align}
    \label{3ptinteg}
    &\int\limits_{AdS_{d+1}}d^{d+1}z \ K_{\Delta_{1}}(x_{1},z)K_{\Delta_{2}}(x_{2},z)K_{\frac{d}{2}+i\nu}(x,z) = \lambda_{12\underline{\Delta}} \langle \mathcal{O}_{\Delta_{1}}(x_{1})\mathcal{O}_{\Delta_{2}}(x_{2})\mathcal{O}_{\frac{d}{2}+i\nu}(x)\rangle 
\end{align}

\noindent where $\langle \mathcal{O}_{\Delta_{1}}\mathcal{O}_{\Delta_{2}}\mathcal{O}_{\frac{d}{2}+i\nu}\rangle$ is the $3$-point function conformal structure for scalar operators in a CFT without defects. The coefficient $\lambda_{12\underline{\Delta}}$ is
\begin{align}
    \label{3ptcoeff}
    \lambda_{12\underline{\Delta}} =\pi^{\frac{d}{2}}\frac{\Gamma\left(\frac{\Delta_{1}+\Delta_{2}-\underline{\Delta}}{2}\right)\Gamma\left(\frac{\Delta_{12}+\underline{\Delta}}{2}\right)\Gamma\left(\frac{\underline{\Delta}-\Delta_{12}}{2}\right)\Gamma\left(\frac{\Delta_{1}+\Delta_{2}-d+\underline{\Delta}}{2}\right)}{2\Gamma(\Delta_{1})\Gamma(\Delta_{2})\Gamma\left(\underline{\Delta}\right)}, \quad  \underline{\Delta}=\frac{d}{2}+i\nu. 
\end{align}

\noindent The $AdS_{p+1}$ integral in Eq.~\eqref{bulkchspin0exch1} yields
\begin{align}
    \label{1ptinteg}
    \int\limits_{AdS_{p+1}}d^{p+1}\hat{w} \ K_{\frac{d}{2}-i\nu}(x,\hat{w})= a_{\underline{\widetilde{\Delta}}} \langle \hspace{-0.09cm}\langle \mathcal{O}_{\frac{d}{2}-i\nu}(x)\rangle\hspace{-0.09cm}\rangle
\end{align}

\noindent where $\langle \hspace{-0.09cm}\langle \mathcal{O}_{\frac{d}{2}-i\nu}(x)\rangle\hspace{-0.09cm}\rangle$ is the kinematic structure of the one-point function of a scalar CFT operator in the presence of the defect. The one-point function coefficient $a_{\underline{\widetilde{\Delta}}}$ is
\begin{align}
    \label{1ptcoeff}
    a_{\underline{\widetilde{\Delta}}}= \pi^{\frac{p}{2}}\frac{\Gamma\left(\frac{\underline{\widetilde{\Delta}}}{2}\right)\Gamma\left(\frac{\underline{\widetilde{\Delta}}-p}{2}\right)}{2\Gamma\left(\underline{\widetilde{\Delta}}\right)}, \quad \underline{\widetilde{\Delta}} = \frac{d}{2}-i\nu. 
\end{align}

\noindent Gathering the above results, the bulk exchange Witten diagram can then be written as
\begin{align}
\label{bulkchspin0exch2}
 W_{\Delta,0}(x_{1},x_{2}) &= \int_{-\infty}^{\infty} d\nu \ P(\nu,\Delta)  \lambda_{12\underline{\Delta}} a_{\underline{\widetilde{\Delta}}}\int d^{d}x \ \langle \mathcal{O}_{\Delta_{1}}(x_{1})\mathcal{O}_{\Delta_{2}}(x_{2})\mathcal{O}_{\frac{d}{2}+i\nu}(x)\rangle \langle \hspace{-0.09cm}\langle \mathcal{O}_{\frac{d}{2}-i\nu}(x)\rangle\hspace{-0.09cm}\rangle\nonumber\\
 & = \int_{-\infty}^{\infty} d\nu \ P(\nu,\Delta)  \lambda_{12\underline{\Delta}} a_{\underline{\widetilde{\Delta}}} \ \mathcal{F}_{\frac{d}{2}+i\nu,0}(x_{1},x_{2})
\end{align}

\noindent where $\mathcal{F}_{\frac{d}{2}+i\nu,0}(x_{1},x_{2})$ is the bulk channel partial wave for scalar exchange, which has the integral representation
\begin{align}
    \label{bulkchpwintrep}
    \mathcal{F}_{\Delta,0}(x_{1},x_{2})= \int d^{d}x \ \langle \mathcal{O}_{\Delta_{1}}(x_{1})\mathcal{O}_{\Delta_{2}}(x_{2})\mathcal{O}_{\Delta}(x)\rangle \langle \hspace{-0.09cm}\langle \mathcal{O}_{d-\Delta}(x)\rangle\hspace{-0.09cm}\rangle. 
\end{align}

\noindent The partial wave $\mathcal{F}_{\Delta,0}(x_{1},x_{2})$ is related to bulk channel conformal blocks as follows
\begin{align}
    \label{bulkchpw2}
    \mathcal{F}_{\Delta,0}(x_{1},x_{2}) = \frac{\xi^{-\frac{\Delta_{1}+\Delta_{2}}{2}}}{|x_{1,\perp}|^{\Delta_{1}}|x_{2,\perp}|^{\Delta_{2}}}\left[\mathcal{K}_{\Delta} f_{\Delta,0}(\xi,\eta)+ \widetilde{\mathcal{K}}_{d-\Delta} f_{d-\Delta,0}(\xi,\eta)\right]
\end{align}

\noindent where the scalar bulk channel block $f_{\Delta,0}(\xi,\eta)$ and the normalization factors $\mathcal{K}_{\Delta}, \widetilde{\mathcal{K}}_{d-\Delta}$ are defined in Appendix \ref{app:bulkblock}. Then, using the $\nu$ to $-\nu$ symmetry of the integrand in Eq.~\eqref{bulkchspin0exch2} to replace the partial wave with the block, we get
\begin{align}
\label{bulkchspin0exch3}
 W_{\Delta,0}(x_{1},x_{2}) & = \frac{\xi^{-\frac{\Delta_{1}+\Delta_{2}}{2}}}{|x_{1,\perp}|^{\Delta_{1}}|x_{2,\perp}|^{\Delta_{2}}}\int_{-\infty}^{\infty} d\nu \ 2P(\nu,\Delta)  \lambda_{12\underline{\Delta}} a_{\underline{\widetilde{\Delta}}}  \ \mathcal{K}_{\frac{d}{2}+i\nu,0} f_{\frac{d}{2}+i\nu,0}(x_{1},x_{2}).
\end{align}

\vskip 4pt
\noindent Now, to recast the above as a sum over bulk channel blocks, we deform the $\nu$ contour and gather the contribution from the following set of simple poles at
\begin{align}
\label{bulkchpoles}
    i\nu=\Delta-\frac{d}{2}, \quad i\nu =\Delta_{1}+\Delta_{2}+2n-\frac{d}{2}, \ n\in \mathbb{Z}_{\ge 0}. 
\end{align}

\noindent In Eq.~\eqref{bulkchpoles}, the first pole, which originates from the measure factor $P(\nu,\Delta)$, signifies the exchange of a CFT bulk scalar primary operator with dimension $\Delta$. The second set of poles comes from the $3$-point function coefficient $\lambda_{12\underline{\Delta}}$. These correspond to the exchange of double-twist operators of the form $\mathcal{O}_{1}(\partial^{2})^{n}\mathcal{O}_{2}$, having dimensions $\Delta_{1}+\Delta_{2}+2n$. Taking the residues at these poles, we obtain
\begin{align}
    \label{bulkchspin0exch4}
    W_{\Delta,0}(x_{1},x_{2}) & = \frac{\mathcal{W}_{\Delta,0}(\xi,\eta)}{|x_{1,\perp}|^{\Delta_{1}}|x_{2,\perp}|^{\Delta_{2}}}
\end{align}

\noindent where $\mathcal{W}_{\Delta,0}(\xi,\eta)$ has the following expansion in terms of bulk channel blocks
\begin{align}
    \label{bulkchspin0exch5}
    \mathcal{W}_{\Delta,0}(\xi,\eta) = \xi^{-\frac{\Delta_{1}+\Delta_{2}}{2}}\left[A_{\Delta}f_{\Delta,0}(\xi,\eta)+ \sum_{n=0}^{\infty} B_{n} f_{\Delta_{1}+\Delta_{2}+2n,0}(\xi,\eta)\right]
\end{align}

\noindent and the coefficients  $A_{\Delta}, B_{n}$ are given by
\begin{align}
    \label{bulchspin0exchcoeffs}
    & A_{\Delta} = \frac{\pi^{\frac{p}{2}}\Gamma\left(\frac{\Delta_{1}+\Delta_{2}-\Delta}{2}\right)\Gamma\left(\frac{\Delta_{1}+\Delta_{2}+\Delta-d}{2}\right)\Gamma\left(\frac{\Delta+\Delta_{12}}{2}\right)\Gamma\left(\frac{\Delta-\Delta_{12}}{2}\right)\Gamma\left(\frac{\Delta-p}{2}\right)\Gamma\left(\frac{\Delta}{2}\right)}{4(2\Delta-d)\Gamma(\Delta_{1})\Gamma(\Delta_{2})\Gamma(\Delta)\Gamma\left(\Delta-\frac{d}{2}\right)}, \\
    & \label{Bn} B_{n} = \frac{(-1)^{n}2^{-\Delta_{1}-\Delta_{2}-2n}\pi^{\frac{p+1}{2}}(\Delta_{1})_{n}(\Delta_{2})_{n}\Gamma\left(\frac{\Delta_{1}+\Delta_{2}+2n-p}{2}\right)}{n!(\Delta_{1}+\Delta_{2}+2n+\Delta-d)(\Delta-\Delta_{1}-\Delta_{2}-2n)\left(\Delta_{1}+\Delta_{2}+n-\frac{d}{2}\right)_{n}\Gamma\left(\frac{\Delta_{1}+\Delta_{2}+2n+1}{2}\right)} .
\end{align}

\section{Spinning exchange}\label{2pt_spin_exch}

In this section, we consider the case of spinning exchange.  A tree-level $2$-point diagram involving transverse spin $s$ exchange can be related to a scalar exchange diagram, but with the external dimensions shifted as $(\Delta_{1}+s,\Delta_{2}+s)$, up to a multiplicative factor of $\eta^{s}$ \cite{Gimenez-Grau:2023fcy, Alday:2024srr}. Here, $\eta$ is the cross-ratio defined in Eq.~\eqref{crossratios}. The defect channel block expansion of a transverse spin $s$ exchange diagram is then straightforward. This is because, when expressed in terms of the cross ratios $(\chi,\eta)$, the $\eta$-dependence of the defect channel blocks, given in Eq.~\eqref{defch1}, enters only through the Gegenbauer polynomials $C_{s}^{(\frac{q}{2}-1)}(\eta)$. Therefore, we will not study this case in further detail here. We will instead consider spinning exchange in the bulk channel. For simplicity, we will only deal with the case of spin $\ell=2$ bulk channel exchange, leaving the generalization to arbitrary spin $\ell$ for future work.

\subsection{Bulk channel: Spin $2$ exchange}

We consider a $2$-point tree-level Witten diagram involving external scalars and the exchange of a massive spin-$2$ field in the bulk channel.  Let the bulk cubic coupling be of the form
\begin{align}
  \int\limits_{\mathrm{AdS}_{d+1}} \phi_{1}\nabla^{\mu}\nabla^{\nu}\phi_{2}h_{\mu\nu} \ ,   
\end{align}

\noindent where $\phi_{1},\phi_{2}$ are scalar fields dual to the CFT operators $\mathcal{O}_{1},\mathcal{O}_{2}$ and $h_{\mu\nu}$ denotes a massive spin $2$ field. The interaction vertex for the absorption of $h_{\mu\nu}$ on the brane will be taken to be of the form
\begin{align}
    \int\limits_{\mathrm{AdS}_{p+1}} g^{\hat{\alpha}\hat{\beta}} h_{\hat{\alpha}\hat{\beta}} \ ,  
\end{align}

\noindent where the  hatted indices $(\hat{\alpha},\hat{\beta})$ run along the AdS$_{p+1}$ brane directions. Then the spin-$2$ exchange bulk channel Witten diagram is given by the following integral
\begin{align}
    \label{bulkchspin2exch}
    W_{\Delta,2}(x_{1},x_{2}) = \int\limits_{\mathrm{AdS}_{d+1}}d^{d+1}z\int\limits_{\mathrm{AdS}_{p+1}} d^{p+1}\hat{w} \ V_{\mu\nu}(x_{1},x_{2}) G^{\mu\nu,\hat{\alpha}\hat{\beta}}_{\Delta,2}(z,\hat{w}) g_{\hat{\alpha}\hat{\beta}} . 
\end{align}

\noindent Here $G^{\mu\nu,\hat{\alpha}\hat{\beta}}_{\Delta,2}(z,\hat{w})$ is the bulk-to-brane propagator for a massive spin-$2$ field. The indices $(\mu,\nu)$ indicate the directions in AdS$_{d+1}$, while $(\hat{\alpha},\hat{\beta})$ signify the directions along the AdS$_{p+1}$ brane. The vertex factor $V_{\mu\nu}(x_{1},x_{2})$ is given by \cite{DHoker:1999kzh, Gimenez-Grau:2023fcy, Chen:2026ium}
\begin{align}
    \label{bulkchspin2exch1}
    V_{\mu\nu}(x_{1},x_{2},z)& = \frac{1}{2}\nabla_{(\mu}K_{\Delta_{1}}(x_{1},z)\nabla_{\nu)}K_{\Delta_{1}}(x_{2},z) -\frac{g_{\mu\nu}}{2}\nabla_{\rho}K_{\Delta_{1}}(x_{1},z)\nabla^{\rho}K_{\Delta_{2}}(x_{2},z)\nonumber\\
    &-\frac{g_{\mu\nu}}{4}\left(\Delta_{1}(\Delta_{1}-d)+\Delta_{2}(\Delta_{2}-d)-\Delta(\Delta-d)\right)K_{\Delta_{1}}(x_{1},z)K_{\Delta_{2}}(x_{2},z) .
\end{align}

\noindent This diagram has been analyzed in \cite{Gimenez-Grau:2023fcy, Chen:2026ium}, where it was shown that it can be evaluated in closed form as a finite sum of contact Witten diagrams if $\Delta_{1}+\Delta_{2}-\Delta\in 2\mathbb{Z}_{\ge0}$. We are interested in obtaining the bulk channel block decomposition of this diagram for general values of the external and exchanged scaling dimensions\footnote{The spin-$2$ bulk channel exchange diagram for generic values of the scaling dimensions has been computed in Mellin space in \cite{Chen:2023yvw}.}. This can be carried out efficiently using the embedding space formalism developed in \cite{Costa:2014kfa} for spinning AdS propagators. We relegate the details of the derivation to Appendix \ref{app:spin2exch}. The final result is
\begin{align}
    \label{bulkchspin2exch2a}
    W_{\Delta,2}(x_{1},x_{2})
     & = \frac{\mathcal{W}_{\Delta,2}(\xi,\eta)}{|x_{1,\perp}|^{\Delta_{1}}|x_{2,\perp}|^{\Delta_{2}}}
\end{align}

\noindent where $\mathcal{W}_{\Delta,2}(\xi,\eta)$ has the following bulk channel block decomposition
\begin{align}
    \label{bulkchspin2exch2}
    \mathcal{W}_{\Delta,2}(\xi,\eta)
     & = \xi^{-\frac{\Delta_{1}+\Delta_{2}}{2}}\big[A_{n,2} + \sum_{\ell=0}^{2} B_{n,\ell} \ f_{\Delta_{1}+\Delta_{2}+2n+\ell,\ell}(\xi,\eta)\big] .
\end{align}

\noindent The coefficients $A_{\Delta,2}$ and $B_{n,\ell}$ are given by
\begin{align}
    \label{spin2bulkchcoeffs}
    & A_{\Delta,2}= \frac{\pi^{\frac{p}{2}}\Delta(d-\Delta+1)\Gamma\left(\frac{\Delta}{2}\right)\Gamma\left(\frac{\Delta-p}{2}\right)\Gamma\left(\frac{\Delta_{1}+\Delta_{2}-\Delta+2}{2}\right)\Gamma\left(\frac{\Delta_{1}+\Delta_{2}-d+\Delta+2}{2}\right)\Gamma\left(\frac{\Delta\pm\Delta_{12}+2}{2}\right)}{4\left(\frac{d-1}{2}\right)_{2}(\Delta-1)(\Delta+1)(\Delta-d+1)\Gamma\left(\Delta-\frac{d}{2}+1\right)\Gamma(\Delta_{1})\Gamma(\Delta_{2})\Gamma(\Delta_{3})},
\end{align}

\begin{align}
    \label{spin2bulkchcoeffs1}
     B_{n,0} &= \frac{(-1)^{n}\pi^{\frac{p+1}{2}}2^{-\Delta_{1}-\Delta_{2}-2n}(p+1)(\Delta_{1}+\Delta_{2}+2n-d)(\Delta_{1})_{n}(\Delta_{2})_{n}\Gamma\left(\frac{\Delta_{1}+\Delta_{2}+2n-p}{2}\right)}{n!(\Delta_{1}+\Delta_{2}+n-\frac{d}{2})_{n}\Gamma\left(\frac{\Delta_{1}+\Delta_{2}+2n+1}{2}\right)} \times\nonumber\\
     & \times \bigg[ (\Delta_{2}+n) (d (\Delta_{2}+n+1)-2 n (\Delta_{1}+n)) (d-2 (\Delta_{1}+\Delta_{2}+n)) \nonumber\\
    &\hspace{0.5cm}-\frac{d(\Delta_{2})_{2}(\Delta_{1}+\Delta_{2}+2n)_{2}}{d+1}\bigg]\alpha_{0}(\nu=-i(\Delta_{1}+\Delta_{2}+2n-h)), 
\end{align}

\noindent where
\begin{align}
    & \alpha_{0}(\nu)= \frac{1}{d(\Delta-1)(d-\Delta-1)(\nu^{2}+(h+1)^{2})}- \frac{1}{d\Delta(d-\Delta)(\nu^{2}+h^{2})},  
\end{align}

\begin{align}
    \label{spin2bulkchcoeffs2}
     B_{n,2} &= \frac{(-1)^{n}2^{-\Delta_{1}-\Delta_{2}-2n-3}\pi^{\frac{p+1}{2}}(\Delta_{1}+\Delta_{2}+2n+1-d)(\Delta_{1}+2)_{n}(\Delta_{2}+2)_{n}}{n!(\Delta_{1}+\Delta_{2}+2n+2+\Delta-d)(\Delta-\Delta_{1}-\Delta_{2}-2n-2)\left(\Delta_{1}+\Delta_{2}+2+n-\frac{d}{2}\right)_{n}} \times \nonumber\\
    & \times \frac{\Gamma\left(\frac{\Delta_{1}+\Delta_{2}+2n+2-p}{2}\right)}{(\Delta_{1}+\Delta_{2}+2n+1)(\Delta_{1}+\Delta_{2}+2n+3-d)\left(\Delta_{1}+\Delta_{2}+n-\frac{d}{2}\right)_{n}\Gamma\left(\frac{\Delta_{1}+\Delta_{2}+2n+5}{2}\right)}. 
\end{align}


\section{Crossed channel decompositions}\label{2pt_exch_cross}

In this section, we study the crossed channel block decompositions of $2$-point contact and tree-level exchange Witten diagrams. This is facilitated by the application of the equation of motion identities that relate exchange diagrams to contact diagrams. We consider the bulk channel block decomposition of the contact diagram in \ref{subsec:croschcontact}. This is followed by the discussion of a recursive method for obtaining the bulk channel decomposition of the defect channel exchange diagram in subsection \ref{subsec:bulktodef}, and the defect channel expansion of the bulk channel diagram in subsection \ref{subsec:deftobulk}.

\subsection{Contact diagram}
\label{subsec:croschcontact}

Here we perform the bulk channel expansion of the contact diagram $W_{c}(x_{1},x_{2})$ given in Eq.~\eqref{contact}. For this, let us first consider the bulk channel scalar exchange diagram $W_{\Delta,0}(x_{1},x_{2})$ given in Eq.~\eqref{bulkchspin0exch}. The exchange diagram can be related to the contact diagram by acting with the bulk channel Casimir operator as follows
\begin{align}
    \label{bulkeom}
    \left(\frac{1}{2}(\mathcal{J}_{1}+\mathcal{J}_{2})^{2}+\Delta(\Delta-d)\right)W_{\Delta,0}(x_{1},x_{2}) = W_{c}(x_{1},x_{2}) 
\end{align}

\noindent where $\mathcal{J}_{1}, \mathcal{J}_{2}$ are conformal symmetry generators which act on the boundary points $x_{1},x_{2}$ respectively. Relations of this form have been derived for Witten diagrams in holographic CFTs without defects in \cite{Zhou:2018sfz} and for BCFTs in \cite{Mazac:2018biw}. The generalization to the case of co-dimension $q>1$ defects that we employ here is straightforward, and we will not derive it here in detail. The above relation arises because applying the Casimir on the exchange diagram amounts to the action of the $AdS_{d+1}$ Laplacian on the bulk-to-brane propagator $G_{\Delta}(z,\hat{w})$ inside the integral in Eq.~\eqref{bulkchspin0exch}. Then, using the equation of motion satisfied by $G_{\Delta}(z,\hat{w})$ yields a delta function which enables us to trivially perform the integral over the $AdS_{d+1}$ bulk point $z$. The remaining integral over the $AdS_{p+1}$ point $\hat{w}$ then gives the contact diagram.     

\vskip 4pt
We can express Eq.~\eqref{bulkeom} in terms of functions of cross-ratios as
\begin{align}
    \label{bulkeom1}
    \mathcal{D}_{\mathrm{bulk}}(\mathcal{W}_{\Delta,0}(\xi,\eta))= \mathcal{W}_{c}(\xi,\eta)
\end{align}

\noindent where $\mathcal{D}_{\mathrm{bulk}}$ is a second order differential operator. The explicit form of this operator is given in Appendix \ref{app:bulkchcasdiff}. Now, to get the bulk channel conformal block expansion of $\mathcal{W}_{c}(\xi,\eta)$, we use the bulk channel decomposition of the exchange diagram obtained in Eq.\eqref{bulkchspin0exch5}. The action of $\mathcal{D}_{\mathrm{bulk}}$ on the blocks appearing in Eq.\eqref{bulkchspin0exch5} is
\begin{align}
    \label{buleom2}
    &\mathcal{D}_{\mathrm{bulk}}\left[\xi^{-\frac{\Delta_{1}+\Delta_{2}}{2}}f_{\Delta,0}(\xi,\eta)\right] =0  \nonumber\\
    & \mathcal{D}_{\mathrm{bulk}}\left[\xi^{-\frac{\Delta_{1}+\Delta_{2}}{2}}f_{\Delta_{1}+\Delta_{2}+2n,0}(\xi,\eta)\right]\nonumber\\
    &= (\Delta(\Delta-d)-(\Delta_{1}+\Delta_{2}+2n)(\Delta_{1}+\Delta_{2}+2n-d)) \xi^{-\frac{\Delta_{1}+\Delta_{2}}{2}}f_{\Delta_{1}+\Delta_{2}+2n,0}(\xi,\eta) .
\end{align}

\noindent where we have used the fact that the bulk channel blocks are eigenfunctions of the bulk channel Casimir. Consequently, the bulk channel decomposition of $\mathcal{W}_{\mathrm{contact}}(\xi,\eta)$ is given by
\begin{align}
    \label{crossedchcontact}
    \mathcal{W}_{c}(\xi,\eta)=\xi^{-\frac{\Delta_{1}+\Delta_{2}}{2}} \sum_{n=0}^{\infty} c_{n} f_{\Delta_{1}+\Delta_{2}+2n,0}(\xi,\eta)
\end{align}

\noindent where the coefficient $c_{n}$ is
\begin{align}
    \label{cn}
    c_{n}=  \frac{(-1)^{n}2^{-\Delta_{1}-\Delta_{2}-2n}\pi^{\frac{p+1}{2}}(\Delta_{1})_{n}(\Delta_{2})_{n}\Gamma\left(\frac{\Delta_{1}+\Delta_{2}+2n-p}{2}\right)}{n!\left(\Delta_{1}+\Delta_{2}+n-\frac{d}{2}\right)_{n}\Gamma\left(\frac{\Delta_{1}+\Delta_{2}+2n+1}{2}\right)} .
\end{align}


\subsection{Defect channel expansion of bulk exchange Witten diagram}
\label{subsec:bulktodef}

We consider now the defect channel decomposition of tree-level bulk channel exchange Witten diagrams. Assuming the external scaling dimensions to be unequal, this expansion takes the following form. 
\begin{align}
    \label{blkexchdefblock}
    \mathcal{W}_{\Delta,\ell}(\xi,\eta) =\sum_{n,s=0}^{\infty} \widehat{A}^{(1)}_{n,s} \widehat{f}_{\Delta_{1}+2n+s,s}(\xi,\eta)+ \sum_{n,s=0}^{\infty} \widehat{A}^{(2)}_{n,s} \widehat{f}_{\Delta_{2}+2n+s,s}(\xi,\eta) .
\end{align}

\noindent An efficient way to compute the coefficients $\widehat{A}^{(1)}_{n,s}, \widehat{A}^{(2)}_{n,s}$ is to apply the recursive method developed in \cite{Zhou:2018sfz} for studying the crossed channel decomposition of Witten diagrams for CFTs without defects and for boundary CFTs in \cite{Mazac:2018biw}. Let us briefly review the essential elements of this method, but now adapted to the case of higher codimension defects. 

\vskip 4pt
We first apply the bulk channel Casimir to the exchange diagram. If the exchanged spin in the bulk channel is non-zero, this will yield in general, a sum of contact diagrams. In terms of cross-ratios, we get a relation of the form
\begin{align}
    \label{bulkeom3}
    \mathcal{D}_{\mathrm{bulk}}(\mathcal{W}_{\Delta,\ell}(\xi,\eta))= \sum_{k} \mathbb{C}_{k}\mathcal{W}_{c,k}(\xi,\eta)
\end{align}

\noindent where the rhs of the above equation involves contact diagrams coming from interaction vertices with $\ell$ derivatives. The above combination of contact diagrams can be expanded in the defect channel as
\begin{align}
    \label{contactdefexp}
    \sum_{k} \mathbb{C}_{k}\mathcal{W}_{c,k}(\xi,\eta) =\sum_{s=0}^{\ell}\sum_{n=0}^{\infty}\left(\widehat{a}^{(1)}_{n,s} \widehat{f}_{\Delta_{1}+2n+s,s}(\xi,\eta)+\widehat{a}^{(2)}_{n,s} \widehat{f}_{\Delta_{2}+2n+s,s}(\xi,\eta)\right)
\end{align}

\noindent In the lhs of Eq.~\eqref{bulkeom3}, we insert the defect channel expansion of $\mathcal{W}_{\Delta,\ell}(\xi,\eta)$. The action of $\mathcal{D}_{\mathrm{bulk}}$ on the defect channel blocks can be written as a finite linear combination of defect channel blocks with shifted dimensions and spins. Then comparing this with the crossed channel expansion of the contact diagrams in Eq.~\eqref{contactdefexp} yields a set of finite term linear recursion relations for coefficients $\widehat{A}^{(1)}_{n,s}, \widehat{A}^{(2)}_{n,s}$ in the defect channel decomposition of the exchange diagram. These recursion relations can be solved provided we are given the coefficients $\widehat{a}^{(1)}_{n,s},\widehat{a}^{(2)}_{n,s}$ for the contact diagrams, and the seed coefficients $\widehat{A}^{(1)}_{0,s}, \widehat{A}^{(2)}_{0,s}$. We will now perform this analysis in more detail in the simpler case of scalar exchange in the bulk channel. 

\subsubsection{Scalar exchange in bulk channel}
\label{subsec:bulkl0todef}

Using the explicit expression for the defect channel block given in Appendix \ref{app:defectblock}, we find that the action of the differential operator $\mathcal{D}_{\mathrm{bulk}}$ on the defect block is  
\begin{align}
\label{bulkCasdefblock}
\mathcal{D}_{bulk} \left[\widehat{f}_{\hat{\Delta},s} (\xi,\eta) \right] & = \widehat{\mathcal{R}}^{(1)}_{\hat{\Delta},s} \widehat{f}_{\hat{\Delta}-1,s-1}(\xi,\eta)+\widehat{\mathcal{R}}^{(2)}_{\hat{\Delta},s}\widehat{f}_{\hat{\Delta}-1,s+1}(\xi,\eta)+ \widehat{\mathcal{R}}^{(3)}_{\hat{\Delta},s}\widehat{f}_{\hat{\Delta}+1,s-1}(\xi,\eta)\nonumber\\
    & + \widehat{\mathcal{R}}^{(4)}_{\hat{\Delta},s} \widehat{f}_{\hat{\Delta}+1,s+1}(\xi,\eta)+\widehat{\mathcal{R}}^{(5)}_{\hat{\Delta},s} \widehat{f}_{\hat{\Delta},s}(\xi,\eta) .
\end{align}

\noindent where the coefficients $\widehat{\mathcal{R}}^{(i)}$ are given by
\begin{align}
\label{ricoeffs}
& \widehat{\mathcal{R}}^{(1)}_{\hat{\Delta},s} =-\frac{s(q+s-3)(\Delta_{1}-\hat{\Delta}+2-q-s)(\Delta_{2}-\hat{\Delta}+2-q-s)}{(q+2s-2)(q+2s-4)}, \nonumber\\
& \mathcal{R}^{(2)}_{\hat{\Delta},s} = -(\Delta_{1}+s-\hat{\Delta})(\Delta_{2}+s-\hat{\Delta}), \nonumber\\
& \mathcal{R}^{(3)}_{\hat{\Delta},s} = -\frac{4 s \hat{\Delta}  (\hat{\Delta} +1-p)(q+s-3)(\Delta_{1}+\hat{\Delta}+2-s-d)(\Delta_{2}+\hat{\Delta}+2-s-d)}{(2 \hat{\Delta} +2-p) (2 \hat{\Delta}-p )(q+2s-2)(q+2s-4)}, \nonumber\\
& \mathcal{R}^{(4)}_{\hat{\Delta},s} = -\frac{4 \hat{\Delta}  (\hat{\Delta} +1-p)(\Delta_{1}+\hat{\Delta}+s-p)(\Delta_{2}+\hat{\Delta}+s-p)}{(2 \hat{\Delta} +2-p) (2 \hat{\Delta}-p )}, \nonumber\\
& \mathcal{R}^{(5)}_{\hat{\Delta},s} = 2\hat{\Delta}(\hat{\Delta}-p) +2s(q+s-2) + d(\Delta_{1}+\Delta_{2})-\Delta^{2}_{1}-\Delta^{2}_{2}+ \Delta(\Delta-d).
\end{align}

\noindent It should be noted that the relation in Eq.~\eqref{bulkCasdefblock} holds for generic values of $\hat{\Delta}$. It is easy to verify that taking $\eta=1,q=1,s=0$ simplifies the above recurrence relation to one which involves only three terms. This case agrees with the action of bulk channel Casimir on boundary channel blocks, which was derived for BCFTs in \cite{Mazac:2018biw}. But for our purposes, we only need to consider the action of $\mathcal{D}_{\mathrm{bulk}}$ on defect blocks with dimensions $\hat{\Delta}=\Delta_{1}+2n+s$, and $\hat{\Delta}=\Delta_{2}+2n+s$. We will henceforth use the following notation
\begin{align}
    \widehat{\mathcal{R}}^{(\alpha,i)}_{n,s} \equiv \widehat{\mathcal{R}}^{(\alpha, i)}_{\Delta_{\alpha}+2n,s}, \quad \alpha=1,2, \quad i=1,\cdots,5.
\end{align}

\noindent Using Eq.~\eqref{bulkCasdefblock}, the action of $\mathcal{D}_{\mathrm{bulk}}$ on the scalar exchange bulk channel diagram can then be written as
    \begin{align}
    \label{blkexchl0defblock}
    \mathcal{D}_{\mathrm{bulk}}\left[\mathcal{W}_{\Delta,0}(\xi,\eta)\right]&=\sum_{\alpha=1}^{2}\sum_{n,s=0}^{\infty} \bigg(\widehat{\mathcal{R}}^{(\alpha,1)}_{n,s+1}\widehat{A}^{(\alpha)}_{n,s+1}+\widehat{\mathcal{R}}^{(\alpha,2)}_{n+1,s-1}\widehat{A}^{(\alpha)}_{n+1,s-1}+ \widehat{\mathcal{R}}^{(\alpha,3)}_{n-1,s+1}\widehat{A}^{(\alpha)}_{n-1,s+1}\nonumber\\
 &+\widehat{\mathcal{R}}^{(\alpha,4)}_{n,s-1}\widehat{A}^{(\alpha)}_{n,s-1}+\widehat{\mathcal{R}}^{(\alpha,5)}_{n,s}\widehat{A}^{(\alpha)}_{n,s}\bigg) \widehat{f}_{\Delta_{\alpha}+2n+s,s}(\xi,\eta) .
\end{align}

\noindent Now, as mentioned before, acting with $\mathcal{D}_{\mathrm{bulk}}$ on $\mathcal{W}_{\Delta,0}(\xi,\eta)$ generates a single contact diagram coming from a non-derivative interaction vertex. The defect channel block expansion of this contact diagram is
\begin{align}
\label{contactdefblock}
\mathcal{W}_{c}(\xi,\eta) = \sum_{n=0}^{\infty} \left[\widehat{a}^{(1)}_{n}\widehat{f}_{\Delta_{1}+2n,0}(\xi,\eta)+ \widehat{a}^{(2)}_{n}\widehat{f}_{\Delta_{2}+2n,0}(\xi,\eta)\right]
\end{align}

\noindent where the coefficients $\widehat{a}^{(i)}_{n}$ were obtained in Eq.~\eqref{contactdefchnlcoeff}. Then comparing Eq.~\eqref{blkexchl0defblock} and \eqref{contactdefblock}, we obtain the following recursion relations
\begin{align}
\label{bulktodefectrecrel1}
 & \widehat{\mathcal{R}}^{(\alpha,1)}_{n,1}\widehat{A}^{(\alpha)}_{n,1}+ \widehat{\mathcal{R}}^{(\alpha,3)}_{n-1,1}\widehat{A}^{(\alpha)}_{n-1,1}+\widehat{\mathcal{R}}^{(\alpha,5)}_{n,0}\widehat{A}^{(\alpha)}_{n,0}= \widehat{a}^{(\alpha)}_{n}, \quad \forall n\ge 0, \ \alpha =1,2
\end{align}

\noindent and
\begin{align}
\label{bulktodefectrecrel2}
&\widehat{\mathcal{R}}^{(\alpha,1)}_{n,s+1}\widehat{A}^{(\alpha)}_{n,s+1}+\widehat{\mathcal{R}}^{(\alpha,2)}_{n+1,s-1}\widehat{A}^{(\alpha)}_{n+1,s-1}+ \widehat{\mathcal{R}}^{(\alpha,3)}_{n-1,s+1}\widehat{A}^{(\alpha)}_{n-1,s+1}+\widehat{\mathcal{R}}^{(\alpha,4)}_{n,s-1}\widehat{A}^{(\alpha)}_{n,s-1}\nonumber\\
 &+\widehat{\mathcal{R}}^{(\alpha,5)}_{n,s}\widehat{A}^{(\alpha)}_{n,s}= 0, \quad \forall s\ge 1, \ n\ge 0,  \ \alpha=1,2. 
\end{align}

\noindent We need to supplement the above recursion relations with the boundary conditions 
\begin{align}
    \label{bdycond}
    \widehat{A}^{(\alpha)}_{-1,s}=0, \ \widehat{A}^{(\alpha)}_{n,-1}=0.
\end{align}

\noindent To solve the above recursion relations, we require a set of seed coefficients which are $\widehat{A}^{(\alpha)}_{0,s}$ for all values of the transverse spin $s$\footnote{In the analogous recursion relations for boundary CFTs, there is just a single seed coefficient, due to the fact there is no non-zero transverse spin the boundary channel expansion.}. To see this consider the homogeneous equations in Eq.~\eqref{bulktodefectrecrel2} for $n=0$ which gives,
\begin{align}
\label{bulktodefectrecrel3}
&\widehat{\mathcal{R}}^{(\alpha,1)}_{0,s+1}\widehat{A}^{(\alpha)}_{0,s+1}+\widehat{\mathcal{R}}^{(\alpha,2)}_{1,s-1}\widehat{A}^{(\alpha)}_{1,s-1} +\widehat{\mathcal{R}}^{(\alpha,4)}_{0,s-1}\widehat{A}^{(\alpha)}_{0,s-1}+\widehat{\mathcal{R}}^{(\alpha,5)}_{0,s}\widehat{A}^{(\alpha)}_{0,s}= 0 
\end{align}

\noindent where we have also used the boundary conditions in Eq.~\eqref{bdycond}. Now,  if we are given the coefficients $(\widehat{A}^{(\alpha)}_{0,s-1}, \widehat{A}^{(\alpha)}_{0,s},\widehat{A}^{(\alpha)}_{0,s+1})$, we can solve for $\widehat{A}^{(\alpha)}_{1,s}$ using Eq.~\eqref{bulktodefectrecrel3} for $s\ge 0$. Next, we can consider the homogeneous equations with $n=1$. From these equations, we can solve for $\widehat{A}^{(\alpha)}_{2,s}$ for all $s\ge 0$, since the previous step already determines $(\widehat{A}^{(\alpha)}_{1,s-1}, \widehat{A}^{(\alpha)}_{1,s},\widehat{A}^{(\alpha)}_{1,s+1})$ and $\widehat{A}^{(\alpha)}_{0,s+1}$ is assumed to be given as input. Therefore it is evident that we can recursively solve all the coefficients $\widehat{A}^{(\alpha)}_{n,s}$ for $n\ge 1, s\ge 0$, if we are provided with the set of seed coefficients $\widehat{A}^{(\alpha)}_{0,s}$ with $s\ge 0$. The inhomogeneous equations Eq.~\eqref{bulktodefectrecrel1} impose constraints on the allowed values of the seed coefficients. In particular, the $n$-th inhomogeneous equation constrains the values of  $\widehat{A}^{(\alpha)}_{0,s}$ for $s=0,1,\ldots, n+1$.

\subsubsection{Computation of seed coefficients}
\label{subsec:bulktodefseeds}


Here, we describe how the seed coefficients can be obtained using the Mellin representation of the bulk-channel Witten diagram. For the case of scalar exchange, this is given by
\begin{align}
    \label{bulkchMellin}
    \mathcal{W}_{\Delta,0}(\xi,\eta)= \int \frac{d\delta d\rho}{(2\pi i)^{2}} \ \xi^{-\delta}(2\eta)^{\delta-\rho} \ \Gamma(\delta)\Gamma(\rho-\delta)\Gamma\left(\frac{\Delta_{1}-\rho}{2}\right)\Gamma\left(\frac{\Delta_{2}-\rho}{2}\right)\ M_{\Delta,0}(\delta,\rho) ,
\end{align}

\noindent where the Mellin amplitude $M_{\Delta,0}(\delta,\rho)$ has been computed in \cite{Gimenez-Grau:2023fcy} and is given by\footnote{Mellin formalism for boundary and defect CFTs has been developed in \cite{Goncalves:2018fwx,Rastelli:2017ecj}.}
\begin{align}
    \label{bulkchMellinamp}
    M_{\Delta,0}(\delta,\rho)&= \frac{\pi^{\frac{p}{2}}\Gamma\left(\frac{\Delta-p}{2}\right)\Gamma\left(\frac{\Delta_{1}+\Delta_{2}-\Delta-d}{2}\right)}{16\Gamma(\Delta_{1})\Gamma(\Delta_{2})\Gamma\left(\frac{2\Delta-d+2}{2}\right)\left(\delta-\frac{\Delta_{1}+\Delta_{2}-\Delta}{2}\right)} \times\nonumber\\
    & \ {}_3F_2\left(
\begin{array}{c}
1+\frac{\Delta-d+p}{2},\, 1+\frac{\Delta-\Delta_{1}-\Delta_{2}}{2},\, \delta+\frac{\Delta-\Delta_{1}-\Delta_{2}}{2} \\
\Delta+1-\frac{d}{2},\, \delta+1+\frac{\Delta-\Delta_{1}-\Delta_{2}}{2}
\end{array}
; 1\right) .
\end{align}

\noindent  Note that for scalar exchange, $M_{\Delta,0}(\delta,\rho)$ is independent of the Mellin variable $\rho$.  The Mellin amplitude in Eq.\eqref{bulkchMellinamp} has poles in $\delta$ at $\delta=\frac{\Delta_{1}+\Delta_{2}-\Delta-2k}{2}$, with $k \in \mathbb{Z}_{\ge 0}$. These poles encode the exchange of a scalar primary operator $\mathcal{O}_{\Delta,0}$ and its conformal descendants in the bulk-channel OPE. The contribution of double-twist operators in the bulk-channel block decomposition of the Witten diagram $W_{\Delta,0}(\xi,\eta)$ comes from the poles of the gamma function factor $\Gamma(\delta)$ in Eq.~\eqref{bulkchMellin}. 

\vskip 4pt
Now, from the block expansion in Eq.~\eqref{blkexchdefblock}, we can see that the seed coefficients $\widehat{A}^{(1)}_{0,s}$ and $\widehat{A}^{(2)}_{0,s}$ can be isolated by considering the coefficients of the $\chi^{-\Delta_{1}-s}\eta^{s}$ and $\chi^{-\Delta_{2}-s}\eta^{s}$ terms respectively. Therefore, to extract the seed coefficients from Mellin space, we need to find the coefficients of these terms from the Mellin representation in Eq.~\eqref{bulkchMellin}. We will focus on the $\chi^{-\Delta_{1}-s}\eta^{s}$ term below. The other set of coefficients can be obtained by simply replacing $\Delta_{1}$ by $\Delta_{2}$ in the final result.  

\vskip 4pt
Consider the contribution from the pole of $\Gamma\left(\frac{\Delta_{1}-\rho}{2}\right)$ at $\rho=\Delta_{1}$ in the Mellin integral Eq.~\eqref{bulkchMellin}. This yields
\begin{align}
    \label{bulkchMellin1}
    W_{\Delta,0}(\xi,\eta)= -2\Gamma\left(\frac{\Delta_{21}}{2}\right)\int \frac{d\delta}{2\pi i} \ (\chi-2\eta)^{-\delta}(2\eta)^{\delta-\Delta_{1}} \ \Gamma(\delta)\Gamma(\Delta_{1}-\delta)\ M_{\Delta,0}(\delta)+\cdots
\end{align}

\noindent where we have used $\chi=\xi+2\eta$. The dots denote additional contributions which are not important for our current purpose. Since the Mellin amplitude is independent of $\rho$, we have denoted it here as $M_{\Delta,0}(\delta)$. Now, we are interested in the large $\chi$ behaviour. We can then series expand the $(\chi-2\eta)^{-\delta}$ as
\begin{align}
    (\chi-2\eta)^{-\delta} =\chi^{-\delta}\sum_{k=0}^{\infty} \frac{(\delta)_{k}}{k!} \ (2\eta)^{k}\chi^{-k} .
\end{align}

\noindent To extract the coefficient of $\chi^{-\Delta_{1}-s}\eta^{s}$ from Eq.~\eqref{bulkchMellin1}, we need to consider the contribution from the poles at $\delta=\Delta_{1}+m$, for $0\le m\le s$, coming from the $\Gamma(\Delta_{1}-\delta)$ factor. The residues at these poles give the desired seed coefficient to be\footnote{We are assuming here that the scaling dimensions are generic, so that none of the poles of the Mellin amplitude coincides with the poles of $\Gamma(\Delta_{1}-\delta)$.}
\begin{align}
\label{Ahat10s}
    \widehat{A}^{(1)}_{0,s} = 2^{s+1}\Gamma\left(\frac{\Delta_{21}}{2}\right)\Gamma(\Delta_{1}+s) \sum_{m=0}^{s}\frac{(-1)^{m}}{m!(s-m)!}   M_{\Delta,0}(\Delta_{1}+m).
\end{align}

\noindent The result for $\widehat{A}^{(2)}_{0,s}$ is the same with $\Delta_{1}$ replaced by $\Delta_{2}$. The above results for the seed coefficients can also be obtained using the integrated vertex identity, which allows exchange Witten diagrams to be expressed in terms of infinite sums of contact diagrams. In \cite{Mazac:2018biw}, this approach was used to compute the seed coefficients in the crossed channel block decomposition of the exchange Witten diagram for holographic boundary CFTs. We have checked that the generalization of this method to the case of $p$-dimensional defects yield the same results as the ones derived above from the Mellin space approach. As a further consistency check, it is easy to verify that $\widehat{A}^{(1)}_{0,1}$ obtained from Eq.~\eqref{Ahat10s}, and similarly $\widehat{A}^{(2)}_{0,1}$, satisfy the inhomogeneous recursion relation Eq.~\eqref{bulktodefectrecrel1} for $n=0$.

\subsection{Bulk channel expansion of defect exchange Witten diagram}
\label{subsec:deftobulk}

We will now consider the bulk channel expansion of a tree-level defect channel exchange diagram. This can be written as
\begin{align}
    \label{defexchblkblock}
    \widehat{\mathcal{W}}_{\hat{\Delta},s}(\xi,\eta) = \sum_{n,\ell=0}^{\infty} A_{n,\ell} \ \xi^{-\frac{\Delta_{1}+\Delta_{2}}{2}}f_{\Delta_{1}+\Delta_{2}+2n+\ell,\ell}(\xi,\eta). 
\end{align}

\noindent The strategy to compute the coefficients $A_{n,\ell}$ is the similar to the one employed in section \ref{subsec:deftobulk}. In this case, we need to consider the action of the defect channel Casimir on the bulk channel blocks. In the defect channel, we have two Casimir operators \cite{Billo:2016cpy}. One of these corresponds to the $SO(p+1,1)$ conformal symmetry generators and the other is associated to $SO(d-p)$. Their actions can be expressed via the following second-order differential equations \cite{Billo:2016cpy}
\begin{align}
    \label{defcas}
   & \mathcal{D}_{\chi} =(4-\chi^{2})\partial^{2}_{\chi}-(p+1)\chi\partial_{\chi}+\hat{\Delta}(\hat{\Delta}-p), \nonumber\\
   & \mathcal{D}_{\eta} =(1-\eta^{2})\partial^{2}_{\eta}+(1-q)\eta\partial_{\eta} +s(s+q-2).
\end{align}

\noindent In obtaining the above differential operators, we have chosen $(\chi,\eta)$ as the two independent cross ratios rather than $(\xi,\eta)$. Also note that in the Casimir eigenvalues we have set the parallel spin to zero, since we are only considering $2$-point functions of bulk CFT operators, where the exchanged operators in the defect channel cannot have non-zero parallel spin. 

\vskip 4pt
Now we need to know how the action of the defect channel Casimir differential operators in Eq.~\eqref{defcas} on a bulk channel block can be re-expressed in terms of a linear combination of bulk channel blocks. Compared to our previous case, this is more complicated since, in general, the bulk channel blocks do not admit closed-form expressions. Applying the differential operator $\mathcal{D}_{\chi}$ we find the following recurrence relation for the bulk blocks
\begin{align}
    \label{radcasrec}
    &\mathcal{D}_{\chi}\left[\xi^{-\frac{\Delta_{1}+\Delta_{2}}{2}}f_{\Delta,\ell}(\xi,\eta)\right] \nonumber\\
    &= \xi^{-\frac{\Delta_{1}+\Delta_{2}}{2}}\bigg[\mathcal{R}^{(1)}_{\Delta,\ell}f_{\Delta-2,\ell+2}(\xi,\eta)+ \mathcal{R}^{(2)}_{\Delta,\ell}f_{\Delta,\ell}(\xi,\eta)+\mathcal{R}^{(3)}_{\Delta,\ell}f_{\Delta-2,\ell}(\xi,\eta)\nonumber\\
    &+\mathcal{R}^{(4)}_{\Delta,\ell}f_{\Delta+2,\ell}(\xi,\eta)+\mathcal{R}^{(5)}_{\Delta,\ell}f_{\Delta+2,\ell+2}(\xi,\eta)+\mathcal{R}^{(6)}_{\Delta,\ell}f_{\Delta-2,\ell-2}(\xi,\eta)\nonumber\\
    &+ \mathcal{R}^{(7)}_{\Delta,\ell}f_{\Delta+2,\ell-2}(\xi,\eta)+\mathcal{R}^{(8)}_{\Delta,\ell}f_{\Delta,\ell-2}(\xi,\eta)+\mathcal{R}^{(9)}_{\Delta,\ell}f_{\Delta,\ell+2}(\xi,\eta)\bigg] 
\end{align}

\noindent The coefficients $\mathcal{R}^{(i)}_{\Delta,\ell}$ are rather cumbersome and so we will not display them here\footnote{For deriving the recursion relation and to compute the coefficients $\mathcal{R}^{(i)}_{\Delta,\ell}$, we employed the lightcone expansion for bulk channel blocks.}. Their expressions are given in the ancillary Mathematica file attached to this paper. We get a similar relation from the action of the differential operator $\mathcal{D}_{\eta}$. It can be checked that setting $\eta=1, q=1, \ell=0$ in Eq.~\eqref{radcasrec} drastically simplifies it to a three-term relation. This special case also matches with the result obtained in \cite{Mazac:2018biw} for the action of the boundary channel Casimir on bulk channel blocks in the case of BCFT. 

\vskip 4pt
The defect exchange diagram can be related to contact diagrams by the action of the defect channel Casimir operators. Then expanding these contact diagrams in terms of bulk channel blocks leads to relations of the form
\begin{align}
    \label{radcasdefexch}
    \mathcal{D}_{\chi}\widehat{\mathcal{W}}_{\hat{\Delta},s}(\xi,\eta) =\sum_{\ell=0}^{2s}\sum_{n=0}^{\infty} c_{n,\ell} \ \xi^{-\frac{\Delta_{1}+\Delta_{2}}{2}}f_{\Delta_{1}+\Delta_{2}+2n+\ell,\ell}(\xi,\eta) . 
\end{align}


\noindent Now we plug in the bulk channel expansion of $\widehat{\mathcal{W}}_{\hat{\Delta},\ell}(\xi,\eta)$ in the lhs of Eq.~\eqref{radcasdefexch} and use Eq.~\eqref{radcasrec} for the action of $\mathcal{D}_{\chi}$ on bulk channel blocks with double-twist dimensions $\Delta_{1}+\Delta_{2}+2n+\ell$. Let us denote the corresponding coefficients $\mathcal{R}^{(i)}_{\Delta,\ell}$ as $\mathcal{R}^{(i)}_{n,\ell}$. Then comparing with the rhs of Eq.~\eqref{radcasdefexch}, we obtain the following recursion relations for the coefficients $A_{n,\ell}$
\begin{align}
    \label{radcasrecrel0}
   & \mathcal{R}^{(1)}_{n+2,\ell-2}A_{n+2,\ell-2}+\mathcal{R}^{(2)}_{n,\ell}A_{n,\ell}+ \mathcal{R}^{(3)}_{n+1,\ell}A_{n+1,\ell}+\mathcal{R}^{(4)}_{n-1,\ell}A_{n-1,\ell}\nonumber\\
   & +\mathcal{R}^{(5)}_{n,\ell-2}A_{n,\ell-2}+\mathcal{R}^{(6)}_{n,\ell+2}A_{n,\ell+2}+\mathcal{R}^{(7)}_{n-2,\ell+2}A_{n-2,\ell+2}+\mathcal{R}^{(8)}_{n-1,\ell+2}A_{n-1,\ell+2}\nonumber\\
   &+ \mathcal{R}^{(9)}_{n+1,\ell-2}A_{n+1,\ell-2}=c_{n,\ell}, \quad \forall n\ge 0, \ell \ge 0.
\end{align}

\noindent We further impose the boundary conditions $A_{n,-2}=0$ and $A_{-2,\ell}=A_{-1,\ell}=0$ for all $n,\ell \ge 0$. Since the bulk channel expansion in the rhs of Eq.~\eqref{radcasdefexch} contains only blocks with spins up to $\ell=2s$, the relations in Eq.~\eqref{radcasrecrel0} are homogeneous for $\ell >2s$. We will also get similar recursion relations by applying the differential operator $\mathcal{D}_{\eta}$ on the defect channel exchange diagram. We will not write them down explicitly here. 

\vskip 4pt
\noindent Using the relations Eq.~\eqref{radcasrecrel0}, we can recursively solve for the coeffecients $A_{n,\ell}$ with $n >0$, if we input the seed coefficients $A_{0,\ell}$ for all values of $\ell \ge 0$ and the contact diagram coefficients $c_{n,\ell}$. We will evaluate $A_{0,\ell}$ in the next section. 




\subsubsection{Computation of seed coefficients}
\label{subsec:deftoblkseeds}

In this section, we compute the seed coefficients $A_{0,\ell}$ by employing the Mellin representation of the defect channel exchange diagram. For this, let us first consider the leading terms as $\xi\rightarrow 0$ in the bulk channel expansion of $\widehat{\mathcal{W}}_{\hat{\Delta},s}(\xi,\eta)$. Using the lightcone expansion of bulk blocks \cite{Billo:2016cpy}, this is given by
\begin{align}
\label{seedblocks}
     \widehat{\mathcal{W}}_{\hat{\Delta},s}(\xi,\eta) =\sum_{\ell=0}^{\infty} A_{0,\ell} \ g_{\tau_{0},\ell}(\eta) + \cdots ,
\end{align}

\noindent where the dots denote higher order terms in $\xi$ and $\tau_{0}=\Delta_{1}+\Delta_{2}$. The function $g_{\tau_{0},\ell}(\eta)$ is 
\begin{align}
    g_{\tau_{0},\ell}(\eta)=(1-\eta^{2})^{\frac{\ell}{2}} \ _{2}F_{1}\left(\frac{2\ell+\tau_{0}+\Delta_{12}}{4},\frac{2\ell+\tau_{0}-\Delta_{12}}{4},\frac{2\ell+\tau_{0}+1}{2},1-\eta^{2}\right),
\end{align}

\noindent where $\Delta_{12}=\Delta_{1}-\Delta_{2}$. The Mellin transform of $g^{(0)}_{\tau_{0},\ell}(\eta)$ is related to a class of orthogonal polynomials, referred to as continuous Hahn polynomials, which play an important role in the Mellin space approach to conformal bootstrap \cite{Gopakumar:2016wkt, Gopakumar:2016cpb, Gopakumar:2018xqi, Penedones:2019tng, Bissi:2022mrs}. The orthogonality property of these polynomials will allow us to efficiently extract $A_{0,\ell}$ in Mellin space.

\vskip 4pt
Now we will compute $A_{0,\ell}$ via Mellin space as follows. Let us consider the following Mellin representation of $\widehat{\mathcal{W}}_{\hat{\Delta},s}(\xi,\eta)$
\begin{align}
\label{defchMellin}
    \widehat{\mathcal{W}}_{\hat{\Delta},s}(\xi,\eta)= \int \frac{d\delta d\rho}{(2\pi i)^{2}} \ \frac{\Gamma(\delta)\Gamma(\rho-\delta)\Gamma\left(\frac{\Delta_{1}-\rho}{2}\right)\Gamma\left(\frac{\Delta_{2}-\rho}{2}\right)}{\xi^{\delta}(2\eta)^{\rho-\delta}} \ \widehat{M}_{\hat{\Delta},s}(\delta,\rho).
\end{align}

\noindent The Mellin amplitude $\widehat{M}_{\hat{\Delta},s}(\delta,\rho)$ is related to the Mellin amplitude for scalar exchange but with shifted external dimensions as follows \cite{Gimenez-Grau:2023fcy, Alday:2024srr}
\begin{align}
\label{spindefMellin}
  \widehat{M}^{\Delta_{1},\Delta_{2}}_{\hat{\Delta},s}(\delta,\rho) = 2^{s}(\Delta_{1})_{s}(\Delta_{2})_{s} (\rho-\delta)_{s}\widehat{M}^{\Delta_{1}+s,\Delta_{2}+s}_{\hat{\Delta},0}(\delta,\rho+s) , 
\end{align}

\noindent where the superscripts in Eq.~\eqref{spindefMellin} denote the external scaling dimensions. The scalar exchange amplitude $\widehat{M}^{\Delta_{1},\Delta_{2}}_{\hat{\Delta},0}(\delta,\rho)$ was derived in \cite{Gimenez-Grau:2023fcy} and is given by, 
\begin{align}
    \label{defectMellin}
   \widehat{M}^{\Delta_{1},\Delta_{2}}_{\hat{\Delta},0}(\delta,\rho) & = -\frac{\pi^{\frac{p}{2}}\Gamma\left(\frac{\Delta_{1}+\hat{\Delta}-p}{2}\right)\Gamma\left(\frac{\Delta_{2}+\hat{\Delta}-p}{2}\right)}{8\Gamma(\Delta_{1})\Gamma(\Delta_{2})\Gamma\left(\hat{\Delta}+1-\frac{p}{2}\right)(\rho-\hat{\Delta})}  \ {}_3F_2\left(
\begin{array}{c}
1-\frac{\Delta_{1}-\hat{\Delta}}{2},\, 1-\frac{\Delta_{2}-\hat{\Delta}}{2},\, \frac{\hat{\Delta}-\rho}{2} \\
\hat{\Delta}+1-\frac{p}{2},\, 1+\frac{\hat{\Delta}-\rho}{2}
\end{array}
; 1\right) .
\end{align}

\noindent Note that for scalar exchange, the Mellin amplitude is independent of $\delta$.  $\widehat{M}_{\hat{\Delta},s}(\delta,\rho)$ has simple poles in $\rho$ at $\rho+s=\hat{\Delta}+2k$, where $k\in \mathbb{Z}_{\ge 0}$.  

\vskip 4pt
Now we are interested in the leading $\xi^{0}$ term in the bulk channel expansion of $\widehat{\mathcal{W}}_{\hat{\Delta},s}(\xi,\eta)$. This must come from the contribution of the pole of $\Gamma(\delta)$ at $\delta=0$ in the above Mellin integral. Evaluating the residue at this pole, we get the relation
\begin{align}
\label{blkseedcoeffeq}
   \sum_{\ell=0}^{\infty} A_{0,\ell} \ g_{\tau_{0},\ell}(\eta)=  \int \frac{d\rho}{2\pi i} \ (2\eta)^{-\rho} \ \Gamma(\rho)\Gamma\left(\frac{\Delta_{1}-\rho}{2}\right)\Gamma\left(\frac{\Delta_{2}-\rho}{2}\right) \ \widehat{M}_{\hat{\Delta},s}(\delta=0,\rho). 
\end{align}

\noindent We now use the Mellin representation of $g_{\tau_{0},\ell}(\eta)$ derived in Appendix \ref{app:hahnpol}. This is given by
\begin{align}
\label{gt0Mellin}
   g_{\tau_{0},\ell}(\eta)= \int\frac{d\rho}{2\pi i} \ (2\eta)^{-\rho}\ \Gamma\left(\rho\right)\Gamma\left(\frac{\Delta_{1}-\rho}{2}\right)\Gamma\left(\frac{\Delta_{2}-\rho}{2}\right) \ \mathbb{Q}_{\tau_{0},\ell}(\rho)
\end{align}

\noindent where $\mathbb{Q}_{\tau_{0},\ell}(\rho)$ is a polynomial in $\rho$ of degree $\ell/2$ and is given by
\begin{align}
    \label{mackpol}
    \mathbb{Q}_{\tau_{0},\ell}(\rho) = \mathcal{N}_{\tau_{0},\ell} \left(\frac{\Delta_{1}}{2}\right)_{\frac{\ell}{2}}\left(\frac{\Delta_{2}}{2}\right)_{\frac{\ell}{2}} {} _{3}F_{2}\left(-\frac{\ell}{2},\frac{\tau_{0}+\ell-1}{2},\frac{\rho}{2}, \frac{\Delta_{1}}{2},\frac{\Delta_{2}}{2},1 \right)
\end{align}

\noindent and the normalization factor $\mathcal{N}_{\tau_{0},\ell}$ is
\begin{align}
    \mathcal{N}_{\tau_{0},\ell}=\frac{2^{\tau_{0}+2l-2}\Gamma\left(\frac{\tau_{0}+2\ell+1}{2}\right)}{\sqrt{\pi} \Gamma(\Delta_{1}+\ell)\Gamma(\Delta_{2}+\ell)}. 
\end{align}

\noindent $\mathbb{Q}_{\tau_{0},\ell}(\rho)$ is related to the continuous Hahn polynomial as shown in Appendix \ref{app:hahnpol}. Now using the orthogonality property of these polynomials from Eq.~\eqref{orthnorm2}, we can invert Eq.~\eqref{blkseedcoeffeq} and extract $A_{0,\ell}$. We thus get
\begin{align}
\label{def2blkseedMellin}
   A_{0,\ell} = \frac{1}{\kappa_{\ell}(\tau_{0})}\int_{-i\infty}^{i\infty}  \frac{d\rho}{2\pi i} \ 2^{-\rho}\Gamma(\rho) \Gamma\left(\frac{\Delta_{1}-\rho}{2}\right)\Gamma\left(\frac{\Delta_{2}-\rho}{2}\right) \mathbb{Q}_{\tau_{0},\ell}(\rho) \widehat{M}_{\hat{\Delta},s}(\delta=0,\rho) 
\end{align}

\noindent where
\begin{align}
   \kappa_{\ell}(\tau_{0}) = (-1)^{\frac{\ell}{2}}(\ell/2)! \frac{\Gamma\left(\frac{\tau_{0}+2\ell-1}{2}\right)}{\Gamma\left(\frac{\tau_{0}+\ell-1}{2}\right)}\mathcal{N}_{\tau_{0},\ell}. 
\end{align}

\noindent We compute the above integral in closed form in Appendix \ref{app:A0linteg}. The final result is

\begin{align}
\label{a0lfinalres}
    A_{0,\ell}
    &= \frac{(-1)^{\ell/2}\left(\frac{\Delta_{1}}{2}\right)_{\ell/2}\left(\frac{\Delta_{2}}{2}\right)_{\ell/2}}{(\ell/2)! \left(\frac{\Delta_{1}+\Delta_{2}+\ell-1}{2}\right)_{\ell/2}}\ \frac{\pi^{\frac{p+1}{2}}\Gamma\left(\frac{\Delta_{1}+s+\hat{\Delta}-p}{2}\right)\Gamma\left(\frac{\Delta_{2}+s+\hat{\Delta}-p}{2}\right)\Gamma\left(\frac{\hat{\Delta}+1}{2}\right)}{2^{2+\Delta_{1}+\Delta_{2}}\Gamma\left(\hat{\Delta}+1-\frac{p}{2}\right)}  \times \nonumber\\
    & \sum_{k=0}^{\ell/2} \frac{\left(-\frac{\ell}{2}\right)_{k}\left(\frac{\Delta_{1}+\Delta_{2}+\ell-1}{2}\right)_{k}}{k!\Gamma\left(\frac{\Delta_{1}+\Delta_{2}+1+2k+2s}{2}\right)}\bigg[\frac{(\Delta_{1}+\Delta_{2}+2k+2s-1)\Gamma\left(\frac{\Delta_{1}+\Delta_{2}+2s-\hat{\Delta}-1}{2}\right)}{(\hat{\Delta}+2k)\Gamma\left(\frac{\Delta_{1}+s+1}{2}\right)\Gamma\left(\frac{\Delta_{2}+s+1}{2}\right)} \nonumber\\
     & \ {}_4F_3\left(
\begin{array}{c}
1-\frac{\Delta_{1}+s-\hat{\Delta}}{2},\, 1-\frac{\Delta_{2}+s-\hat{\Delta}}{2},\, \frac{\hat{\Delta}+1-p}{2}, \, \frac{\hat{\Delta}+2k}{2} \\
\frac{\hat{\Delta}-\Delta_{1}-\Delta_{2}-2s+3}{2},\, \hat{\Delta}+1-\frac{p}{2}, \frac{\hat{\Delta}+2k+2}{2} \, 
\end{array}
; 1\right) \nonumber \\
& + \frac{\Gamma\left(\frac{\hat{\Delta}-\Delta_{1}-\Delta_{2}-2s+1}{2}\right)\Gamma\left(\hat{\Delta}+1-\frac{p}{2}\right)\Gamma\left(\frac{\Delta_{1}+\Delta_{2}+2s-p}{2}\right)}{\Gamma\left(1+\frac{\hat{\Delta}-s-\Delta_{1}}{2}\right)\Gamma\left(1+\frac{\hat{\Delta}-s-\Delta_{2}}{2}\right)\Gamma\left(\frac{\hat{\Delta}+1-p}{2}\right)\Gamma\left(\frac{\Delta_{1}+\Delta_{2}+\hat{\Delta}+2s+1-p}{2}\right)} \times \nonumber\\
& \ {}_4F_3\left(
\begin{array}{c}
\frac{\Delta_{1}+s+1}{2},\, \frac{\Delta_{2}+s+1}{2},\, \frac{\Delta_{1}+\Delta_{2}+2s-p}{2}, \, \frac{\Delta_{1}+\Delta_{2}+2k+2s-1}{2} \\
\frac{1+\Delta_{1}+\Delta_{2}+2s-\hat{\Delta}}{2},\, \frac{\Delta_{1}+\Delta_{2}+\hat{\Delta}+2s-p+1}{2}, \frac{\Delta_{1}+\Delta_{2}+2k+2s+1}{2} \, 
\end{array}
; 1\right)\bigg]. 
\end{align}

\noindent The $ _{4}F_{3}(1)$ hypergeometric functions appearing in the above expression are balanced. It is also worth noting that for $\ell=0,s=0,p=d-1$, our result for $A_{0,0}$ agrees with the seed coefficient, computed using the integrated vertex identity, in \cite{Mazac:2018biw} for the bulk channel expansion of a boundary exchange Witten diagram. However, our representation is manifestly symmetric in $\Delta_{1},\Delta_{2}$ unlike the result given in \cite{Mazac:2018biw}. 


\section{$2$-point Loop diagrams}
\label{2pt_loop}

In this section, we derive the conformal partial wave expansion of some $1$-loop Witten diagrams which contribute to the $2$-point function of bulk CFT operators. 

\subsection{Bulk exchange}
\label{subsec:loop1}

\begin{figure}[htp]
    \centering
    \includegraphics[width=3.8cm]{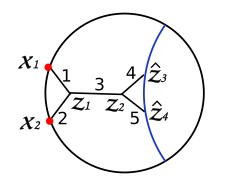}
    \caption{Bulk exchange diagram involving $1$-loop corrected one-point function.}
    \label{fig:loop1}
\end{figure}

Consider the Witten exchange diagram in Fig.~\ref{fig:loop1}. For simplicity, we will only study the case of scalar exchanges. This diagram is given by the following integral 
\begin{align}
\label{bulkexch1}
 W(x_{1},x_{2}) &= \int\limits_{AdS_{d+1}}d^{d+1}z_{1}d^{d+1}z_{2}\int\limits_{AdS_{p+1}}d^{p+1}\hat{z}_{3}d^{p+1}\hat{z}_{4} \ K_{\Delta_{1}}(x_{1},z_{1})K_{\Delta_{2}}(x_{2},z_{1}) \times \nonumber \\
 & \hspace{5.0cm}G_{\Delta_{3}}(z_{1},z_{2}) G_{\Delta_{4}}(z_{2},\hat{z}_{3}) G_{\Delta_{4}}(z_{2},\hat{z}_{4}). 
\end{align}

\noindent We want to obtain the bulk-channel partial wave expansion of this diagram. Let us then apply the split representations of the propagators labelled as $4$ and $5$. This factorises the bulk AdS integrals into a tree-level $4$-point exchange diagram without the $AdS_{p+1}$ brane and two tree-level $1$-point Witten diagrams in the presence of the brane. We thus get,
\begin{align}
\label{bulkexch1c}
     W(x_{1},x_{2}) &= \int_{-\infty}^{\infty} d\nu_{4}d\nu_{5} \ P(\nu_{4},\Delta_{4}) P(\nu_{5},\Delta_{5}) \times \nonumber \\
     &\int\limits_{\partial AdS_{d+1}} d^{d}x_{4}d^{d}x_{5} \ \mathcal{A}^{12\underline{4}\underline{5}}_{\Delta_{3},\mathrm{exch}}(x_{1},x_{2},x_{4},x_{5}) \mathbf{A}_{\underline{\widetilde{4}}}(x_{4})\mathbf{A}_{\underline{\widetilde{5}}}(x_{5})
\end{align}

\noindent where $\mathcal{A}^{12\underline{4}\underline{5}}_{\Delta_{3},\mathrm{exch}}(x_{1},x_{2},x)$ is a $4$-point tree-level exchange Witten diagram computed in the absence of the defect. $\mathbf{A}_{\underline{\widetilde{4}}}(x_{4}), \mathbf{A}_{\underline{\widetilde{5}}}(x_{5}) $ denote the tree-level $1$-point Witten diagrams, encountered in Section \ref{2pt_tree_direct}. Now, we can use the conformal partial wave expansion of the $4$-point exchange diagram, which is given by
\begin{align}
\label{4ptexchcpw}
    \mathcal{A}^{12\underline{4}\underline{5}}_{\Delta_{3},\mathrm{exch}}(x_{1},x_{2},x) & = \int_{-\infty}^{\infty} d\nu_{3} \ P(\nu_{3},\Delta_{3}) \ \lambda_{12\underline{3}} \lambda_{\underline{\widetilde{3}}\underline{4}\underline{5}} \ \Psi^{1\hspace{0.02cm}2\hspace{0.02cm}\underline{4}\hspace{0.02cm}\underline{5}}_{\frac{d}{2}+i\nu_{3},0} (x_{1},x_{2},x_{4},x_{5})
\end{align}

\noindent where $\Psi^{1\hspace{0.02cm}2\hspace{0.02cm}\underline{4}\hspace{0.02cm}\underline{5}}_{\frac{d}{2}+i\nu_{3},0} (x_{i})$ is the $4$-point scalar exchange conformal partial wave in a CFT without defects. Then Eq.~\eqref{bulkexch1c} becomes
\begin{align}
\label{loop1alt}
     W(x_{1},x_{2}) &= \int_{-\infty}^{\infty} \prod_{i=3}^{5} d\nu_{i} \ P(\nu_{i},\Delta_{i}) \ \lambda_{12\underline{3}} \lambda_{\underline{\widetilde{3}}\underline{4}\underline{5}} \ a_{\underline{\widetilde{4}}}a_{\underline{\widetilde{5}}} \times \nonumber\\
     &\int\limits_{\partial AdS_{d+1}} d^{d}x_{4}d^{d}x_{5} \ \Psi^{1\hspace{0.02cm}2\hspace{0.02cm}\underline{4}\hspace{0.02cm}\underline{5}}_{\frac{d}{2}+i\nu_{3}} (x_{1},x_{2},x_{4},x_{5}) \langle\hspace{-0.09cm}\langle \mathcal{O}_{\frac{d}{2}-i\nu_{4}}(x_{4})\rangle\hspace{-0.09cm}\rangle \langle\hspace{-0.09cm}\langle \mathcal{O}_{\frac{d}{2}-i\nu_{5}}(x_{5})\rangle\hspace{-0.09cm}\rangle.
\end{align}

\noindent Now to see how the bulk channel partial wave arises, we use the integral representation of the $4$-point conformal partial wave
\begin{align}
\label{4ptpw}
   &\Psi^{12\underline{4}\underline{5}}_{\frac{d}{2}+\nu_{3},0} (x_{1},x_{2},x_{4},x_{5})\nonumber\\
   &= \int d^{d}x_{3} \ \langle \mathcal{O}_{\Delta_{1}}(x_{1})\mathcal{O}_{\Delta_{2}}(x_{2}) \mathcal{O}_{\frac{d}{2}+i\nu_{3}}(x)\rangle \langle\mathcal{O}_{\frac{d}{2}-i\nu_{3}}(x)\mathcal{O}_{\frac{d}{2}+i\nu_{4}}(x_{4}) \mathcal{O}_{\frac{d}{2}+i\nu_{5}}(x_{5})\rangle 
\end{align}

\noindent We then have the following integral over the $AdS_{d+1}$ boundary points 
\begin{align}
\label{x3x4x5integ}
    & \int d^{d}x_{3} \ \langle \mathcal{O}_{\Delta_{1}}(x_{1})\mathcal{O}_{\Delta_{2}}(x_{2}) \mathcal{O}_{\frac{d}{2}+i\nu_{3}}(x)\rangle\int d^{d}x_{4}d^{d}x_{5} \ \langle\mathcal{O}_{\frac{d}{2}-i\nu_{3}}(x)\mathcal{O}_{\frac{d}{2}+i\nu_{4}}(x_{4}) \mathcal{O}_{\frac{d}{2}+i\nu_{5}}(x_{5})\rangle \times \nonumber\\
    & \hspace{7.0cm}\langle\hspace{-0.09cm}\langle \mathcal{O}_{\frac{d}{2}-i\nu_{4}}(x_{4})\rangle\hspace{-0.09cm}\rangle \langle\hspace{-0.09cm}\langle \mathcal{O}_{\frac{d}{2}-i\nu_{5}}(x_{5})\rangle\hspace{-0.09cm}\rangle
\end{align}

\noindent Due to conformal symmetry, the integral over $x_{4},x_{5}$ is proportional to the one-point function $\langle \hspace{-0.09cm}\langle\mathcal{O}_{\frac{d}{2}-i\nu_{3}}(x)\rangle\hspace{-0.09cm}\rangle$. The remaining integral over $x_{3}$ which now involves the $3$-point function and $\langle \hspace{-0.09cm}\langle\mathcal{O}_{\frac{d}{2}-i\nu_{3}}(x)\rangle\hspace{-0.09cm}\rangle$ will yield the bulk channel partial wave for the exchange of $\mathcal{O}_{\underline{\Delta_{3}}}$. The spectral integrals $\nu_{4},\nu_{5}$ can be done using the following identity
\begin{align}
\label{ident1}
    & a^{1-\mathrm{loop}}_{\underline{\widetilde{3}}}\langle \hspace{-0.09cm}\langle\mathcal{O}_{\frac{d}{2}-i\nu_{3}}(x)\rangle\hspace{-0.09cm}\rangle \nonumber\\
    &=  \int_{-\infty}^{\infty} d\nu_{4}d\nu_{5} \ P(\nu_{4},\Delta_{4}) P(\nu_{5},\Delta_{5})  \  \lambda_{\underline{\widetilde{3}}\underline{4}\underline{5}} \ a_{\underline{\widetilde{4}}}a_{\underline{\widetilde{5}}} \times\nonumber\\
    &\int\limits_{\partial AdS_{d+1}} d^{d}x_{4}d^{d}x_{5} \ \langle\mathcal{O}_{\frac{d}{2}-i\nu_{3}}(x)\mathcal{O}_{\frac{d}{2}+i\nu_{4}}(x_{4}) \mathcal{O}_{\frac{d}{2}+i\nu_{5}}(x_{5})\rangle \langle\hspace{-0.09cm}\langle \mathcal{O}_{\frac{d}{2}-i\nu_{4}}(x_{4})\rangle\hspace{-0.09cm}\rangle \langle\hspace{-0.09cm}\langle \mathcal{O}_{\frac{d}{2}-i\nu_{5}}(x_{5})\rangle\hspace{-0.09cm}\rangle
\end{align}

\noindent where $a^{1-\mathrm{loop}}$ denotes a $1$-loop corrected one-point function coefficient. We compute this coefficient in Appendix \ref{app:1ptloop}. The identity in  Eq.~\eqref{ident1} follows from applying the split and spectral representation of the bulk propagators appearing in the Witten diagram in Fig.~\ref{fig:1ptloop1}. Gathering the above results, we get, 
\begin{align}
\label{bulkexch1b}
     W(x_{1},x_{2}) & = \int_{-\infty}^{\infty} d\nu_{3} \ P(\nu_{3},\Delta_{3}) \ \lambda_{12\underline{3}} \ a^{1-\mathrm{loop}}_{\underline{\widetilde{3}}} \int d^{d}x_{3} \ \langle \mathcal{O}_{\Delta_{1}}(x_{1})\mathcal{O}_{\Delta_{2}}(x_{2}) \mathcal{O}_{\frac{d}{2}+i\nu_{3}}(x)\rangle \langle \hspace{-0.09cm}\langle\mathcal{O}_{\frac{d}{2}-i\nu_{3}}(x)\rangle\hspace{-0.09cm}\rangle \nonumber\\
     & = \int_{-\infty}^{\infty} d\nu_{3} \ P(\nu_{3},\Delta_{3}) \ \lambda_{12\underline{3}} \ a^{1-\mathrm{loop}}_{\underline{\widetilde{3}}}\ \mathcal{F}_{\frac{d}{2}+i\nu_{3},0} (x_{1},x_{2}). 
\end{align}

Let us now note the poles in $\nu_{3}$ in the above integrand, which are relevant for recasting the partial wave expansion into a sum over bulk channel conformal blocks. These poles are at  
\begin{align}
\label{polesloop1}
   & \underline{\Delta_{3}}= \Delta_{3}, \ \underline{\Delta_{3}}= \Delta_{1}+\Delta_{2}+2n, \ \underline{\Delta_{3}} = \Delta_{4}+\Delta_{5}+2m, \ (n,m)\in \mathbb{Z}_{\ge 0}.
\end{align}

\noindent The first pole in Eq.~\eqref{polesloop1} as usual comes from the measure factor $P(\nu_{3},\Delta_{3})$ and corresponds to the exchange of the operator $\mathcal{O}_{3}$ in the bulk channel block expansion. The second set of poles is due to the $3$-point factor $\lambda_{12\underline{3}}$. This indicates the presence of double-twist operators $[\mathcal{O}_{1}\mathcal{O}_{2}]_{n,0}$ in the block expansion. Finally, the last set of poles in Eq.~\eqref{polesloop1} comes from the $1$-point function coefficient in Eq.~\eqref{bulkexch1b}. This can be seen from its expression given in Appendix \ref{app:1ptloop}. These poles give rise to the exchange of double-twist operators $[\mathcal{O}_{4}\mathcal{O}_{5}]_{n,0}$ in the bulk channel block decomposition.

\subsection{Box diagram}

\begin{figure}[htp]
    \centering
    \includegraphics[width=4.0cm]{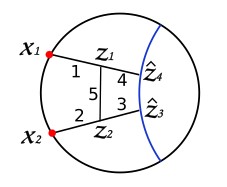}
    \caption{Box diagram with bulk cubic exchange}
    \label{fig:1loopbox}
\end{figure}

Here we will derive the bulk channel partial wave expansion of the loop diagram shown in Fig.~\ref{fig:1loopbox}. Applying the split representation of the propagators labelled as $3,4$, we get 
\begin{align}
\label{box1}
   & W(x_{1},x_{2})\nonumber\\
   &= \int_{-\infty}^{\infty} \prod_{i=3}^{4}d\nu_{i} \ P(\nu_{i},\Delta_{i}) \hspace{-0.3cm}\int\limits_{\partial AdS_{d+1}} \hspace{-0.3cm}d^{d}x_{3}d^{d}x_{4} \ \mathcal{A}^{1\underline{4}2\underline{3}}_{\Delta_{5},\mathrm{exch}}(x_{1},x_{4},x_{2},x_{5}) \mathbf{A}_{\underline{\tilde{3}}}(x_{3})\mathbf{A}_{\underline{\tilde{4}}}(x_{4}), 
\end{align}

\noindent where $\mathcal{A}^{1\underline{4}2\underline{3}}_{\Delta_{5},\mathrm{exch}}(x_{i})$ is a $4$-point tree-level t-channel exchange Witten diagram in the absence of the defect. $\mathbf{A}_{\underline{\tilde{3}}}(x_{3}), \mathbf{A}_{\underline{\tilde{4}}}(x_{4})$ are tree-level one-point function diagrams in the presence of the defect. We can now expand $\mathcal{A}^{1\underline{4}2\underline{3}}_{\Delta_{5},\mathrm{exch}}(x_{i})$ in terms of $t$-channel $4$-point conformal partial waves. This results in the following expression
\begin{align}
\label{box2}
     W(x_{1},x_{2}) &= \int_{-\infty}^{\infty} \left(\prod_{i=3}^{5}d\nu_{i} \ P(\nu_{i},\Delta_{i})\right)    \lambda_{1\hspace{0.02cm}\underline{4}\hspace{0.02cm}\underline{5}} \lambda_{2\hspace{0.02cm}\underline{3}\hspace{0.02cm}\underline{5}}  a_{\underline{\widetilde{3}}} a_{\underline{\widetilde{4}}} \times\nonumber\\
     &\int\limits_{\partial AdS_{d+1}} d^{d}x_{3}d^{d}x_{4} \ \Psi^{1\underline{4}2\underline{3}}_{\frac{d}{2}+i\nu_{5},0}(x_{i}) \langle\hspace{-0.09cm}\langle \mathcal{O}_{\frac{d}{2}-i\nu_{3}}(x_{3})\rangle\hspace{-0.09cm}\rangle \langle\hspace{-0.09cm}\langle \mathcal{O}_{\frac{d}{2}-i\nu_{4}}(x_{4})\rangle\hspace{-0.09cm}\rangle .
\end{align}

\noindent Now the integral over the boundary points in Eq.~\eqref{box2} should generate the the bulk channel partial wave for the exchange of $\mathcal{O}_{5}$. In order to show this, it is convenient to decompose the $t$-channel partial wave in terms of $s$-channel partial waves using the $s$-channel partial waves using the $6$-$j$ symbol of the $SO(d+1,1)$ conformal group. This takes the form
\begin{align}
    \label{6j4pt}
    \Psi^{1423}_{\Delta',J'}(x_{1},x_{2},x_{3},x_{4})= \sum_{J=0}^{\infty} \int_{\frac{d}{2}}^{\frac{d}{2}+i\infty}\frac{d\Delta}{2\pi i} \ \frac{1}{n_{\Delta,J}} \left\{
\begin{array}{ccc}
1 & 2 & \mathcal{O}' \\
3 & 4 & \mathcal{O}
\end{array}
\right\} \Psi^{1234}_{\Delta,J}(x_{1},x_{2},x_{3},x_{4}), 
\end{align}

\noindent where $\Psi^{1234}_{\Delta,J}(x_{i})$ is the $s$-channel partial wave. The factor involving the curly brackets denotes the $6$-$j$ symbol and it is defined by taking the inner product between a $t$-channel and a $s$-channel partial wave. Here we are using the notation and conventions of \cite{Meltzer:2019nbs} for the $6$-$j$ symbol. Then using Eq.~\eqref{6j4pt}, we can write Eq.~\eqref{box2} as
\begin{align}
\label{box5}
 W(x_{1},x_{2}) &= \sum_{J=0}^{\infty} \int_{\frac{d}{2}}^{\frac{d}{2}+i\infty}\frac{d\Delta}{2\pi i} \ \frac{1}{n_{\Delta,J}} \int \left(\prod_{i=3}^{5}d\nu_{i} \ P(\nu_{i},\Delta_{i})\right) \lambda_{1\hspace{0.02cm}\underline{4}\hspace{0.02cm}\underline{5}} \lambda_{2\hspace{0.02cm}\underline{3}\hspace{0.03cm}\underline{5}}  a_{\underline{\widetilde{3}}} a_{\underline{\widetilde{4}}} \left\{
\begin{array}{ccc}
1 & 2 & \underline{5} \\
\underline{3} & \underline{4} & \mathcal{O}
\end{array}
\right\} \times\nonumber\\
& \int\limits_{\partial AdS_{d+1}} d^{d}x_{3}d^{d}x_{4} \ \Psi^{12\underline{3}\hspace{0.04cm}\underline{4}}_{\Delta,J}(x_{1},x_{2},x_{3},x_{4}) \langle\hspace{-0.09cm}\langle \mathcal{O}_{\frac{d}{2}-i\nu_{3}}(x_{3})\rangle\hspace{-0.09cm}\rangle \langle\hspace{-0.09cm}\langle \mathcal{O}_{\frac{d}{2}-i\nu_{4}}(x_{4})\rangle\hspace{-0.09cm}\rangle.
\end{align}

\noindent Now, the integral over $x_{3},x_{4}$ is of the same form as we encountered in Eq.~\eqref{loop1alt} in the previous section. As stated there, conformal symmetry ensures that this integral must be proportional to a bulk channel partial wave. We thus have, 
\begin{align}
\label{box4}
\int d^{d}x_{3}d^{d}x_{4} \ \Psi^{12\hspace{0.02cm}\underline{3}\hspace{0.02cm}\underline{4}}_{\Delta,J}(x_{i})  \langle\hspace{-0.09cm}\langle \mathcal{O}_{\frac{d}{2}-i\nu_{3}}(x_{3})\rangle\hspace{-0.09cm}\rangle \langle\hspace{-0.09cm}\langle \mathcal{O}_{\frac{d}{2}-i\nu_{4}}(x_{4})\rangle\hspace{-0.09cm}\rangle  = \mathcal{C}^{\underline{3}\hspace{0.02cm}\underline{4}}_{\Delta,J}\mathcal{F}^{1,2}_{\Delta,J} (x_{1},x_{2}).
\end{align}

\noindent We will not compute the coefficient $\mathcal{C}^{\underline{3}\hspace{0.02cm}\underline{4}}_{\Delta,J}$ here explicitly. Then using Eq.~\eqref{box4} in Eq.~\eqref{box5} we obtain
\begin{align}
\label{box5a}
& W(x_{1},x_{2}) \nonumber\\
&= \int \left(\prod_{i=3}^{5}d\nu_{i} \ P(\nu_{i},\Delta_{i})\right) \lambda_{1\hspace{0.02cm}\underline{4}\hspace{0.02cm}\underline{5}} \lambda_{2\hspace{0.02cm}\underline{3}\hspace{0.03cm}\underline{5}}  a_{\underline{\widetilde{3}}} a_{\underline{\widetilde{4}}}\sum_{J=0}^{\infty} \int_{\frac{d}{2}}^{\frac{d}{2}+i\infty}\frac{d\Delta}{2\pi i} \ \frac{1}{n_{\Delta,J}}  \left\{
\begin{array}{ccc}
1 & 2 & \underline{5} \\
\underline{3} & \underline{4} & \mathcal{O}
\end{array}
\right\} \mathcal{C}^{\underline{3}\hspace{0.02cm}\underline{4}}_{\Delta,J}\mathcal{F}^{1,2}_{\Delta,J} (x_{1},x_{2}).
\end{align}

We can further perform the spectral integral over $\nu_{5}$. This can be done using the same reasoning applied in \cite{Meltzer:2019nbs} to the analysis of the $s$-channel decomposition of a tree-level $t$-channel $4$-point Witten exchange diagram. The $6$-$j$ symbol can be split as follows
\begin{align}
    \label{6jsplit}
    \left\{
\begin{array}{ccc}
1 & 2 & \underline{5} \\
\underline{3} & \underline{4} & \mathcal{O}
\end{array}
\right\} = K^{1\underline{4}}_{\underline{\tilde{5}}} \left(
\begin{array}{ccc}
1 & 2 & \underline{5} \\
\underline{3} & \underline{4} & \mathcal{O}
\end{array}
\right) + K^{2\underline{3}}_{\underline{5}} \left(
\begin{array}{ccc}
1 & 2 & \underline{\tilde{5}} \\
\underline{3} & \underline{4} & \mathcal{O}
\end{array}
\right)
\end{align}

\noindent where the terms denoted by angular brackets in the rhs of the above equation denote the inner product between a $s$-channel partial wave and a single $t$-channel conformal block. As noted in \cite{Meltzer:2019nbs}, these terms do not have any poles in $\nu_{5}$ towards the right and left sides of the principal series contour, respectively. Thus, the $\nu_{5}$ contour can be closed towards the right for the first term and towards the left for the second term. Now note that the $3$-point function factors $\lambda_{1\hspace{0.02cm}\underline{4}\hspace{0.02cm}\underline{5}},  \lambda_{2\hspace{0.02cm}\underline{3}\hspace{0.02cm}\underline{5}}$ have poles when $\underline{\Delta_{5}}=\Delta_{1}+\underline{\Delta_{3}}+2n$ and $\underline{\Delta_{5}}=\Delta_{1}+\underline{\Delta_{4}}+2n$. The first angular bracket factor in Eq.~\eqref{6jsplit} has zeroes precisely at these locations. Consequently, contribution from the poles of the $3$-point factors to the $\nu_{5}$ integral gets cancelled by these zeroes. Thus, we only get a contribution from the pole at $\underline{\Delta_{5}}=\Delta_{5}$. The bulk channel partial wave expansion of the diagram in Fig.~\ref{fig:1loopbox} is then given by
\begin{align}
\label{box5b}
W(x_{1},x_{2}) &= \int \left(\prod_{i=3}^{4}d\nu_{i} \ P(\nu_{i},\Delta_{i})\right) (d-2\Delta_{5})\lambda_{1\hspace{0.02cm}\underline{4}5} \lambda_{2\hspace{0.02cm}\underline{3}5}  a_{\underline{\widetilde{3}}} a_{\underline{\widetilde{4}}} \times \nonumber\\
& \hspace{0.5cm}\sum_{J=0}^{\infty} \int_{\frac{d}{2}}^{\frac{d}{2}+i\infty}\frac{d\Delta}{2\pi i} \ \frac{K^{1\underline{4}}_{5}}{n_{\Delta,J}}  \left(
\begin{array}{ccc}
1 & 2 & 5 \\
\underline{3} & \underline{4} & \mathcal{O}
\end{array}
\right) \mathcal{C}^{\underline{3}\hspace{0.02cm}\underline{4}}_{\Delta,J}\mathcal{F}^{1,2}_{\Delta,J} (x_{1},x_{2}).
\end{align}

\noindent The $6$-$j$ symbol factor in the above equation has poles at $\Delta=\Delta_{1}+\Delta_{2}+2n+J$ and $\Delta=\underline{\Delta_{3}}+\underline{\Delta_{4}}+2n+J$. After performing the $\nu_{3},\nu_{4}$ spectral integrals, these poles give rise to the exchange of double-twist operators $[\mathcal{O}_{1}\mathcal{O}_{2}]_{n,J}$ and $[\mathcal{O}_{3}\mathcal{O}_{4}]_{n,J}$ in the bulk channel block decomposition of this diagram. 



\subsection{Bulk-to-brane propagator correction}

\begin{figure}[htp]
    \centering
    \includegraphics[width=3.8cm]{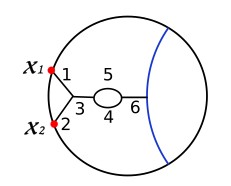}
    \caption{Loop correction to bulk-to-brane propagator}
    \label{fig:1loopmasscorr}
\end{figure}

Let us now consider the Witten diagram in Fig.~\ref{fig:1loopmasscorr}, involving a $1$-loop correction to the bulk-to-brane propagator. To obtain the bulk-channel conformal partial wave expansion of this diagram, let us split the propagators $4,5$ appearing in the loop. Then we get,
\begin{align}
\label{blkbranepropcorr}
    W(x_{1},x_{2}) =& \int d\nu_{4}d\nu_{5}  \ P(\nu_{4},\Delta_{4})P(\nu_{5},\Delta_{5})\times \nonumber\\
    &\int\limits_{\partial AdS_{d+1}}d^{d}x_{4}d^{d}x_{5} \ \mathcal{A}^{12\underline{5}\underline{4}}_{\Delta_{3},\mathrm{exch}}(x_{1},x_{2},x_{4},x_{5})  \mathbf{A}^{\underline{\widetilde{5}}\underline{\widetilde{4}}}_{\Delta_{6},\mathrm{exch}}(x_{4},x_{5}),
\end{align}

\noindent where $\mathcal{A}^{12\underline{5}\underline{4}}_{\Delta_{3},\mathrm{exch}}$ is a $4$-point tree-level exchange diagram without the defect and $\mathbf{A}^{\underline{\widetilde{5}}\underline{\widetilde{4}}}_{\Delta_{6},\mathrm{exch}}$ is tree-level $2$-point bulk-exchange diagram in the presence of the defect. Now we can insert the respective partial wave expansions of these sub-diagrams in Eq.~\eqref{blkbranepropcorr}. This yields,
\begin{align}
\label{blkbranepropcorr1}
    W(x_{1},x_{2}) =& \int d\nu_{3}d\nu_{4}d\nu_{5}d\nu_{6}  \ P(\nu_{3},\Delta_{3})P(\nu_{4},\Delta_{4})P(\nu_{5},\Delta_{5})P(\nu_{6},\Delta_{6}) \ \lambda_{12\underline{3}} \lambda_{\underline{\widetilde{3}}\underline{4}\underline{5}}\times \nonumber\\
    & \lambda_{\underline{\widetilde{5}}\underline{\widetilde{4}}\underline{6}} a_{\underline{\widetilde{6}}}\int\limits_{\partial AdS_{d+1}}d^{d}x_{4}d^{d}x_{5} \ \Psi^{12\underline{5}\underline{4}}_{\frac{d}{2}+i\nu_{3},0}(x_{1},x_{2},x_{4},x_{5})  \mathcal{F}^{\underline{\widetilde{5}}\underline{\widetilde{4}}}_{\frac{d}{2}+i\nu_{6},0}(x_{4},x_{5}) .
\end{align}

\noindent Now, using the integral representation of the $4$-point conformal partial waves, we can write the conformal integral in Eq.~\eqref{blkbranepropcorr1} as
\begin{align}
    \label{blkbranepropcorr2}
    & \int d^{d}x_{4}d^{d}x_{5} \ \Psi^{12\underline{5}\underline{4}}_{\frac{d}{2}+i\nu_{3},0}(x_{1},x_{2},x_{4},x_{5})  \mathcal{F}^{\underline{\widetilde{5}}\underline{\widetilde{4}}}_{\frac{d}{2}+i\nu_{6},0}(x_{4},x_{5}) \nonumber\\
    &= \int \left(\prod_{i=3}^{6}d^{d}x_{i}\right) \left\langle \mathcal{O}_{1}(x_{1})\mathcal{O}_{2}(x_{2})\mathcal{O}_{\underline{3}}(x_{3})\right\rangle \langle \mathcal{O}_{\underline{\widetilde{3}}}(x_{3})\mathcal{O}_{\underline{4}}(x_{4})\mathcal{O}_{\underline{5}}(x_{5})\rangle \times \nonumber\\
    & \hspace{3.0cm}\langle \mathcal{O}_{\underline{\widetilde{5}}}(x_{5})\mathcal{O}_{\underline{\widetilde{4}}}(x_{4})\mathcal{O}_{\underline{6}}(x_{6})\rangle \langle\hspace{-0.09cm}\langle \mathcal{O}_{\underline{\widetilde{6}}}(x_{6})\rangle\hspace{-0.09cm}\rangle  
\end{align}

\noindent We can then apply the following bubble identity \cite{Meltzer:2019nbs, Karateev:2018oml}
\begin{align}
    \label{bubbleid}
     & \int d^{d}x_{4}d^{d}x_{5} \ \langle \mathcal{O}_{\underline{\widetilde{3}}}(x_{3})\mathcal{O}_{\underline{4}}(x_{4})\mathcal{O}_{\underline{5}}(x_{5})\rangle  \langle \mathcal{O}_{\underline{\widetilde{5}}}(x_{5})\mathcal{O}_{\underline{\widetilde{4}}}(x_{4})\mathcal{O}_{\underline{6}}(x_{6})\rangle = B_{\underline{\Delta_{3}}} \delta^{d}(x_{3}-x_{6}) \delta(\nu_{3}-\nu_{6})
\end{align}

\noindent in Eq.~\eqref{blkbranepropcorr2} to get,
\begin{align}
    \label{blkbranepropcorr3}
    & \int d^{d}x_{4}d^{d}x_{5} \ \Psi^{12\underline{5}\underline{4}}_{\frac{d}{2}+i\nu_{3},0}(x_{1},x_{2},x_{4},x_{5})  \mathcal{F}^{\underline{\widetilde{5}}\underline{\widetilde{4}}}_{\frac{d}{2}+i\nu_{6},0}(x_{4},x_{5}) \nonumber\\
    &= B_{\underline{\Delta_{3}}} \delta(\nu_{3}-\nu_{6})\int d^{d}x_{3} \left\langle \mathcal{O}_{1}(x_{1})\mathcal{O}_{2}(x_{2})\mathcal{O}_{\underline{3}}(x_{3})\right\rangle  \left\langle \mathcal{O}_{\underline{\widetilde{3}}}(x_{3})\right\rangle  \nonumber\\
    & = B_{\underline{\Delta_{3}}} \delta(\nu_{3}-\nu_{6}) \mathcal{F}^{1,2}_{\frac{d}{2}+i\nu_{3},0}(x_{1},x_{2}), 
\end{align}

\noindent where the scalar bubble-coefficient $B_{\underline{\Delta_{3}}}$ is given by \cite{Meltzer:2019nbs, Karateev:2018oml}
\begin{align}
B_{\underline{\Delta_{3}}} =\frac{2\pi^{\frac{3d}{2}}}{\Gamma(\frac{d}{2})} \ \frac{\Gamma(\underline{\Delta_{3}}-\frac{d}{2})\Gamma(\frac{d}{2}-\underline{\Delta_{3}})}{\Gamma(\underline{\Delta_{3}})\Gamma(d-\underline{\Delta_{3}})} .
\end{align}

\noindent Plugging the above result into the expression in Eq.~\eqref{blkbranepropcorr1} and performing the $\nu_{6}$ spectral integral using the delta function in Eq.~\eqref{blkbranepropcorr3}, we obtain the following bulk channel partial wave expansion.
\begin{align}
\label{blkbranepropcorr4}
    W(x_{1},x_{2}) =& \int_{-\infty}^{\infty} \left(\prod_{i=3}^{5} d\nu_{i}\   P(\nu_{i},\Delta_{i})\right) P(\nu_{3},\Delta_{6})  \lambda_{12\underline{3}} \lambda_{\underline{\widetilde{3}}\hspace{0.02cm}\underline{4}\hspace{0.02cm}\underline{5}} \lambda_{\underline{\widetilde{5}}\hspace{0.02cm}\underline{\widetilde{4}}\hspace{0.02cm}\underline{3}} a_{\underline{\widetilde{3}}} B_{\underline{\Delta_{3}}}  \mathcal{F}^{1,2}_{\frac{d}{2}+i\nu_{3},0}(x_{1},x_{2}). 
\end{align}

\noindent From the above result, let us take note of the following set of poles in $\underline{\Delta_{3}}$
\begin{align}
    \label{polesb2brcorr}
    &\underline{\Delta_{3}} =\Delta_{3}, \ \underline{\Delta_{3}} =\Delta_{6}, \ \underline{\Delta_{3}} =\Delta_{1}+\Delta_{2}+2n, \nonumber\\
    & \underline{\Delta_{3}} =\underline{\Delta_{4}}+\underline{\Delta_{5}}+2n, \ \underline{\Delta_{3}} =\underline{{\widetilde{\Delta}_{4}}}+\underline{\widetilde{\Delta}_{5}}+2n. 
\end{align}

\noindent Closing the $\nu_{3}$ contour, the poles at $\Delta_{3}, \Delta_{6}$ give rise to the exchange of the operators $\mathcal{O}_{3}, \mathcal{O}_{6}$ in the bulk channel block expansion. We also get the exchange of double-twist operators $[\mathcal{O}_{1}\mathcal{O}_{2}]_{n,0}$ from the poles at $\Delta_{1}+\Delta_{2}+2n$. The poles at $\underline{\Delta_{4}}+\underline{\Delta_{5}}+2n$ $  \underline{{\widetilde{\Delta}_{4}}}+\underline{\widetilde{\Delta}_{5}}+2n$, lead to the exchange of the double-twist operators $[\mathcal{O}_{4}\mathcal{O}_{5}]_{n,0}$, after performing the $\nu_{4},\nu_{5}$ integrals by picking up the poles in the measure factors $P(\nu_{4},\Delta_{4}), P(\nu_{5},\Delta_{5})$ at $\frac{d}{2}+i\nu_{4}=\Delta_{4},\frac{d}{2}+i\nu_{5}=\Delta_{5}$ and $\frac{d}{2}-i\nu_{4}=\Delta_{4},\frac{d}{2}-i\nu_{5}=\Delta_{5}$, respectively.


\section{Three point correlators}
\label{3pt_corr}

In this section, we turn our attention to $3$-point correlation functions involving one CFT bulk operator and two defect operators. Such correlators will be referred to as form factors \cite{Girault:2025kzt}. We will consider the contribution of some tree-level Witten diagrams to this observable, and decompose it in terms of the so-called form factor blocks, which we define below. 

\vskip 4pt
Let us denote the correlation function of interest to us here as  $\langle \hspace{-0.09cm}\langle \widehat{\mathcal{O}}_{\hat{\Delta}_{1}}(\hat{x}_{1})\widehat{\mathcal{O}}_{\hat{\Delta}_{2}}(\hat{x}_{2})\mathcal{O}_{\Delta_{3}}(x_{3})\rangle\hspace{-0.09cm}\rangle$, where $\widehat{\mathcal{O}}_{\hat{\Delta}}$ denotes a scalar primary operator on the defect, and $\mathcal{O}$ is a scalar primary operator in the bulk. This $3$-point function can be written as 
\begin{align}
\label{formfactor}
    \langle \hspace{-0.09cm}\langle \widehat{\mathcal{O}}_{\hat{\Delta}_{1}}(\hat{x}_{1})\widehat{\mathcal{O}}_{\hat{\Delta}_{2}}(\hat{x}_{2})\mathcal{O}_{\Delta_{3}}(x_{3})\rangle\hspace{-0.09cm}\rangle= \frac{|x_{3,\perp}|^{-\Delta_{3}}}{|\hat{x}_{12}|^{\hat{\Delta}_{1}+\hat{\Delta}_{2}}}\left(\frac{|x_{3,\perp}|^{2}+\hat{x}_{23}^{2}}{|x_{3,\perp}|^{2}+\hat{x}_{13}^{2}}\right)^{\frac{\hat{\Delta}_{12}}{2}} \mathcal{H}(w), 
\end{align}

\noindent where $\hat{\Delta}_{12}=\hat{\Delta}_{1}-\hat{\Delta}_{2}$ and $w$ is a cross-ratio defined as
\begin{align}
    w=\frac{|x_{3,\perp}|^{2}\hat{x}^{2}_{12}}{(|x_{\perp,3}|^{2}+\hat{x}_{13}^{2})(|x_{3,\perp}|^{2}+\hat{x}_{23}^{2})} \ .
\end{align}

\noindent The function $\mathcal{H}(w)$ is not fixed by conformal symmetry. Using either the bulk-to-defect operator expansion or the OPE between the two defect primary operators, $\mathcal{H}(w)$ can be decomposed as follows \cite{Girault:2025kzt}
\begin{align}
\label{formfacope}
    \mathcal{H}(w) = \sum_{\widehat{\mathcal{O}}} \hat{\lambda}_{12\widehat{\mathcal{O}}} b_{\widehat{\mathcal{O}}\mathcal{O}_{3}} H_{\hat{\Delta}}(w), 
\end{align}

\noindent where $\hat{\lambda}_{12\widehat{\mathcal{O}}}$ is the OPE coefficient involving three defect operators, $b_{\widehat{\mathcal{O}}\mathcal{O}_{3}}$ is the bulk-defect coefficient. The kinematic function $H_{\hat{\Delta}}(w)$ is the form-factor block which is determined by $SO(p+1,1)\times SO(d-p)$ symmetry and is given by
\begin{align}
\label{formfacblock}
    H_{\hat{\Delta}}(w)= w^{\frac{\hat{\Delta}}{2}} \ _{2}F_{1}\left(\frac{\hat{\Delta}+\hat{\Delta}_{12}}{2}, \frac{\hat{\Delta}-\hat{\Delta}_{12}}{2},\hat{\Delta}+1-\frac{p}{2},w\right) .
\end{align}

\noindent Note that only scalar defect operators can appear in the block expansion in Eq.~\eqref{formfacope}. This is due to the fact that the $3$-point function involving two scalar defect primaries and a defect primary with transverse spin-$s$ vanishes. 

\subsection{Witten diagram for form factor}

\begin{figure}[htp]
    \centering
    \includegraphics[width=3.5cm]{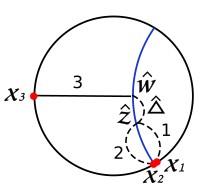}
    \caption{Form Factor diagram involving integral over two brane points.}
    \label{fig:3ptform}
\end{figure}

We now consider the contribution of a tree-level Witten diagram involving the exchange of a brane scalar field, to the form-factor defined in Eq.\eqref{formfactor}. This is shown in figure \ref{fig:3ptform} and is given by the following integral 
\begin{align}\label{formfacdiag1}
    &W_{\hat{\Delta}}(\hat{x}_{1},\hat{x}_{2},x_{3})= \int\limits_{AdS_{p+1}} d^{p+1}\hat{z} d^{p+1}\hat{w} \ \widehat{K}_{\hat{\Delta}_{1}}(\hat{x}_{1},\hat{z})\widehat{K}_{\hat{\Delta}_{2}}(\hat{x}_{2},\hat{z})\widehat{G}_{\hat{\Delta}}(\hat{z},\hat{w}) K_{\Delta_{3}}(x_{3},\hat{w}) \ , 
\end{align}

\noindent where the various propagators involved in the above expression have been defined in Section \ref{sec:prelim}. Now, to derive the conformal block expansion of this Witten diagram, we apply the spectral and split representation of the brane-to-brane propagator. This allows us to factorize the two $AdS_{p+1}$ integrals in Eq.~\eqref{formfacdiag1} as,
\begin{align}
    \label{formfacdiag2}
    &W_{\hat{\Delta}}(\hat{x}_{1},\hat{x}_{2},x_{3})\nonumber\\
    &= \int_{-\infty}^{\infty} d\nu \ \widehat{P}(\nu,\hat{\Delta}) \int\limits_{\partial AdS_{p+1}} d^{p}\hat{x} \bigg[\int\limits_{AdS_{p+1}} d^{p+1}\hat{z} \ \widehat{K}_{\hat{\Delta}_{1}}(\hat{x}_{1},\hat{z})\widehat{K}_{\hat{\Delta}_{2}}(\hat{x}_{2},\hat{z})\widehat{K}_{\frac{p}{2}+i\nu}(\hat{x},\hat{z})\bigg] \times \nonumber\\
    & \hspace{5.0cm}\bigg[\int\limits_{AdS_{p+1}} d^{p+1}\hat{w} \ \widehat{K}_{\frac{p}{2}+i\nu}(\hat{x},\hat{z}) K_{\Delta_{3}}(x_{3},\hat{w})\bigg]. 
\end{align}

\noindent In Eq.~\eqref{formfacdiag2}, the $AdS_{p+1}$ integral that involves three brane-to-defect propgators yields
\begin{align}
    \label{defect3ptdiag}
    & \int\limits_{AdS_{p+1}} d^{p+1}\hat{z}_{1} \ \widehat{K}_{\hat{\Delta}_{1}}(\hat{x}_{1},\hat{z}_{1})\widehat{K}_{\hat{\Delta}_{2}}(\hat{x}_{2},\hat{z}_{1})\widehat{K}_{\frac{p}{2}+i\nu}(\hat{x},\hat{z}_{1})= \hat{\lambda}_{12\underline{\hat{\Delta}}} \langle\hspace{-0.09 cm}\langle \widehat{\mathcal{O}}_{\hat{\Delta}_{1}}(\hat{x}_{1})\widehat{\mathcal{O}}_{\hat{\Delta}_{2}}(\hat{x}_{2})\widehat{\mathcal{O}}_{\frac{p}{2}+i\nu}(\hat{x})\rangle\hspace{-0.09 cm}\rangle, 
\end{align}

\noindent where $\langle\hspace{-0.09 cm}\langle \widehat{\mathcal{O}}_{\hat{\Delta}_{1}}\widehat{\mathcal{O}}_{\hat{\Delta}_{2}}\widehat{\mathcal{O}}_{\frac{p}{2}+i\nu}\rangle\hspace{-0.09 cm}\rangle$ denotes the $3$-point conformal structure for scalar operators in $p$-dimensional CFT. The coefficient $\hat{\lambda}_{12\underline{\hat{\Delta}}}$ takes the same form as given in Eq.\eqref{3ptcoeff} with $(\Delta_{1},\Delta_{2},\underline{\Delta},d)$ replaced by $(\hat{\Delta}_{1},\hat{\Delta}_{2},\underline{\hat{\Delta}},p)$. We have already encountered the second $AdS_{p+1}$ integral in Section \ref{sec:contactdiag}. This generates a two-point function involving a bulk and a defect CFT operator. Thus, we obtain, 
\begin{align}
\label{formfacdiag3}
    W_{\hat{\Delta}}(\hat{x}_{1},\hat{x}_{2},x_{3}) &= \int_{-\infty}^{\infty} d\nu \ \widehat{P}(\nu,\hat{\Delta}) \ \hat{\lambda}_{12\underline{\hat{\Delta}}} b_{3\underline{\widetilde{\hat{\Delta
    }}}}\times \nonumber \\
    &\int\limits_{\partial AdS_{p+1}} d^{p}\hat{x} \ \langle\hspace{-0.09cm}\langle \widehat{\mathcal{O}}_{\hat{\Delta}_{1}}(\hat{x}_{1})\widehat{\mathcal{O}}_{\hat{\Delta}_{2}}(\hat{x}_{2})\widehat{\mathcal{O}}_{\frac{p}{2}+\nu}(\hat{z})\rangle \hspace{-0.09cm}\rangle \langle \hspace{-0.09cm}\langle \widehat{\mathcal{O}}_{\frac{p}{2}-\nu}(\hat{x})\mathcal{O}_{\Delta_{3}}(x_{3})\rangle\hspace{-0.09cm}\rangle .
\end{align}


\noindent Now, the integral over the boundary of $AdS_{p+1}$ containing the product of the $3$-point and $2$-point conformal structures can be identified with the form-factor partial wave, which we denote as
\begin{align}
    \label{formfacpw}
    \mathbb{H}_{\hat{\Delta}}(\hat{x}_{1},\hat{x}_{2},x_{3}) = \frac{|x_{3,\perp}|^{-\Delta_{3}}}{|\hat{x}_{12}|^{\hat{\Delta}_{1}+\hat{\Delta}_{2}}}\left(\frac{|x_{3,\perp}|^{2}+\hat{x}_{23}^{2}}{|x_{3,\perp}|^{2}+\hat{x}_{13}^{2}}\right)^{\frac{\hat{\Delta}_{12}}{2}} \left[\mathcal{N}_{\hat{\Delta}} H_{\hat{\Delta}}(w)+ \mathcal{N}_{p-\hat{\Delta}} H_{p-\hat{\Delta}}(w)\right] ,
\end{align}

\noindent where $H_{\hat{\Delta}}(w)$ is the form factor block defined in Eq.~\eqref{formfacblock}. The normalization factor $\mathcal{N}_{\hat{\Delta}}$ is given in Appendix \ref{app:formfac}, where we also demonstrate explicitly that the conformal integral in Eq.~\eqref{formfacdiag3} gives the partial wave defined above. Using these results, we thus get the following representation of the form-factor Witten diagram,
\begin{align}
\label{formfacdiag4}
    & W_{\hat{\Delta}}(\hat{x}_{1},\hat{x}_{2},x_{3}) =\frac{|x_{3,\perp}|^{-\Delta_{3}}}{|\hat{x}_{12}|^{\hat{\Delta}_{1}+\hat{\Delta}_{2}}}\left(\frac{|x_{3,\perp}|^{2}+\hat{x}_{23}^{2}}{|x_{3,\perp}|^{2}+\hat{x}_{13}^{2}}\right)^{\frac{\hat{\Delta}_{12}}{2}} \mathcal{H}_{\hat{\Delta}}(w), 
\end{align}

\noindent where $\mathcal{H}_{\hat{\Delta}}(w)$ is given by
\begin{align}
\label{formfacdiag5}
   \mathcal{H}_{\hat{\Delta}}(w) &= 2\int_{-\infty}^{\infty} d\nu \ \widehat{P}(\nu,\hat{\Delta}) \ \hat{\lambda}_{12\underline{\hat{\Delta}}} b_{3\underline{\widetilde{\hat{\Delta
    }}}} \ \mathcal{N}_{\frac{p}{2}+i\nu} \ H_{\frac{p}{2}+i\nu}(w) .
\end{align}

\noindent To arrive at Eq.~\eqref{formfacdiag5}, we have used the $\nu\leftrightarrow -\nu$ shadow symmetry of the spectral integral to substitute the partial wave with the block, thereby introducing the overall factor of $2$ and the normalization $\mathcal{N}_{\frac{p}{2}+i\nu}$. We can now write the above result as a sum over form-factor blocks by deforming the $\nu$-contour and picking up the simple pole contributions at
\begin{align}
    \label{formfacpoles}
    i\nu =\hat{\Delta}-\frac{p}{2}, \quad i\nu=\hat{\Delta}_{1}+\hat{\Delta}_{2}+2n-\frac{p}{2}, \quad i\nu=\Delta_{3}+2m-\frac{p}{2}
\end{align}

\noindent where $n,m$ are non-negative integers. The first pole in Eq.~\eqref{formfacpoles} is from the spectral measure $\widehat{P}(\nu,\hat{\Delta})$ and gives the exchange of a defect primary operator with dimension $\hat{\Delta}$ in the block expansion. The second set of poles originate from the $3$-point OPE coefficient $\hat{\lambda}_{12\underline{\hat{\Delta}}}$. These correspond to the exchange of defect double-twist operators $\widehat{\mathcal{O}}_{1}(\partial^{2})^{n}\widehat{\mathcal{O}}_{2}$. The last set of poles in Eq.\eqref{formfacpoles} are due to the coefficient $b_{3\underline{\widetilde{\hat{\Delta
    }}}}$. Their contribution denotes the exchange of transverse derivative operators of the type $\partial_{i}\partial^{i} \mathcal{O}_{3}$. The block decomposition of the form-factor Witten diagram in Eq.~\eqref{formfacdiag5} is then given by
\begin{align}
    \label{formfacblockexp}
    \mathcal{H}_{\hat{\Delta}}(w) &= c_{\hat{\Delta}}H_{\hat{\Delta}}(w)+\sum_{n=0}^{\infty} \alpha_{n} H_{\hat{\Delta}_{1}+\hat{\Delta}_{2}+2n}(w)+\sum_{m=0}^{\infty} \beta_{m} H_{\Delta_{3}+2m}(w), 
\end{align}   

\noindent where the coefficients $c_{\hat{\Delta}}, \alpha_{n},\beta_{m}$ are given by 

\begin{align}
    & c_{\hat{\Delta}}= \frac{\pi^{\frac{p}{2}}}{32}\frac{\Gamma\left(\frac{\hat{\Delta}_{1}+\hat{\Delta}_{2}-\hat{\Delta}}{2}\right)\Gamma\left(\frac{\hat{\Delta}_{1}+\hat{\Delta}_{2}+\hat{\Delta}-p}{2}\right)\Gamma\left(\frac{\hat{\Delta}_{12}+\hat{\Delta}}{2}\right)\Gamma\left(\frac{\hat{\Delta}-\hat{\Delta}_{12}}{2}\right)\Gamma\left(\frac{\Delta_{3}-\hat{\Delta}}{2}\right)\Gamma\left(\frac{\Delta_{3}+\hat{\Delta}-p}{2}\right)}{(2\hat{\Delta}-p)\Gamma\left(\hat{\Delta}-\frac{p}{2}\right)\Gamma(\hat{\Delta}_{1})\Gamma(\hat{\Delta}_{2})\Gamma(\Delta_{3})}, \\
    &  \alpha_{n} = \frac{(-1)^{n+1}\pi^{\frac{p}{2}}\Gamma(\hat{\Delta}_{1}+n)\Gamma(\hat{\Delta}_{2}+n)\Gamma\left(\frac{\hat{\Delta}_{1}+\hat{\Delta}_{2}+\Delta_{3}+2n-p}{2}\right)\Gamma\left(\frac{\Delta_{3}-\hat{\Delta}_{1}-\hat{\Delta}_{2}-2n}{2}\right)}{16 n!(\hat{\Delta}_{1}+\hat{\Delta}_{2}+2n-\hat{\Delta})(\hat{\Delta}_{1}+\hat{\Delta}_{2}+2n+\hat{\Delta}-p)\left(\hat{\Delta}_{1}+\hat{\Delta}_{2}+n-\frac{p}{2}\right)_{n} \Gamma(\hat{\Delta}_{1})\Gamma(\hat{\Delta}_{2})\Gamma(\Delta_{3})}, \\
    & \beta_{m} = \frac{(-1)^{m+1}\pi^{\frac{p}{2}}}{16m!}\frac{\Gamma\left(\frac{\hat{\Delta}_{1}+\hat{\Delta}_{2}-\Delta_{3}-2m}{2}\right)\Gamma\left(\frac{\hat{\Delta}_{1}+\hat{\Delta}_{2}+\Delta_{3}+2m-p}{2}\right)\Gamma\left(\frac{\hat{\Delta}_{12}+\Delta_{3}+2m}{2}\right) }{(\Delta_{3}+2m-\hat{\Delta})(\Delta_{3}+2m+\hat{\Delta}-p)\Gamma\left(\Delta_{3}+2m-\frac{p}{2}\right)}\times \nonumber\\
    & \hspace{1.0cm}\times \frac{\Gamma\left(\frac{\Delta_{3}+2m-\hat{\Delta}_{12}}{2}\right)\Gamma\left(\Delta_{3}+m-\frac{p}{2}\right)}{\Gamma(\hat{\Delta}_{1})\Gamma(\hat{\Delta}_{2})\Gamma(\Delta_{3})}.
\end{align}


\section{Crossing kernel: defect-to-bulk channel}
\label{sec:def6j}

In this section, we will consider the crossing kernel for defect CFTs. In CFTs without defects, the crossing kernel or the $6j$ symbol of the conformal group encodes crossing symmetry by determining how a conformal partial wave in one channel can be decomposed in terms of partial waves in another channel. Explicit expressions for the $6j$ symbol have been computed in two and four dimensions in \cite{Liu:2018jhs}, using the Lorentzian inversion formula \cite{Caron-Huot:2017vep}. The $6j$ symbol for the $3$-dimensional conformal group has been studied in \cite{Albayrak:2020rxh}. Crossing kernels for boundary and cross-cap CFTs have been computed in \cite{Hogervorst:2017kbj}, using a generalization of the $\alpha$ space formalism developed in \cite{Hogervorst:2017sfd}. 

\vskip 4pt
For CFTs in the presence of defects, we can have two types of crossing kernels, which we will refer to as the defect-to-bulk channel kernel and the bulk-to-defect channel kernel. This terminology is used to distinguish whether we are decomposing the defect channel partial wave in terms of the bulk channel partial waves or vice versa. In this section, we will only consider the crossing kernel in the defect-to-bulk channel. More precisely, we are interested in evaluating the kernel $\mathcal{J}$ that appears in the following equation.
\begin{align}
\label{deftobulk6j}
\widehat{\mathcal{F}}_{\widehat{\Delta},s}(x_1,x_2) = \sum_{\ell=0}^{\infty}\int_{\frac{d}{2}}^{\frac{d}{2}+i \infty} \frac{d \Delta}{2\pi i} \,\mathcal{J}_{d,p}(\Delta,\ell;\hat{\Delta},s|\Delta_{1},\Delta_{2})  \, \mathcal{F}_{\Delta,\ell}(x_1,x_2) , 
\end{align}

\noindent where, $\widehat{\mathcal{F}}_{\widehat{\Delta},s}(x_1,x_2)$ is the defect channel partial wave defined in Eq.~\eqref{defparwave1} and $\mathcal{F}_{\Delta,\ell}(x_1,x_2)$ is the bulk-channel partial wave given in Eq.~\eqref{bulkpw}. The external operators in our case are scalars with dimensions $\Delta_{1},\Delta_{2}$. Here we suppressed the explicit dependence of the bulk channel partial wave on $\Delta_{1},\Delta_{2}$. Note that the defect channel partial wave is independent of the external scaling dimensions. 

\vskip 4pt
Using the orthogonality property of the bulk channel partial waves belonging to the principal series representation, we can get a Euclidean inversion formula for the crossing kernel in Eq.~\eqref{deftobulk6j}. However, following \cite{Liu:2018jhs}, we will employ the bulk channel Lorentzian inversion formula derived in \cite{Liendo:2019jpu}. This yields the following integral formula for the kernel
\begin{align}
\label{deftobulk6j1}
\mathcal{J}_{d,p}(\Delta,\ell;\hat{\Delta},s|\Delta_{1},\Delta_{2})= \kappa_{\Delta+\ell} \int_{0}^1\int_{0}^1 dz d \bar{z} \, \mu(z,\bar{z}) \, f_{\ell+d-1,\Delta-d+1}(z,\bar{z}) \, \mathrm{dDisc} \left[\widehat{\mathcal{F}}_{\widehat{\Delta},s}(x_1,x_2)\right], 
\end{align}

\noindent where the constant $\kappa_{\beta}$ and the measure $\mu(z,\bar{z})$ are
\begin{align}
\kappa_{\beta} &= \frac{\Gamma \left( \frac{\beta}{2} \right)\Gamma \left( \frac{\beta}{2}+a \right) \Gamma \left( \frac{\beta}{2}-a \right)}{2 \pi^2 \Gamma(\beta) \Gamma(\beta-1)}, \quad a=\frac{\Delta_{2}-\Delta_{1}}{2},\\
\mu(z,\bar{z}) &= \left(\frac{(1-z)(1-\bar{z})}{\sqrt{z \bar{z}}} \right)^{\frac{\Delta_1+\Delta_2}{2}} \frac{|z-\bar{z}|^{d-p-2} |1-z \bar{z}|^{p}}{[(1-z)(1-\bar{z})]^d} .
\end{align}

\noindent In Eq.~\eqref{deftobulk6j1}, $\mathrm{dDisc}[F(z,\bar{z})]$ is the double discontinuity of the function $F(z,\bar{z})$, and is defined as follows
\begin{align}
\mathrm{dDisc}[F(z,\bar{z})] = \cos(\pi a)\, F(z,\bar{z}) 
- \frac{1}{2} e^{- \frac{i \pi (\Delta_1+\Delta_2)}{2}} F^\circlearrowleft(z,\bar{z}) 
- \frac{1}{2} e^{\frac{i \pi (\Delta_1+\Delta_2)}{2}} F^\circlearrowright(z,\bar{z}) ,
\end{align}

\noindent where $F^\circlearrowleft$ and $F^\circlearrowright$ indicate that $\bar{z}$ is analytically continued around $0$ in the clockwise and counterclockwise direction, respectively, while $z$ is held fixed. To compute the dDisc of the defect channel partial wave, we first write the partial wave in terms of the defect channel blocks $\widehat{f}_{\hat{\Delta},s}(z,\bar{z})$ as given in Eq.~\eqref{defparwave1}. Then using the following expression for the blocks from Eq.~\eqref{rlt1}
\begin{align}
\label{defblock}
\widehat{f}_{\hat{\Delta},s}(z,\bar{z})= 2^{-s}z^{\frac{\hat{\Delta}-s}{2}} \bar{z}^{\frac{\hat{\Delta}+s}{2}} \ _{2}F_{1}\left(\hat{\Delta},\frac{p}{2},\hat{\Delta}+1-\frac{p}{2},z \bar{z}\right) {}_{2}F_{1}\left(-s,\frac{q}{2}-1,2-s-\frac{q}{2}, \frac{z}{\bar{z}}\right), 
\end{align}

\noindent we get the double discontinuity of the block to be,
\begin{align}
    \label{dDiscdefectblock}
     \mathrm{dDisc}\left[\widehat{f}_{\hat{\Delta},s}(z,\bar{z})\right]= 2\sin\left[\frac{\pi}{2}(\hat{\Delta}-s-\Delta_{1})\right]\sin\left[\frac{\pi}{2}(\hat{\Delta}-s-\Delta_{2})\right]\widehat{f}_{\hat{\Delta},s}(z,\bar{z}).
\end{align}

\vskip 4pt
The bulk channel block $f_{\ell+d-1,\Delta-d+1}(x,\bar{x})$ in equation \eqref{deftobulk6j1} does not have a closed form expression in general. However, there are some special cases where they can be written in terms of $4$-point conformal blocks \cite{Isachenkov:2018pef, Liendo:2019jpu}. In particular, we will consider $p=0$ and $p=2$ dimensional defects, where we have
\begin{align}
\label{bulk4ptblockrels1}
f^{(d,p=0)}_{\Delta,\ell}(z,\bar{z}) &= (z \bar{z})^{\frac{a}{2}} g^{(d,a,0)}_{\Delta,\ell}(1-z,1-\bar{z}), \\
f^{(d,p=2)}_{\Delta,\ell}(z,\bar{z}) &= \frac{(1-z)(1-\bar{z})}{1-z \bar{z}}(z \bar{z})^{\frac{a}{2}} g^{(d-2,a,0)}_{\Delta-1,\ell+1}(1-z,1-\bar{z}) \label{bulk4ptblockrels2},   
\end{align}

\noindent where $g^{(d,a,b)}_{\Delta,\ell}(z,\bar{z})$ is the s-channel 4-pt conformal block with $a=\frac{\Delta_2-\Delta_1}{2}$ and $b=\frac{\Delta_3-\Delta_4}{2}$. Their closed form expressions in $d=2$ and $d=4$ are known to be,
\begin{align}
g^{(d=2,a,b)}_{\Delta,\ell}(z,\bar{z}) &= \frac{k_{\Delta-\ell}(z)k_{\Delta+\ell}(\bar{z}) + k_{\Delta+\ell}(z)k_{\Delta-\ell}(\bar{z})}{1+ \delta_{\ell,0}}, \\
g^{(d=4,a,b)}_{\Delta,l}(z,\bar{z}) &= \frac{z \bar{z}}{\bar{z}-z} \left( k_{\Delta-\ell-2}(z)k_{\Delta+\ell}(\bar{z}) - k_{\Delta+\ell}(z)k_{\Delta-\ell-2}(\bar{z}) \right),  \\
\textrm{with, } \ \ k_{\beta}(z) &= z^{\frac{\beta}{2}} \, {}_2F_1 \left( \frac{\beta}{2}+a, \frac{\beta}{2}+b,\beta,z \right) . \notag
\end{align}

Evaluating the inversion formula integral analytically in the above two cases for general $d$ is still difficult. Therefore, we focus on four representative cases: $d=2,\,p=0$; $d=4,\,p=0$; $d=4,\,p=2$; and $d=6,\,p=2$. These cases are simpler because, using Eqs.~\eqref{defblock}, \eqref{bulk4ptblockrels1}, \eqref{bulk4ptblockrels2}, and the expressions for the $d=2,4$ conformal blocks, we can see that the integrand in the inversion formula factorizes into a product of a function of $z$ and a function of $\bar{z}$ alone. As a result, the double integral over the unit square reduces to the product of two single integrals over the unit interval. The crossing kernel in the defect-to-bulk channel can then be written as
\begin{align}
\label{deftobulk6j2}
\mathcal{J}_{d,p}(\Delta,\ell;\hat{\Delta},s|\Delta_{1},\Delta_{2})& =   \widehat{K}_{\widehat{\Delta}}  \mathcal{B}_{d,p}(\Delta,\ell;\hat{\Delta},s|\Delta_{1},\Delta_{2})\nonumber\\
&+ \widehat{K}_{p-\widehat{\Delta}}  \mathcal{B}_{d,p}(\Delta,\ell;p-\hat{\Delta},s|\Delta_{1},\Delta_{2}), 
\end{align}

\noindent where the function $\mathcal{B}_{d,p}(\Delta,\ell;\hat{\Delta},s|\Delta_{1},\Delta_{2})$ arises from inverting the defect channel block. The second term in the right-hand side of Eq.~\eqref{deftobulk6j2} is due to the inversion of the shadow block. The explicit expressions for $\mathcal{B}_{d,p}(\Delta,\ell;\hat{\Delta},s|\Delta_{1},\Delta_{2})$ in these above mentioned cases are given below. 
\begin{align}
& \mathcal{B}_{d=2,p=0}(\Delta,\ell;\hat{\Delta},s|\Delta_{1},\Delta_{2}) \nonumber\\
&= \frac{A_{\Delta,\ell,\widehat{\Delta},s}^{\Delta_1,\Delta_2}}{1+\delta_{l,0}} \left(\mathcal{Q}_{\frac{\widehat{\Delta}-s-\Delta_{1}}{2},\frac{\Delta_{1}+\Delta_{2}+\Delta+\ell-4}{2}}^{\Delta+\ell,a} \, \mathcal{Q}_{\frac{\widehat{\Delta}+s-\Delta_{1}}{2},\frac{\Delta_{1}+\Delta_{2}-\Delta+\ell-2}{2}}^{2-\Delta+\ell,a} + (s \leftrightarrow -s)\right) \\
& \mathcal{B}_{d=4,p=2}(\Delta,\ell;\hat{\Delta},s|\Delta_{1},\Delta_{2}) \nonumber\\
&= \frac{A_{\Delta,\ell,\widehat{\Delta},s}^{\Delta_1,\Delta_2}}{1+\delta_{l,0}} \left(\mathcal{Q}_{\frac{\widehat{\Delta}-s-\Delta_{1}}{2},\frac{\Delta_{1}+\Delta_{2}+\Delta+\ell-6}{2}}^{\Delta+\ell,a} \, \mathcal{Q}_{\frac{\widehat{\Delta}+s-\Delta_{1}}{2},\frac{\Delta_{1}+\Delta_{2}-\Delta+\ell-2}{2}}^{4-\Delta+\ell,a} + (s \leftrightarrow -s)\right) \\
& \mathcal{B}_{d=4,p=0}(\Delta,\ell;\hat{\Delta},s|\Delta_{1},\Delta_{2}) \nonumber\\
&= A_{\Delta,\ell,\widehat{\Delta},s}^{\Delta_1,\Delta_2} \left(\mathcal{Q}_{\frac{\widehat{\Delta}-s-\Delta_{1}}{2},\frac{\Delta_{1}+\Delta_{2}+\Delta+\ell-6}{2}}^{\Delta+\ell,a} \, \mathcal{Q}_{\frac{\widehat{\Delta}+s-\Delta_{1}+2}{2},\frac{\Delta_{1}+\Delta_{2}-\Delta+\ell-2}{2}}^{4-\Delta+\ell,a} - (s \leftrightarrow -(s+2))\right) \\
&\mathcal{B}_{d=6,p=2}(\Delta,\ell;\hat{\Delta},s|\Delta_{1},\Delta_{2}) \nonumber\\
&= A_{\Delta,\ell,\widehat{\Delta},s}^{\Delta_1,\Delta_2} \left(\mathcal{Q}_{\frac{\widehat{\Delta}-s-\Delta_{1}}{2},\frac{\Delta_{1}+\Delta_{2}+\Delta+\ell-8}{2}}^{\Delta+\ell,a} \, \mathcal{Q}_{\frac{\widehat{\Delta}+s-\Delta_{1}+2}{2},\frac{\Delta_{1}+\Delta_{2}-\Delta+\ell-2}{2}}^{6-\Delta+\ell,a} - (s \leftrightarrow -(s+2))\right) 
\end{align}

\noindent where $a= \frac{\Delta_2-\Delta_1}{2}$, and
\begin{align}
& A_{\Delta,\ell,\widehat{\Delta},s}^{\Delta_1,\Delta_2} = \frac{\kappa_{\Delta+\ell}}{2^{s}}\ \sin\left[\frac{\pi}{2}(\hat{\Delta}-s-\Delta_{1})\right]\sin\left[\frac{\pi}{2}(\hat{\Delta}-s-\Delta_{2})\right] ,   \\
&\mathcal{Q}_{u,v}^{\beta,a} = \int_{0}^1 \, dz \, z^u \, (1-z)^v \, {}_2F_1 \left( \frac{\beta}{2}+a, \frac{\beta}{2}, \beta ,1-z \right) \notag \\
& \hspace{0.5cm}=\frac{\Gamma(u+1) \Gamma(v+1)}{\Gamma(u+v+2)} \, {}_3F_2 \left( \frac{\beta}{2}+a, \frac{\beta}{2}, v+1 ; \beta , u+v+2 ; 1 \right). 
\end{align}

As a simple application of the defect-to-bulk channel crossing kernel, we consider a one-loop one-point function diagram, depicted by Fig.~\ref{fig:1ptloop1}(b), in Appendix \ref{app:1ptloop6j}. We show there how the loop-corrected one-point function coefficient is determined by the defect-to-bulk channel crossing kernel.


\section{Conclusions and Future Directions}
\label{sec:concl}

In this paper, we have studied Witten diagrams for holographic CFTs in the presence of a defect. We derived the direct channel conformal block decomposition of several two-point Witten diagrams, at both tree and one-loop level, and a tree-level three-point function involving one bulk CFT operator and two defect operators. We obtained recursion relations for the coefficients in the crossed channel expansion of tree-level $2$-point exchange diagrams. The seed coefficients for these recursion relations were evaluated using the Mellin representation of Witten diagrams. Applying the Lorentzian inversion formula for defect CFTs, we have also presented results for the crossing kernel in the defect-to-bulk channel, for zero-dimensional defects in $d=2,4$ and surface defects in $d=4,6$ dimensions. We now outline some future research avenues that we hope to pursue.

A natural extension of the work done in this paper would be to consider Witten diagrams for two-point functions of operators with spin. The kinematics of spinning two-functions in general defect CFTs has been studied in detail in \cite{Kobayashi:2018okw,Lauria:2018klo,Herzog:2020bqw}. 
It would be interesting to adapt the use of weight-shifting operators \cite{Karateev:2017jgd}, in particular their AdS counterparts \cite{Costa:2018mcg}, to the setup involving defects, in order to leverage the computation of spinning Witten diagrams.

Our analysis of loop diagrams has been restricted to only one-loop level, involving mainly scalar exchanges in the bulk channel. We intend to perform a more systematic analysis of loop diagrams in future work by incorporating defect channel exchanges and also higher loops. It would be useful to further develop the application of AdS unitarity methods \cite{Meltzer:2019nbs} to study loop diagrams in holographic defect CFTs \cite{Chen:2024orp}. 

Higher-point functions are also worth considering in more detail. In BCFTs, Wiiten diagrams for three-point functions have recently been studied in \cite{Chen:2023oax}. Their analysis is amenable to extension to higher codimension defects. In this paper, we have already considered the case of the three-point form factor that involves two defect operators and one bulk operator. This correlator is a function of only a single cross-ratio. It would be interesting to analyze the block decomposition of Witten diagrams for three-point functions involving two bulk operators and one defect operator \cite{Lauria:2020emq, Girault:2025kzt, Bianchi:2026orb}. This correlator is more complicated since it, in general, depends on three cross-ratios. 

The coefficients in the block decomposition of tree-level exchange Witten diagrams can be related to the action of analytic bootstrap functionals. This connection requires further elaboration in order to develop the Polyakov approach for bootstrapping two-point functions of bulk operators in general defect CFTs. For this purpose, we would also need to extend our results for the decomposition coefficients in the bulk channel to general spinning exchanges. In this context, it could prove fruitful to consider the case of zero-dimensional defects, recently studied in \cite{Chen:2026ium}. 

In this paper, we have been able to compute closed-form expressions for the defect-to-bulk channel crossing kernel, only for certain specific values of the defect and ambient spacetime dimensions. It is desirable to develop methods to compute the crossing kernel analytically in more general cases. The bulk-to-defect channel crossing kernel has not been studied in this work. The defect channel Lorentzian inversion formula derived in \cite{Lemos:2017vnx} could be used for the purpose of evaluating this kernel. For CFTs without defects, computations of the crossing kernel have also been carried out in Mellin space \cite{Gopakumar:2018xqi, Sleight:2018epi,  Sleight:2018ryu,Ferrero:2019luz}. We expect that a similar analysis can be performed for defect CFTs. These crossing kernels can be applied to extract CFT data such as anomalous dimensions. Another interesting direction would be to study Witten diagrams for correlators in setups such as composite defects \cite{Drukker:2026nvv} and crosscap defects \cite{Shimamori:2024yms}. The latter are higher co-dimension versions of CFTs on real projective space $\mathbb{RP}^{d}$. Witten diagrams and their relation to analytic functionals for CFTs on $\mathbb{RP}^{d}$ have been studied in \cite{Giombi:2020xah}.

\section*{Acknowledgments}

We thank Riccardo Ciccone and Xinan Zhou for useful discussions. S.G. would like to thank the Kavli Institute for Theoretical Sciences, University of Chinese Academy of Sciences, Beijing, and the Department of Theoretical Physics, CERN, Geneva, for hospitality during the completion of this work. T.S. would like to thank the Center for Cosmology and Particle Physics, New York University, New York, for hospitality during the completion of this work. The work of the authors is supported by the Israeli Science Foundation (ISF) Grant No. $1487/21$, by the MOST NSF/BSF physics grant number $202272$, and by the BSF grant number $2024187$. 


\appendix

\section{Conformal blocks}
\label{app:blocks}

In this appendix, we review some details regarding the blocks which appear in the defect and bulk channel decompositions of $2$-point functions of scalar bulk CFT primary operators. We also note here the relation between the partial waves and blocks in both channels, which are useful for performing the block expansion of Witten diagrams considered in the main text.

\subsection{Defect channel}
\label{app:defectblock}

Consider the $2$-point function $\mathcal{G}(\xi,\eta)$ of bulk scalar primary operators defined in Eq.~\eqref{2ptfunction}. The bulk channel expansion of this is given by
\begin{align}
    \label{defch}
    \mathcal{G}(\xi,\eta)= \sum_{\widehat{\mathcal{O}}} b_{1\widehat{\mathcal{O}}} b_{2\widehat{\mathcal{O}}} \  \widehat{f}_{\hat{\Delta},s}(\xi,\eta) .
\end{align}


\noindent Here $\widehat{f}_{\hat{\Delta},s}(\xi,\eta)$ is the conformal block in the defect channel and is given by \cite{Billo:2016cpy}
\begin{align}
\label{defch1}
    \widehat{f}_{\hat{\Delta},s}(\xi,\eta)=\alpha_{q,s} \ _{2}F_{1}\bigg(\frac{q+s}{2}-1,-\frac{s}{2},\frac{q-1}{2},1-\eta^{2}\bigg) f_{\hat{\Delta}}(\chi)
\end{align}

\noindent where
\begin{align}
\label{defch2}
    \alpha_{q,s}=\frac{2^{-s}\Gamma(q+s-2)\Gamma\left(\frac{q}{2}-1\right)}{\Gamma\left(\frac{q}{2}+s-1\right)\Gamma(q-2)}, \ f_{\hat{\Delta}}(\chi)= \chi^{-\hat{\Delta}} \ _{2}F_{1}\left(\frac{\hat{\Delta}}{2},\frac{\hat{\Delta}+1}{2},\hat{\Delta}+1-\frac{p}{2},\frac{4}{\chi^{2}}\right)
\end{align}

\noindent and $\chi=\xi+2\eta$. An equivalent representation of the block in Eq.~\eqref{defch1} for integer $s$ is
\begin{align}
    \label{defch3}
    \widehat{f}_{\hat{\Delta},s}(\xi,\eta)=N_{q,s} C_{s}^{(\frac{q}{2}-1)}(\eta) f_{\hat{\Delta}}(\chi)
\end{align}

\noindent where 
\begin{align}
\label{defblocknorm}
N_{q,s}= \frac{2^{-s}\Gamma\left(\frac{q}{2}-1\right)\Gamma(s+1)}{\Gamma\left(\frac{q}{2}+s-1\right)}
\end{align} 

\noindent and $C_{s}^{(\frac{q}{2}-1)}(\eta)$ denotes the Gegenbauer polynomial. 

\vskip 4pt
It is sometimes useful to work with a different set of cross-ratios $(z,\bar{z})$ which are related to the $(\chi,\eta)$ variables above as
\begin{align}
    \chi=\frac{1+z\bar{z}}{\sqrt{z \bar{z}}}, \quad \eta=\frac{z+\bar{z}}{2\sqrt{z\bar{z}}} .
\end{align}

\noindent In order to rewrite the block $\widehat{f}_{\hat{\Delta},s}(\chi,\eta)$ in terms of the $(z,\bar{z})$ variables, we use the following identities for hypergeometric functions.
\begin{align}
\label{2f1id1}
    &\frac{2^{3-q}\sqrt{\pi}\Gamma(q+s-2)}{\Gamma\left(\frac{q-1}{2}\right)\Gamma\left(\frac{q}{2}+s-1\right)} \ _{2}F_{1}\left(\frac{q+s}{2}-1,-\frac{s}{2},\frac{q-1}{2},1-\eta^{2}\right)\nonumber\\
    &= \left(\frac{z}{\bar{z}}\right)^{-\frac{s}{2}} \ _{2}F_{1}\left(-s,\frac{q}{2}-1,2-\frac{q}{2}-s,\frac{z}{\bar{z}}\right)
\end{align}

\noindent and
\begin{align}
\label{2f1id2}
    f_{\hat{\Delta}}(\chi) = \left\{
    \begin {aligned}
         & (z\bar{z})^{\frac{\hat{\Delta}}{2}} \ _{2}F_{1}\left(\hat{\Delta},\frac{p}{2},\hat{\Delta}+1-\frac{p}{2},z\bar{z}\right) \quad & z\bar{z} < 1 \\
         & (z\bar{z})^{-\frac{\hat{\Delta}}{2}} \ _{2}F_{1}\left(\hat{\Delta},\frac{p}{2},\hat{\Delta}+1-\frac{p}{2},z\bar{z}\right) \quad & z\bar{z} > 1                  \end{aligned}
\right.
\end{align}

\noindent The identity in Eq.~\eqref{2f1id1} holds for any complex $w=(z/\bar{z})^{1/2}$ and integer values of $s$. Defining $r=(z\bar{z})^{1/2}$ and $\hat{\tau}=\hat{\Delta}-s$, we thus get  
\begin{align}
\label{rlt1}
     \widehat{f}_{\hat{\Delta},s}(z,\bar{z}) = 2^{-s} z^{\frac{\hat{\tau}}{2}}\bar{z}^{\frac{\hat{\tau}}{2}+s} \ _{2}F_{1}\left(\hat{\Delta},\frac{p}{2},\hat{\Delta}+1-\frac{p}{2},z\bar{z}\right)  \ _{2}F_{1}\left(-s,\frac{q}{2}-1,2-\frac{q}{2}-s,\frac{z}{\bar{z}}\right)
\end{align}

\noindent for $r<1$ and

\begin{align}
\label{rgt1}
     \hspace{-0.3cm}\widehat{f}_{\hat{\Delta},s}(z,\bar{z}) = 2^{-s} z^{-\frac{\hat{\tau}}{2}-s}\bar{z}^{-\frac{\hat{\tau}}{2}} \ _{2}F_{1}\left(\hat{\Delta},\frac{p}{2},\hat{\Delta}+1-\frac{p}{2},z\bar{z}\right)  \ _{2}F_{1}\left(-s,\frac{q}{2}-1,2-\frac{q}{2}-s,\frac{z}{\bar{z}}\right)
\end{align}

\noindent for $r>1$.

\subsection*{Partial wave}

The defect channel partial wave can be defined as the following integral involving the product of bulk-defect $2$-point functions  
\begin{align}
    \label{defpwave}
    \widehat{\mathcal{F}}_{\Delta,s}(x_{1},x_{2})= \int d^{p}\widehat{x} \ \langle \hspace{-0.09cm}\langle\mathcal{O}_{\Delta_{1}}(x_{1})\widehat{\mathcal{O}}^{i_{1}\cdots i_{s}}_{\widehat{\Delta},s}(\widehat{x})\rangle \hspace{-0.09cm}\rangle \langle \hspace{-0.09cm}\langle \widehat{\mathcal{O}}^{i_{1}\cdots i_{s}}_{p-\hat{\Delta},s}(\widehat{x})\mathcal{O}_{\Delta_{2}}(x_{2})\rangle \hspace{-0.09cm}\rangle .
\end{align}

\noindent Here the external operators $\mathcal{O}_{1},\mathcal{O}_{2}$ are scalars. The superscripts $(i_{1} \cdots i_{s})$ denote the transverse spin indices of the defect operator $\widehat{\mathcal{O}}$.

\vskip 4pt
The partial wave $\widehat{\mathcal{F}}_{\Delta,s}(x_{1},x_{2})$ can be expressed as a linear combination of the block $\widehat{f}_{\hat{\Delta},s}(\chi)$ and it's shadow as follows
\begin{align}
\label{defparwave1}
\widehat{\mathcal{F}}_{\hat{\Delta},s}(x_{1},x_{2}) = \frac{1}{|x_{1,\perp}|^{\Delta_{1}}|x_{2,\perp}|^{\Delta_{2}}}\left[\widehat{K}_{\hat{\Delta}} \widehat{f}_{\hat{\Delta},s}(\chi)+ \widehat{K}_{p-\hat{\Delta}} \widehat{f}_{p-\hat{\Delta},s}(\chi)\right]
\end{align}

\noindent where the normalization factor $\widehat{K}_{\hat{\Delta}}$ is given by
\begin{align}
\label{defpwblocknorm1}
\widehat{K}_{\hat{\Delta}}= \frac{2^{2\hat{\Delta}+1-p}\pi^{\frac{p+1}{2}}\Gamma(p-2\hat{\Delta})}{\Gamma(p-\hat{\Delta})\Gamma\left(\frac{p+1}{2}-\hat{\Delta}\right)} .
\end{align}

\subsection{Bulk channel}
\label{app:bulkblock}

The bulk channel decomposition of $\mathcal{G}(\xi,\eta)$ is of the form
\begin{align}
    \label{bulkch}
    \mathcal{G}(\xi,\eta)= \xi^{-\frac{\Delta_{1}+\Delta_{2}}{2}}\sum_{\mathcal{O}} c_{12\mathcal{O}} a_{\mathcal{O}} \  f_{\Delta,J}(\xi,\eta) .
\end{align}

\noindent The bulk channel block $f_{\Delta,J}(\xi,\eta)$ in general does not admit a closed form expression. Using the relation between conformal Casimir equations and eigenvalue equations for certain integrable systems, an explicit series representation of $f_{\Delta,J}(\xi,\eta)$ was obtained in \cite{Isachenkov:2018pef,Liendo:2019jpu}. In these papers, relations between bulk blocks and $4$-point conformal blocks in defect-free CFTs have also been derived for specific values of the defect and ambient space-time dimensions. The lightcone expansion has been studied in \cite{Billo:2016cpy,Gimenez-Grau:2022ebb}. The series expansion in terms of radial coordinates was developed in \cite{Lauria:2017wav}.

\subsection*{Partial wave}

The bulk channel partial has the following integral representation, which involves the product of a $3$-point conformal structure in a defect-free CFT and the $1$-point function in the presence of the defect.
\begin{align}
    \label{bulkchpwave}
    \mathcal{F}_{\Delta,\ell}(x_{1},x_{2})= \int d^{d}x \ \langle \mathcal{O}_{\Delta_{1}}(x_{1})\mathcal{O}_{\Delta_{2}}(x_{2})\mathcal{O}^{\mu_{1}\cdots \mu_{\ell}}_{\Delta,\ell}(x)\rangle \langle \hspace{-0.09cm}\langle \mathcal{O}^{\mu_{1}\cdots \mu_{\ell}}_{d-\Delta,\ell}(x)\rangle\hspace{-0.09cm}\rangle.
\end{align}

\noindent This can be expressed in terms of the block $f_{\Delta,J}(\xi,\eta)$ and the shadow block $f_{d-\Delta,J}(\xi,\eta)$ as
\begin{align}
    \label{bulkpw}
    \mathcal{F}_{\Delta,\ell}(x_{1},x_{2}) = \frac{\xi^{-\frac{\Delta_{1}+\Delta_{2}}{2}}}{|x_{1,\perp}|^{\Delta_{1}}|x_{2,\perp}|^{\Delta_{2}}}\left[\mathcal{K}_{\Delta,\ell} f_{\Delta,\ell}(\xi,\eta)+ \widetilde{\mathcal{K}}_{d-\Delta,\ell} f_{d-\Delta,\ell}(\xi,\eta)\right]
\end{align}

\noindent where the factors $\mathcal{K}_{\Delta,J}, \widetilde{\mathcal{K}}_{d-\Delta,J}$ are given by
\begin{align}
    \label{bulkpwnorm}
    & \mathcal{K}_{\Delta,\ell} = \frac{2\pi^{(d+1)/2}(\Delta-1)_{\ell/2}(\Delta-d-1)_{\ell/2}\Gamma\left(\frac{\Delta-1}{2}\right)\Gamma\left(\frac{\Delta-p}{2}\right)\Gamma\left(\frac{d-2\Delta}{2}\right)}{2^{\Delta+\ell}\Gamma\left(\frac{d-\Delta+\ell}{2}\right)\Gamma\left(\frac{\Delta+\ell+1}{2}\right)\Gamma\left(\frac{\Delta+\ell-1}{2}\right)\Gamma\left(\frac{q-\Delta}{2}\right)},  \\
    & \widetilde{\mathcal{K}}_{d-\Delta,\ell} = \frac{2\pi^{(d+1)/2}\Gamma\left(\frac{d-\Delta+\ell+\Delta_{12}}{2}\right)\Gamma\left(\frac{d-\Delta+\ell-\Delta_{12}}{2}\right)\Gamma\left(\frac{2\Delta-d}{2}\right)}{2^{d-\Delta+\ell}\Gamma\left(\frac{d-\Delta+\ell+1}{2}\right)\Gamma\left(\frac{d-\Delta+\ell-1}{2}\right)\Gamma\left(\frac{\Delta-p}{2}\right)} .
\end{align}

\subsection{Bulk channel Casimir differential operator}
\label{app:bulkchcasdiff}

The action of the bulk channel Casimir on a function of the cross ratios $(\xi,\eta)$ is given by the following differential operator
\begin{align}
\label{Dbulk1}
        \mathcal{D}_{bulk}\mathcal{G}(\xi,\eta)&=-\xi^{2}(\xi\eta+2+2\eta^{2})\partial^{2}_{\xi}\mathcal{G}(\xi,\eta)+ (1-\eta^{2})(\xi\eta-2(1-\eta^{2}))\partial^{2}_{\eta}\mathcal{G}(\xi,\eta) \nonumber\\
       &-\xi(\xi\eta+2(1+\eta^{2})-2d+(\xi\eta+2+2\eta^{2})(\Delta_{1}+\Delta_{2}))\partial_{\xi}\mathcal{G}(\xi,\eta)\nonumber\\
       &+(-\xi(q-2+\eta^{2})+2\eta(1-\eta^{2})+(\Delta_{1}+\Delta_{2})(1-\eta^{2})(2\eta+\xi))\partial_{\eta}\mathcal{G}(\xi,\eta)\nonumber\\
       &+\xi(1-\eta^{2})(2\xi+4\eta)\partial_{\xi}\partial_{\eta}\mathcal{G}(\xi,\eta) -\Delta_{1}\Delta_{2}(\xi\eta+2\eta^{2})\Delta_{i}\nonumber\\
       &+\sum_{i=1}^{2}\Delta_{i}(d-\Delta_{i})+C_{\Delta,\ell}\mathcal{G}(\xi,\eta)
   \end{align}

\noindent where $C_{\Delta,\ell}=\Delta(\Delta-d)+\ell(\ell+d-2)$. In section \ref{subsec:bulkl0todef}, for deriving Eq.~\eqref{bulkCasdefblock}, it is useful to recast the action of $\mathcal{D}_{bulk}$ in terms of the cross-ratios $(\chi,\eta)$. This can be easily done by using $\xi=\chi-2\eta$ in Eq.~\eqref{Dbulk1}.


\section{Details of spin $2$ bulk channel decomposition}
\label{app:spin2exch}

Here, we derive Eq.~\eqref{bulkchspin2exch2} using the embedding space formalism. Let us briefly review the essential elements of this formalism that will be needed for our purpose here. 

\vskip 4pt
A bulk point in Euclidean AdS$_{d+1}$ can be represented as a point $X$ in $(d+2)$-dimensional Minkowski space $\mathbb{R}^{1,d+1}$, such that $X^{2}=-1$. Points on the boundary of AdS$_{d+1}$ are identified with projective null vectors $P\in \mathbb{R}^{1,d+1}$, so that $P^{2}=0, P \equiv \lambda P, \lambda \in \mathbb{R}^{+}$. To account for the presence of a defect, it is convenient to split the embedding space as $\mathbb{R}^{1,p+1}\times \mathbb{R}^{d-p}$ \cite{Billo:2016cpy}. Then a coordinate index $M$ for a vector in  $\mathbb{R}^{1,d+1}$ can be split as
\begin{align}
    M=(A,I), \ \text{with} \ A= 1,\ldots, p+2, \ I =1,\ldots, q. 
\end{align}

\noindent The scalar product between two vectors $V_{1},V_{2} \in \mathbb{R}^{1,d+1}$ can then be written as
\begin{align}
    V_{1}\cdot V_{2} = V_{1}\bullet V_{2} + V_{1}\circ V_{2}
\end{align}

\noindent where $V_{1}\bullet V_{2}= V_{1,A}V_{2,B}\eta^{AB}$ and $V_{1}\circ V_{2}= V_{1,I}V_{2,J}\delta^{IJ}$. We will denote points on the AdS$_{p+1}$ brane in embedding space as $\hat{X}$, such that $\hat{X}\in \mathbb{R}^{1,p+1}$ and $\hat{X}^{2}=-1$. Equivalently, taking an AdS$_{d+1}$ bulk point $X$ and setting $X_{I}=0$ defines a point on the brane.  

\vskip 4pt
\noindent We will also employ auxiliary polarization vectors to represent a symmetric traceless and transverse rank $J$ AdS tensor in embedding space in index-free notation as follows 
\begin{align}
    F(X;U) = U^{M_{1}} \cdots U^{M_{J}} F_{M_{1}\ldots M_{J}}(X)
\end{align}

\noindent where $U^{2}=0=U.X$. The indices can be restored by applying the following projection operator 
\begin{align}
    \mathbb{K}_{M} = \left(\frac{d-1}{2}+ U\cdot \frac{\partial}{\partial U}\right)\frac{\partial}{\partial U^{M}}. 
\end{align}

\noindent Similarly, a symmetric traceless rank $J$ tensor defined on the lightcone $P^{2}=0$ can be encoded via the polynomial
\begin{align}
    F(P;Z) = Z^{M_{1}} \cdots Z^{M_{J}} F_{M_{1}\ldots M_{J}}(P), 
\end{align}

\noindent where $Z^{2}=0=P\cdot Z$. To recover the indices of the tensor, we apply the projection operator
\begin{align}
    \mathcal{D}_{Z}^{M} = \left(\frac{d-2}{2}+ Z\cdot \frac{\partial}{\partial Z}\right)\frac{\partial}{\partial Z_{M}} -\frac{1}{2}Z^{M}\frac{\partial^{2}}{\partial Z\cdot \partial Z}. 
\end{align}

\vskip 4pt
Now let us return to the main object of interest, which is the bulk channel massive spin $2$ exchange diagram in Eq.~\eqref{bulkchspin2exch}. In embedding space, this can be written as
\begin{align}
    \label{spin2derva}
     &W_{\Delta,2}(P_{1},P_{2})\nonumber\\
     &= \int\limits_{\mathrm{AdS}_{d+1}}dX \hspace{-0.3cm}\int\limits_{\mathrm{AdS}_{p+1}}d\widehat{X} \ K_{\Delta_{1},0}(P_{1},X) (\mathbb{K}\cdot \nabla)^{2}K_{\Delta_{2},0}(P_{1},X) \ \mathcal{D}^{2}_{\widehat{U}} G_{\Delta,2}(X,\widehat{X},U,\widehat{U}) .
\end{align}

\noindent In the above expression, the operator $\mathcal{D}^{2}_{\widehat{U}}$ is defined as
\begin{align}
    \mathcal{D}^{2}_{\widehat{U}} = \frac{1}{2}\widehat{G}^{AB}\frac{\partial}{\partial \widehat{U}^{A}} \frac{\partial}{\partial \widehat{U}^{B}}, 
\end{align}

\noindent where $\widehat{G}^{AB}=\eta^{AB} + \widehat{X}^{A}\widehat{X}^{B}$ is the AdS$_{p+1}$ induced metric. The purpose of $\mathcal{D}^{2}_{\widehat{U}}$ is to implement, in embedding space, the contraction $G_{\Delta,2}^{\mu\nu,\hat{\alpha}\beta} g_{\hat{\alpha}\hat{\beta}}$ in Eq.~\eqref{bulkchspin2exch} that arises from the interaction vertex on the brane. We can now employ the split representation of the bulk-to-brane propagator for a spin $2$ massive field. This can be obtained from the expression for the bulk-to-bulk propagator given in \cite{Costa:2014kfa} and restricting one of the bulk points to lie on the brane. 
\begin{align}
\label{splitrepspin2}
    G_{\Delta,2}(X,\hat{X},U,\hat{U}) &= \sum_{\ell=0}^{2}\int_{-\infty}^{\infty} d\nu \ \alpha_{\ell}(\nu) \frac{\nu^{2} C_{\frac{d}{2}+i\nu,\ell}C_{\frac{d}{2}-i\nu,\ell}}{\pi \ell!\left(\frac{d-2}{2}\right)_{\ell}} \ \times \nonumber\\
    &\int\limits_{\partial\mathrm{AdS}_{d+1}} dP \ ((U.\nabla)(\widehat{U}.\widehat{\nabla)})^{2-\ell} K_{\frac{d}{2}+i\nu,\ell}(P,X;U,\mathcal{D}_{z}) K_{\frac{d}{2}-i\nu,\ell}(P,\hat{X};\widehat{U},Z) , 
\end{align}

\noindent where the spectral coefficients $\alpha_{\ell}(\nu)$ are given by
\begin{align}
\label{alpha0}
    & \alpha_{0}(\nu) = \frac{1}{d(\Delta-1)(d-\Delta-1)(\nu^{2}+(h+1)^{2})}- \frac{1}{d\Delta(d-\Delta)(\nu^{2}+h^{2})},\\
    & \alpha_{1}(\nu) =-\frac{2}{\Delta(d-\Delta)(\nu^{2}+(h+1)^{2})}, \quad \alpha_{2}(\nu) =\frac{1}{\nu^{2}+(\Delta-h)^{2}}, \quad h=\frac{d}{2}, 
\end{align}

\noindent and 
\begin{align}
    C_{\Delta,\ell} =\frac{(\ell+\Delta-1)\Gamma(\Delta)}{2\pi^{\frac{d}{2}}(\Delta-1)\Gamma(\Delta+1-\frac{d}{2})}. 
\end{align}

\noindent Now, inserting the above split representation into Eq.~\eqref{spin2derva}, we can factorize the AdS integrals. The AdS$_{d+1}$ integral then gives a $3$-point function for two scalars and a spin $\ell$ operator in a defect-free CFT. The AdS$_{p+1}$ integral generates the one-point function of a spin $\ell$ operator in the presence of the defect.  It is easy to check that the AdS$_{p+1}$ integral vanishes for the $\ell=0$ term in the sum present in Eq.~\eqref{splitrepspin2}. This reflects the fact that in a parity-invariant theory, only even spin bulk CFT operators can have a non-zero one-point function\footnote{For $p=0$ defects, we can have a non-zero one-point function for a spin $1$ operator. Here we are assuming $p>0$.}. Therefore, we get,
\begin{align}
\label{spin2dervb}
   & W_{\Delta,2}(P_{1},P_{2}) \nonumber\\
   &= \sum_{\ell=0,2}\int_{-\infty}^{\infty}d\nu  \ \widetilde{\alpha}_{\ell}(\nu) \hspace{-0.3cm}\int\limits_{\partial \mathrm{AdS}_{d+1}} dP \ \langle \mathcal{O}_{\Delta_{1}}(P_{1}) \mathcal{O}_{\Delta_{2}}(P_{2}) \mathcal{O}_{\frac{d}{2}+i\nu,\ell}(P,\mathcal{D}_{Z}\rangle  \langle\hspace{-0.09cm}\langle \mathcal{O}_{\frac{d}{2}-i\nu,\ell}(P,Z)\rangle\hspace{-0.09cm}\rangle
\end{align}

\noindent where $\widetilde{\alpha}_{\ell}(\nu)$ is
\begin{align}
    \widetilde{\alpha}_{\ell}(\nu) =  \frac{\nu^{2} C_{\frac{d}{2}+i\nu,\ell}C_{\frac{d}{2}-i\nu,\ell}}{\pi \ell!\left(\frac{d-2}{2}\right)_{\ell}}  \alpha_{\ell}(\nu) \Lambda_{\Delta_{1},\Delta_{2},\underline{\Delta};\ell} \Omega_{\underline{\Delta};\ell}, \quad \underline{\Delta}=\frac{d}{2}+i\nu.
\end{align}

\noindent Here $\Lambda_{\Delta_{1},\Delta_{2},\Delta;\ell} $ is the coefficient that multiplies the CFT $3$-point function obtained after performing the AdS$_{d+1}$ integral. This coefficient was computed for general spinning exchanges in \cite{Costa:2014kfa}. For $\ell=0,2$, these are given by
\begin{align}
\label{3ptcfs}
    & \Lambda_{\Delta_{1},\Delta_{2},\Delta;0}  =  (\Delta_{2})_{2}(\Delta)_{2}\left[4(\lambda_{\Delta_{1},\Delta_{2}+2,\Delta+2}-\lambda_{\Delta_{1},\Delta_{2}+1,\Delta+1})+\frac{d \  \lambda_{\Delta_{1},\Delta_{2},\Delta}}{(d+1)}\right],\\
    & \Lambda_{\Delta_{1},\Delta_{2},\Delta;2} = \frac{4(\Delta_{2})_{2}}{(\Delta)_{2}} \left(\frac{\Delta+\Delta_{1}-\Delta_{2}-2}{2}\right)_{2} \lambda_{\Delta_{1},\Delta_{2}+2,\Delta}
\end{align}

\noindent where $\lambda_{\Delta_{1},\Delta_{2},\Delta}$ is
\begin{align}
  \lambda_{\Delta_{1},\Delta_{2},\Delta} =   \pi^{\frac{d}{2}}\frac{\Gamma\left(\frac{\Delta_{1}+\Delta_{2}-\Delta}{2}\right)\Gamma\left(\frac{\Delta_{12}+\Delta}{2}\right)\Gamma\left(\frac{\Delta-\Delta_{12}}{2}\right)\Gamma\left(\frac{\Delta_{1}+\Delta_{2}-d+\Delta}{2}\right)}{2\Gamma(\Delta_{1})\Gamma(\Delta_{2})\Gamma\left(\Delta\right)}. 
\end{align}

\noindent The coefficient $\Omega_{\underline{\Delta};\ell}$ appears in front of the one-point structure $\langle\hspace{-0.09cm}\langle \mathcal{O}_{\frac{d}{2}-i\nu,\ell}(P,Z)\rangle\hspace{-0.09cm}\rangle$ after doing the AdS$_{p+1}$ integral. Here we only need the expression of $\Omega_{\underline{\Delta};\ell}$ for $\ell=0,2$. These can be extracted from the following integrals
\begin{align}
    & \int\limits_{\mathrm{AdS}_{p+1}}d\widehat{X} \ \mathcal{D}_{\widehat{W}}^{2}\ (\widehat{W}.\widehat{\nabla})^{2}\Pi_{\frac{d}{2}-i\nu,0}(P,\widehat{X})  = \Omega_{\underline{\Delta};0}  \langle\hspace{-0.09cm}\langle \mathcal{O}_{d-\Delta,0}(P)\rangle\hspace{-0.09cm}\rangle ,\\
    & \int\limits_{\mathrm{AdS}_{p+1}}d\widehat{X} \ \mathcal{D}_{\widehat{W}}^{2}\ \Pi_{\frac{d}{2}-i\nu,2}(P,\widehat{X};\widehat{W},Z) = \Omega_{\underline{\Delta};2} \langle\hspace{-0.09cm}\langle \mathcal{O}_{d-\Delta,2}(P,Z)\rangle\hspace{-0.09cm}\rangle, 
\end{align}

\noindent  where the spin $\ell$ one-point function $\langle\hspace{-0.09cm}\langle \mathcal{O}_{\Delta,\ell}(P,Z)\rangle\hspace{-0.09cm}\rangle $ is 
\begin{align}
    \langle\hspace{-0.09cm}\langle \mathcal{O}_{\Delta,\ell}(P,Z)\rangle\hspace{-0.09cm}\rangle = \frac{\left[(P\circ Z)^{2}-(P\circ P)(Z\circ Z)\right]^{\ell/2}}{(P\circ P)^{\frac{\Delta+\ell}{2}}}
\end{align}

\noindent and the coefficients $\Omega_{\Delta;\ell}$ are given by
\begin{align}
\label{oneptcoeffs}
    & \Omega_{\Delta;0} = (d-\Delta)_{2}(a_{d-\Delta}-4 a_{d-\Delta+2}), \ \Omega_{\Delta;2} =\frac{(d-\Delta)a_{d-\Delta}}{(d-\Delta+1)}, \ a_{\Delta}=\frac{\pi^{\frac{p}{2}}\Gamma\left(\frac{\Delta}{2}\right)\Gamma\left(\frac{\Delta-p}{2}\right)}{2\Gamma(\Delta)} . 
\end{align}

\noindent We can now identify the conformal integral in Eq.~\eqref{spin2dervb} as the bulk channel partial wave for spin $\ell$ exchange. Using the relation between the partial wave and the block given in Eq.~\eqref{bulkpw} and the $\nu$ to $-\nu$ symmetry of the integrand, we obtain
\begin{align}
\label{spin2pwexp}
     W_{\Delta,2}(P_{1},P_{2}) 
     & = \frac{\xi^{-\frac{\Delta_{1}+\Delta_{2}}{2}}}{(P_{1}\circ P_{1})^{\frac{\Delta_{1}}{2}}(P_{2}\circ P_{2})^{\frac{\Delta_{2}}{2}}}\sum_{\ell=0,2}\int_{-\infty}^{\infty} d\nu \ 2\widetilde{\alpha}_{\ell}(\nu)  \ \mathcal{K}_{\frac{d}{2}+i\nu,\ell} f_{\frac{d}{2}+i\nu,\ell}(\xi,\eta) . 
\end{align}

\noindent In order to write Eq.~\eqref{spin2pwexp} as a sum over bulk channel blocks, we can deform $\nu$ contour and gather the contribution from the poles of the spectral integrand. The single-trace contribution now comes from the pole of the coefficient $\alpha_{2}(\nu)$ at $i\nu=\Delta-\frac{d}{2}$. $\alpha_{0}(\nu)$ does not have poles in $\nu$ which can lead to physical operator exchanges. We also get the following two sets of double-trace poles
\begin{align}
\label{spin2dbltrpoles}
    i\nu=\Delta_{1}+\Delta_{2}+2n-\frac{d}{2}, \quad  i\nu=\Delta_{1}+\Delta_{2}+2n+2-\frac{d}{2}, \ n\in \mathbb{Z}_{\ge 0}. 
\end{align}

\noindent These poles comes from the factors $\Lambda_{\Delta_{1},\Delta_{2},\underline{\Delta};0}$ and $ \Lambda_{\Delta_{1},\Delta_{2},\underline{\Delta};2}$ given in Eq.~\eqref{3ptcfs}, respectively. After evaluating the residues at the above single-trace and double-trace poles, we obtain the block expansion shown in Eq.~\eqref{bulkchspin2exch2}.




\section{Bulk channel Hahn polynomials}
\label{app:hahnpol}

In this appendix, we derive the Mellin representation of the function $g_{\tau,\ell}(\eta)$, which appears in section \ref{subsec:deftoblkseeds} and give its relation to continuous Hahn polynomials.  $g_{\tau,\ell}(\eta)$ is given by
\begin{align}
    \label{collnrblock}
    g_{\tau,\ell}(\eta) = (1-\eta^{2})^{\frac{\ell}{2}} \ _{2}F_{1}\left(\frac{2\ell+\tau+\Delta_{12}}{4},\frac{2\ell+\tau-\Delta_{12}}{4},\frac{2\ell+\tau+1}{2},1-\eta^{2}\right).
\end{align}

\noindent In section \ref{subsec:deftoblkseeds}, we have taken the twist $\tau=\Delta_{1}+\Delta_{2}$, but here we will consider general values of $\tau$. Now using the Mellin-Barnes representation of the $\ _{2}F_{1}$ hypergeometric function we get
\begin{align}
    \label{collnrblockmellin}
    g_{\tau,\ell}(\eta) = \frac{\mathcal{N}_{\tau,\ell}}{\sqrt{\pi}}\int_{-i\infty}^{i\infty} \frac{ds}{2\pi i} \ \Gamma(s)\Gamma(c-a-b+s)\Gamma(a-s)\Gamma(b-s) \ (1-\eta^{2})^{\frac{\ell}{2}}\eta^{-2s}
\end{align}

\noindent where 
\begin{align}
    & a=\frac{2\ell+\tau+\Delta_{12}}{4},\ b=\frac{2\ell+\tau-\Delta_{12}}{4}, \ c=\frac{2\ell+\tau+1}{2}, \  \mathcal{N}_{\tau,\ell}= \frac{2^{2\ell+\tau-2}\Gamma\left(\frac{2\ell+\tau+1}{2}\right)}{\sqrt{\pi} \Gamma\left(\frac{2\ell+\tau\pm \Delta_{12}}{2}\right)}
\end{align}

\noindent and we have used the notation $\Gamma(x\pm y)=\Gamma(x)\Gamma(y)$. Since in our case $\ell$ is an even integer, we can expand the $(1-\eta^{2})^{\frac{\ell}{2}}$ factor in binomial series. We then have
\begin{align}
\label{collnrblockmellin1}
     g_{\tau,\ell}(\eta) &=  \frac{\mathcal{N}_{\tau,\ell}}{\sqrt{\pi}}\sum_{k=0}^{\frac{\ell}{2}}(-1)^{k}\binom{\frac{\ell}{2}}{k}\int \frac{ds}{2\pi i} \ \Gamma(s)\Gamma(c-a-b+s)\Gamma(a-s)\Gamma(b-s) \ \eta^{2k-2s} .
\end{align}

\noindent Now for the $k$-th term in the above sum, we perform the change of variables $\rho=2s-2k$. This yields,
\begin{align}
\label{collnrblockmellin2}
     g_{\tau,\ell}(\eta) &=  \frac{\mathcal{N}_{\tau,\ell}}{\sqrt{\pi}}\sum_{k=0}^{\frac{\ell}{2}}(-1)^{k}\binom{\frac{\ell}{2}}{k}\int \frac{d\rho}{2\pi i} (2\eta)^{-\rho} \ \Gamma(\rho) \Gamma\left(\frac{\tau+\Delta_{12}}{4}-\frac{\rho}{2}\right)\Gamma\left(\frac{\tau-\Delta_{12}}{4}-\frac{\rho}{2}\right) \times \nonumber\\
     & \bigg[2^{\rho-1} \frac{\Gamma\left(\frac{\rho+2k}{2}\right)\Gamma\left(\frac{\rho+2k+1}{2}\right)}{\Gamma\left(\rho\right)} \ \frac{\Gamma\left(\frac{2\ell+\tau+\Delta_{12}}{4}-\frac{\rho}{2}-k\right)\Gamma\left(\frac{2\ell+\tau-\Delta_{12}}{4}-\frac{\rho}{2}-k\right)}{\Gamma\left(\frac{\tau+\Delta_{12}}{4}-\frac{\rho}{2}\right)\Gamma\left(\frac{\tau-\Delta_{12}}{4}-\frac{\rho}{2}\right)}\bigg]. 
\end{align}

\noindent Simplifying the expression inside the brackets and performing the sum over $k$, we get
\begin{align}
     & \sum_{k=0}^{\frac{\ell}{2}}(-1)^{k}\binom{\frac{\ell}{2}}{k} \bigg[ 2^{-2k}\sqrt{\pi} \ (\rho)_{2k}\left(\frac{\tau+\Delta_{12}-2\rho}{4}\right)_{\frac{\ell}{2}-k}\left(\frac{\tau-\Delta_{12}-2\rho}{4}\right)_{\frac{\ell}{2}-k}\bigg] \nonumber\\
     & = \sqrt{\pi}\left(\frac{\tau+\Delta_{12}-2\rho}{4}\right)_{\frac{\ell}{2}}\left(\frac{\tau-\Delta_{12}-2\rho}{4}\right)_{\frac{\ell}{2}} \times \nonumber\\
     &\ _{3}F_{2}\left(-\frac{\ell}{2},\frac{\rho+1}{2},\frac{\rho}{2}, -\frac{2\ell+\tau+\Delta_{12}-2\rho}{4}+1,-\frac{2\ell+\tau-\Delta_{12}-2\rho}{4}+1,1 \right)
\end{align}

\noindent Employing a further change of the Mellin integration variable $\rho$, we can finally write the Mellin representation of $g_{\tau,\ell}(\eta)$ as
\begin{align}
\label{collnrblockmellin3}
    & g_{\tau,\ell}(\eta) = \int\frac{d\rho}{2\pi i} \ (2\eta)^{-\rho} \Gamma\left(\rho\right)\Gamma\left(\frac{\tau+\Delta_{12}-2\rho}{4}\right)\Gamma\left(\frac{\tau-\Delta_{12}-2\rho}{4}\right)  \mathbb{Q}_{\tau,\ell}(\rho)
\end{align}

\noindent where $\mathbb{Q}_{\tau,\ell}(\rho)$ is given by
\begin{align}
\label{hahnpol}
      \mathbb{Q}_{\tau,\ell}(\rho) &=\mathcal{N}_{\tau,\ell} \left(\frac{\tau+\Delta_{12}-2\rho}{4}\right)_{\frac{\ell}{2}}\left(\frac{\tau-\Delta_{12}-2\rho}{4}\right)_{\frac{\ell}{2}}\times\nonumber\\
     &\ _{3}F_{2}\left(-\frac{\ell}{2},\frac{\rho+1}{2},\frac{\rho}{2}, -\frac{2\ell+\tau+\Delta_{12}-2\rho}{4}+1,-\frac{2\ell+\tau-\Delta_{12}-2\rho}{4}+1,1 \right)
\end{align}

\noindent Note that $\mathbb{Q}^{(\tau)}_{\ell,0}(\rho)$ is a polynomial of degree $\ell/2$ in $\rho$. Using the Sheppard identity, we can write an equivalent form of $\mathbb{Q}_{\tau,\ell}(\rho)$ which is more convenient for our purposes in section \ref{subsec:deftoblkseeds}. This yields, 
\begin{align}
\label{hahnpol1}
     & \mathbb{Q}_{\tau,\ell}(\rho)= \mathcal{N}_{\tau,\ell} \left(\frac{\tau+\Delta_{12}}{4}\right)_{\frac{\ell}{2}}\left(\frac{\tau-\Delta_{12}}{4}\right)_{\frac{\ell}{2}} \ _{3}F_{2}\left(-\frac{\ell}{2},\frac{\rho}{2},\frac{\tau+\ell-1}{2}, \frac{\tau+\Delta_{12}}{4},\frac{\tau-\Delta_{12}}{4},1 \right)
\end{align}

\noindent $\mathbb{Q}_{\tau,\ell}(\rho)$ is related to the so-called continuous Hahn polynomial as follows. A degree $n$ continuous Hahn polynomial is defined as
\begin{align}
    p_{n}(x;ab,c,d) = i^{n}\frac{(a+c)_{n}(a+d)_{n}}{n!} \ {}_3F_2\left(
\begin{array}{c}
-n,\, n+a+b+c+d-1,\, a+ix \\
a+c,\, a+d
\end{array}
; 1\right) 
\end{align}

\noindent These polynomials satisfy the orthogonality relation
\begin{align}
\label{orthnorm2}
    &\int_{-\infty}^{\infty} \frac{dx}{2\pi} \ \Gamma(a+ix)\Gamma(b+ix)\Gamma(c-ix)\Gamma(d-ix) \ p_{m}(x;a,b,c,d)p_{n}(x;a,b,c,d) \nonumber\\
    & = \frac{\Gamma(n+a+c)\Gamma(n+a+d)\Gamma(n+b+c)\Gamma(n+b+d)}{n!(2n+a+b+c+d-1)\Gamma(n+a+b+c+d-1)} \delta_{m,n}
\end{align}

\noindent for $\mathbb{R}(a)>0,\mathbb{R}(b)>0, \mathbb{R}(c)>0,\mathbb{R}(d)>0$ and $c=\bar{a}, d=\bar{b}$. Now, choosing the following values for the parameters 
\begin{align}
   & n=\frac{\ell}{2}, \ a+ix=\frac{\rho}{2}, \ b+ix=\frac{\rho+1}{2}, \ a+\bar{a}=\frac{\tau+\Delta_{12}}{4}, \ a+\bar{b}=\frac{\tau-\Delta_{12}}{4},  \nonumber\\
   & 2\mathbb{R}(a+b)+n-1=\frac{\ell+\tau-1}{2}
\end{align}

\noindent we see that $\mathbb{Q}^{(\tau)}_{\ell,0}(\rho)$ is proportional to the continuous Hahn polynomial of degree $\ell/2$ 
\begin{align}
    \mathbb{Q}_{\tau,\ell}(\rho) = i^{-\ell/2}(\ell/2)! \ \mathcal{N}_{\tau,\ell} \ p_{\ell/2}(\rho). 
\end{align}

\noindent From Eq.~\eqref{orthnorm2} we then get the orthogonality relation for $\mathbb{Q}^{(\tau)}_{\ell,0}(\rho)$ to be
\begin{align}
    & \int_{-i\infty}^{i\infty} \frac{d\rho}{2\pi i} \ 2^{-\rho}\Gamma(\rho)\Gamma\left(\frac{\tau+\Delta_{12}-2\rho}{4}\right)\Gamma\left(\frac{\tau-\Delta_{12}-2\rho}{4}\right) \mathbb{Q}_{\tau,\ell}(\rho) \mathbb{Q}_{\tau,\ell'}(\rho) = \kappa_{\ell}(\tau)\delta_{\ell,\ell'}
\end{align}

\noindent where the factor $\kappa_{\ell}(\tau)$ is 
\begin{align}
     \kappa_{\ell}(\tau)& =  (-1)^{\frac{\ell}{2}}(\ell/2)! \frac{\Gamma\left(\frac{\tau+2\ell-1}{2}\right)}{\Gamma\left(\frac{\tau+\ell-1}{2}\right)}\mathcal{N}_{\tau,\ell}.
\end{align}

\section{Computation of $A_{0,\ell}$}
\label{app:A0linteg}

Our objective here is to compute the following integral, which appears in section \ref{subsec:deftoblkseeds}. 
\begin{align}
\label{def2blkseedMellin1}
   A_{0,\ell} = \frac{1}{\kappa_{\ell}(\tau_{0})}\int_{-i\infty}^{i\infty}  \frac{d\rho}{2\pi i} \ 2^{-\rho}\Gamma(\rho) \Gamma\left(\frac{\Delta_{1}-\rho}{2}\right)\Gamma\left(\frac{\Delta_{2}-\rho}{2}\right) \mathbb{Q}_{\tau_{0},\ell}(\rho) \widehat{M}_{\hat{\Delta},s}(\delta=0,\rho) 
\end{align}

\noindent The Mellin amplitude $\widehat{M}_{\hat{\Delta},s}(\delta=0,\rho)$ in the above equation contains a $\ _{3}F_{2}$ hypergeometric function\footnote{This $\ _{3}F_{2}$ appears with argument $1$ and the condition for it to be well-defined is $\Delta_{1}+\Delta_{2}-p>0$. When this condition is not satisfied, we need to analytically continue the $\ _{3}F_{2}$}. To evaluate Eq.~\eqref{def2blkseedMellin1}, we find it convenient to use the integral representation of $\ _{3}F_{2}$ in terms of $\ _{2}F_{1}$ which is as follows
\begin{align}
    \label{3f22f1rel}
    \ _{3}F_{2}\left(a_{1},a_{2},a_{3}; b_{1},b_{2},1\right) = \frac{\Gamma(b_{1})}{\Gamma(a_{1})\Gamma(b_{1}-a_{1})} \int_{0}^{1} dt \ t^{a_{1}-1} (1-t)^{b_{1}-a_{1}-1}  \ _{2}F_{1}\left(a_{2},a_{3}; b_{2},t\right)
\end{align}

\noindent for $\mathbb{R}(b_{1})>\mathbb{R}(a_{1})>0$. Applying this identity, $\widehat{M}_{\hat{\Delta},s}(\delta=0,\rho)$ can be written as
\begin{align}
    \label{Mellindef2}
    \widehat{M}_{\hat{\Delta},s}(\delta=0,\rho) & = \frac{\pi^{\frac{p}{2}}2^{s}(\Delta_{1})_{s}(\Delta_{2})_{s}\Gamma\left(\frac{\Delta_{1}+s+\hat{\Delta}-p}{2}\right)\Gamma\left(\frac{\Delta_{2}+s+\hat{\Delta}-p}{2}\right)}{16\Gamma(\Delta_{1}+s)\Gamma(\Delta_{2}+s)\Gamma\left(\hat{\Delta}+1-\frac{p}{2}\right)} \ (\rho)_{s}\times \nonumber\\
    &\int_{0}^{1}dt \ t^{\frac{\hat{\Delta}-s-\rho}{2}-1}  \ _{2}F_{1}\left(1-\frac{\Delta_{1}+s-\hat{\Delta}}{2},1-\frac{\Delta_{2}+s-\hat{\Delta}}{2}; \hat{\Delta}+1-\frac{p}{2},t\right) 
\end{align}

\noindent where we have used Eqs.~\eqref{spindefMellin} and \eqref{defectMellin}. Let us also expand $\mathbb{Q}_{\tau_{0},\ell}(\rho)$ as
\begin{align}
    \label{qhahn}
    \mathbb{Q}_{\tau_{0},\ell}(\rho) = \mathcal{N}_{\tau_{0},\ell}\left(\frac{\Delta_{1}}{2}\right)_{\ell/2}\left(\frac{\Delta_{2}}{2}\right)_{\ell/2}\sum_{k=0}^{\ell/2} \frac{\left(-\frac{\ell}{2}\right)_{k}\left(\frac{\ell+\tau_{0}-1}{2}\right)_{k}\left(\frac{\rho}{2}\right)_{k}}{k!\left(\frac{\Delta_{1}}{2}\right)_{k}\left(\frac{\Delta_{2}}{2}\right)_{k}}
\end{align}

\noindent Now let us consider the $\rho$ integral. This evaluates to 
\begin{align}
\label{rhointeg}
    &\int_{-i\infty}^{i\infty}  \frac{d\rho}{2\pi i} \ 2^{-\rho}\Gamma(\rho) \Gamma\left(\frac{\Delta_{1}-\rho}{2}\right)\Gamma\left(\frac{\Delta_{2}-\rho}{2}\right) \left(\frac{\rho}{2}\right)_{k} (\rho)_{s}\ t^{-\frac{\rho}{2}}\nonumber\\
    &=2^{2-\Delta_{1}-\Delta_{2}-s}\sqrt{\pi} \frac{\left(\frac{\Delta_{1}+s}{2}\right)_{k}\left(\frac{\Delta_{2}+s}{2}\right)_{k}\Gamma(\Delta_{1}+s)\Gamma(\Delta_{2}+s)}{\Gamma\left(\frac{\Delta_{1}+\Delta_{2}+1}{2}+s+k\right)} \ t^{\frac{s+1}{2}} \times \nonumber\\
    &\quad \ _{2}F_{1}\left(\frac{\Delta_{1}+s+1}{2}, \frac{\Delta_{2}+s+1}{2}, \frac{\Delta_{1}+\Delta_{2}+1}{2}+s+k, 1-t\right) . 
\end{align}

\noindent We are now left with the $t$-integral, which is given by
\begin{align}
    \label{tinteg}
    \mathbb{J}=\int_{0}^{1} dt \ t^{\frac{\hat{\Delta}+1}{2}-1} &\ _{2}F_{1}\left(1-\frac{\Delta_{1}+s-\hat{\Delta}}{2},1-\frac{\Delta_{2}+s-\hat{\Delta}}{2}; \hat{\Delta}+1-\frac{p}{2},t\right) \times \nonumber\\
    & \ _{2}F_{1}\left(\frac{\Delta_{1}+s+1}{2}, \frac{\Delta_{2}+s+1}{2}, \frac{\Delta_{1}+\Delta_{2}+1}{2}+s+k, 1-t\right). 
\end{align}

\noindent We can perform this integral by series expanding the $\ _{2}F_{1}(\cdots,t)$ around $t=0$ and expanding $\ _{2}F_{1}(\cdots,1-t)$ around $t=1$. Then the $t$-integral gives an Euler beta function. Therefore, we get the following double infinite sum representation of $\mathbb{J}$
\begin{align}
    \label{tintegres}
    \mathbb{J}=\sum_{m,n=0}^{\infty} \frac{\left(1+\frac{\hat{\Delta}-s-\Delta_{1}}{2}\right)_{m}\left(1+\frac{\hat{\Delta}-s-\Delta_{2}}{2}\right)_{m}\left(\frac{\Delta_{1}+s+1}{2}\right)_{n}\left(\frac{\Delta_{2}+s+1}{2}\right)_{n}}{m!\left(\hat{\Delta}+1-\frac{p}{2}\right)_{m}\left(\frac{\Delta_{1}+\Delta_{2}+2k+2s+1}{2}\right)_{n}} \ \frac{\Gamma\left(\frac{\hat{\Delta}+1}{2}+m\right)}{\Gamma\left(\frac{\hat{\Delta}+3}{2}+m+n\right)}. 
\end{align}

\noindent Remarkably, the above series can be re-summed to give the following result, which involves two balanced $\ _{4}F_{3}(1)$ hypergeometric functions. 
\begin{align}
\label{bal4f3res}
     \mathbb{J}& =\Gamma\left(\frac{\hat{\Delta}+1}{2}\right)\bigg[\frac{(\Delta_{1}+\Delta_{2}+2k+2s-1)\Gamma\left(\frac{\Delta_{1}+\Delta_{2}+2s-\hat{\Delta}-1}{2}\right)}{(\hat{\Delta}+2k)\Gamma\left(\frac{\Delta_{1}+s+1}{2}\right)\Gamma\left(\frac{\Delta_{2}+s+1}{2}\right)} \nonumber\\
     & \ {}_4F_3\left(
\begin{array}{c}
1-\frac{\Delta_{1}+s-\hat{\Delta}}{2},\, 1-\frac{\Delta_{2}+s-\hat{\Delta}}{2},\, \frac{\hat{\Delta}+1-p}{2}, \, \frac{\hat{\Delta}+2k}{2} \\
\frac{\hat{\Delta}-\Delta_{1}-\Delta_{2}-2s+3}{2},\, \hat{\Delta}+1-\frac{p}{2}, \frac{\hat{\Delta}+2k+2}{2} \, 
\end{array}
; 1\right) \nonumber \\
& + \frac{\Gamma\left(\frac{\hat{\Delta}-\Delta_{1}-\Delta_{2}-2s+1}{2}\right)\Gamma\left(\hat{\Delta}+1-\frac{p}{2}\right)\Gamma\left(\frac{\Delta_{1}+\Delta_{2}+2s-p}{2}\right)}{\Gamma\left(1+\frac{\hat{\Delta}-s-\Delta_{1}}{2}\right)\Gamma\left(1+\frac{\hat{\Delta}-s-\Delta_{2}}{2}\right)\Gamma\left(\frac{\hat{\Delta}+1-p}{2}\right)\Gamma\left(\frac{\Delta_{1}+\Delta_{2}+\hat{\Delta}+2s+1-p}{2}\right)} \times \nonumber\\
& \ {}_4F_3\left(
\begin{array}{c}
\frac{\Delta_{1}+s+1}{2},\, \frac{\Delta_{2}+s+1}{2},\, \frac{\Delta_{1}+\Delta_{2}+2s-p}{2}, \, \frac{\Delta_{1}+\Delta_{2}+2k+2s-1}{2} \\
\frac{1+\Delta_{1}+\Delta_{2}+2s-\hat{\Delta}}{2},\, \frac{\Delta_{1}+\Delta_{2}+\hat{\Delta}+2s-p+1}{2}, \frac{\Delta_{1}+\Delta_{2}+2k+2s+1}{2} \, 
\end{array}
; 1\right)\bigg] ,
\end{align}

\noindent Gathering all the prefactors, we then get the result presented in Eq. \eqref{a0lfinalres}.

\vskip 4pt
 It is also possible to derive an alternative representation of $\mathbb{J}$ in terms of a linear combination of a balanced $\ _{4}F_{3}(1)$ and a balanced $\ _{3}F_{2}(1)$\footnote{The bulk channel seed coefficient computed for BCFTs in \cite{Mazac:2018biw} was presented in this form. }. This can be achieved by using the Mellin-Barnes representation of the two $\ _{2}F_{1}$s in Eq.~\eqref{tinteg} in the variable $t/(1-t)$. The $t$-integral then yields a delta function involving the Mellin integration variables, using which one of the Mellin-Barnes integrals can be easily performed. The remaining Mellin integral can be evaluated by closing the contour and picking up two sets of residues. This results in a sum of two infinite series, one of which re-sums to yield a balanced $\ _{4}F_{3}(1)$ and the other one gives a balanced $\ _{3}F_{2}$. However,  this representation is not manifestly symmetric in $(\Delta_{1},\Delta_{2})$, unlike the one given in Eq.~\eqref{bal4f3res}.


\section{Form-factor partial wave}
\label{app:formfac}

In this appendix, we compute the following conformal integral, which defines the form-factor partial wave. 
\begin{align}
\label{ffpw}
    \mathbb{H}_{\hat{\Delta}}(\hat{x}_{1},\hat{x}_{2},x_{3}) &=\int d^{p}\hat{x} \ \langle \hspace{-0.09cm}\langle\widehat{\mathcal{O}}_{\hat{\Delta}_{1}}(\hat{x}_{1})\widehat{\mathcal{O}}_{\hat{\Delta}_{2}}(\hat{x}_{2})\widehat{\mathcal{O}}_{\hat{\Delta}_{}}(\hat{x})\rangle\hspace{-0.09cm}\rangle \langle \hspace{-0.09cm}\langle\widehat{\mathcal{O}}_{p-\hat{\Delta}}(\hat{x})\mathcal{O}_{\Delta_{3}}(x_{3})\rangle\hspace{-0.09cm}\rangle .
\end{align}

\noindent Using the expressions for the $3$-point and $2$-point conformal structures appearing in the above integral, we get
\begin{align}
\label{ffpw1}
   & \mathbb{H}_{\hat{\Delta}}(\hat{x}_{1},\hat{x}_{2},x_{3}) \nonumber\\
    & = \frac{|x_{3,\perp}|^{p-\hat{\Delta}-\Delta_{3}}}{|\hat{x}_{12}|^{\hat{\Delta}_{1}+\hat{\Delta}_{2}-\hat{\Delta}}}\int d^{p}\hat{x} \ \frac{1}{|\hat{x}_{1}-\hat{x}|^{\hat{\Delta}_{12}+\hat{\Delta}}|\hat{x}_{2}-\hat{x}|^{\hat{\Delta}-\hat{\Delta}_{12}}(|x_{3,\perp}|^{2}+(\hat{x}_{3}^{a}-\hat{x}^{a})^{2})^{p-\hat{\Delta}}}.
\end{align}

\noindent where $\hat{\Delta}_{12}=\hat{\Delta}_{1}-\hat{\Delta}_{2}$. To compute the above integral, we apply Schwinger parameterization for each of the three factors in the denominator of the integrand. We thus have the integral,
\begin{align}
\label{ffpw2}
     \mathcal{I}(\hat{x}_{1},\hat{x}_{2},x_{3}) &=\int d^{p}\hat{x} \ \frac{1}{|\hat{x}_{1}-\hat{x}|^{\hat{\Delta}_{12}+\hat{\Delta}}|\hat{x}_{2}-\hat{x}|^{\hat{\Delta}-\hat{\Delta}_{12}}(|x_{3,\perp}|^{2}+(\hat{x}_{3}^{a}-\hat{x}^{a})^{2})^{p-\hat{\Delta}}}\nonumber\\
    & = \frac{1}{\Gamma\left(\frac{\hat{\Delta}\pm\hat{\Delta}_{12}}{2}\right)\Gamma(p-\hat{\Delta})}\int_{0}^{\infty} \prod_{i=1}^{3}du_{i} \ u_{1}^{\frac{\hat{\Delta}+\hat{\Delta}_{12}}{2}-1} u_{2}^{\frac{\hat{\Delta}-\hat{\Delta}_{12}}{2}-1} u_{3}^{p-\hat{\Delta}-1} \times \nonumber\\
    & \int d^{p}\hat{x} \exp\left[-u_{1}(\hat{x}_{1}-\hat{x})^{2}-u_{2}(\hat{x}_{2}-\hat{x})^{2}-u_{3}(|x_{3,\perp}|^{2}+(\hat{x}_{3}^{a}-\hat{x}^{a})^{2})\right].
\end{align}

\noindent where in the overall normalization factor, we have used the notation $\Gamma(a\pm b)=\Gamma(a+b)\Gamma(a-b)$. The integral over $\hat{x}$ is now easy to perform, and we get
\begin{align}
\label{ffpw3}
     \mathcal{I}(\hat{x}_{1},\hat{x}_{2},x_{3}) 
    & = \frac{\pi^{\frac{p}{2}}}{\Gamma\left(\frac{\hat{\Delta}\pm\hat{\Delta}_{12}}{2}\right)\Gamma(p-\hat{\Delta})}\int_{0}^{\infty} \prod_{i=1}^{3}du_{i} \ u_{1}^{\frac{\hat{\Delta}+\hat{\Delta}_{12}}{2}-1} u_{2}^{\frac{\hat{\Delta}-\hat{\Delta}_{12}}{2}-1} u_{3}^{p-\hat{\Delta}-1} \times \nonumber\\
    & \hspace{1.0cm}\mathcal{U}^{-\frac{p}{2}}\exp\bigg[-u_{3}|x_{3,\perp}|^{2}-\mathcal{U}^{-1}\sum_{\substack{i,j=1\\ i<j}}^{3}u_{i}u_{j}\hat{x}^{2}_{ij}\bigg],
\end{align}

\noindent where $\mathcal{U}=u_{1}+u_{2}+u_{3}$. To do the integrals over the Schwinger parameters, let us insert the following identity within the integral in Eq.~\eqref{ffpw3}
\begin{align}
    1=\int d\lambda \ \delta(\lambda_{2}-u_{3})
\end{align}

\noindent and then perform the rescaling $u_{i}\rightarrow \lambda u_{i}$. Using $\delta(\lambda-\lambda u_{3})= \lambda^{-1}\delta(1-u_{3})$ the $u_{3}$ integral can be easily done. Thus, we have,
\begin{align}
\label{ffpw4}
    & \int_{0}^{\infty} \prod_{i=1}^{3}du_{i} \ u_{1}^{\frac{\hat{\Delta}+\hat{\Delta}_{12}}{2}-1} u_{2}^{\frac{\hat{\Delta}-\hat{\Delta}_{12}}{2}-1} u_{3}^{p-\hat{\Delta}-1} \mathcal{U}^{-\frac{p}{2}}\exp\bigg[-u_{3}|x_{3,\perp}|^{2}-\mathcal{U}^{-1}\sum_{\substack{i,j=1\\ i<j}}^{3}u_{i}u_{j}\hat{x}^{2}_{ij}\bigg] \nonumber\\
    & = \int_{0}^{\infty} du_{1}du_{2} \ u_{1}^{\frac{\hat{\Delta}+\hat{\Delta}_{12}}{2}-1} u_{2}^{\frac{\hat{\Delta}-\hat{\Delta}_{12}}{2}-1} (u_{1}+u_{2}+1)^{-\frac{p}{2}}\int_{0}^{\infty} d\lambda \ \lambda^{\frac{p}{2}-1} \times \nonumber\\
    &\hspace{2.0cm}\times \exp\left[-\lambda(x_{3}^{i})^{2}-\frac{\lambda}{(u_{1}+u_{2}+1)}(u_{1}u_{2}\hat{x}_{12}^{2}+u_{2}\hat{x}_{23}^{2}+u_{1}\hat{x}_{13}^{2})\right].
\end{align}

\noindent Now we evaluate the $\lambda$ integral which gives
\begin{align}
\label{ffpw4}
    &\int_{0}^{\infty} d\lambda \ \lambda^{\frac{p}{2}-1}  \exp\left[-\lambda(x_{3}^{i})^{2}-\frac{\lambda}{(u_{1}+u_{2}+1)}(u_{1}u_{2}\hat{x}_{12}^{2}+u_{2}\hat{x}_{23}^{2}+u_{1}\hat{x}_{13}^{2})\right]\nonumber\\
    &= \Gamma\left(\frac{p}{2}\right) (1+u_{1}+u_{2})^{\frac{p}{2}}\left[(1+u_{1}+u_{2})(x_{3}^{i})^{2}+u_{1}u_{2}\hat{x}_{12}^{2}+u_{2}\hat{x}_{23}^{2}+u_{1}\hat{x}_{13}^{2}\right]^{-\frac{p}{2}} . 
\end{align}

\noindent We are now left with the following integrals over the remaining two Schwinger parameters 
\begin{align}
\label{ffpw5}
   \int_{0}^{\infty} du_{1}du_{2} \ \frac{u_{1}^{\frac{\hat{\Delta}+\hat{\Delta}_{12}}{2}-1} u_{2}^{\frac{\hat{\Delta}-\hat{\Delta}_{12}}{2}-1}}{\left[(1+u_{1}+u_{2})(x_{3}^{i})^{2}+u_{1}u_{2}\hat{x}_{12}^{2}+u_{2}\hat{x}_{23}^{2}+u_{1}\hat{x}_{13}^{2}\right]^{\frac{p}{2}}} \ .
\end{align}

\vskip 4pt
\noindent These integrals are straightforward to compute. As expected, we find that up to position-dependent kinematic factors, Eq.~\eqref{ffpw5} evaluates to a linear combination of the form-factor block and its shadow. Finally, the result for the integral in Eq.~\eqref{ffpw} is
\begin{align}
    \label{ffpw6}
   \mathbb{H}_{\hat{\Delta}}(\hat{x}_{1},\hat{x}_{2},x_{3}) = \frac{|x_{3,\perp}|^{-\Delta_{3}}}{|\hat{x}_{12}|^{\hat{\Delta}_{1}+\hat{\Delta}_{2}}}\left(\frac{|x_{3,\perp}|^{2}+\hat{x}_{23}^{2}}{|x_{3,\perp}|^{2}+\hat{x}_{13}^{2}}\right)^{\frac{\hat{\Delta}_{12}}{2}} \left[\mathcal{N}_{\hat{\Delta}} H_{\hat{\Delta}}(w)+ \mathcal{N}_{p-\hat{\Delta}} H_{p-\hat{\Delta}}(w)\right], 
\end{align}

\noindent where $H_{\hat{\Delta}}(w)$ is the form-factor block
\begin{align}
    H_{\hat{\Delta}}(w)= w^{\frac{\hat{\Delta}}{2}} \ _{2}F_{1}\left(\frac{\hat{\Delta}+\hat{\Delta}_{12}}{2}, \frac{\hat{\Delta}-\hat{\Delta}_{12}}{2},\hat{\Delta}+1-\frac{p}{2},w\right)
\end{align}

\noindent and the factor $\mathcal{N}_{\hat{\Delta}}$ is given by
\begin{align}
    \mathcal{N}_{\hat{\Delta}}= \frac{\Gamma\left(\frac{p}{2}-\hat{\Delta}\right)\Gamma\left(\frac{\hat{\Delta}+\hat{\Delta}_{12}}{2}\right)}{\Gamma\left(\frac{p-\hat{\Delta}+\hat{\Delta}_{12}}{2}\right)} .
\end{align}

\section{Bulk one-point function}
\label{app:bulk1pt}

In this section, we consider a scalar bulk-to-brane propagator $G_{\Delta}(z_{1},\hat{z}_{2})$ and evaluate it's integral over the brane point $\hat{z}_{2}$ in closed form. This result is useful for the computation of certain loop diagrams, in particular the one considered in Appendix \ref{app:1ptloop}. Consider then the following integral
\begin{align}
\label{bulk1ptfnc}
    \mathcal{I}(z_{1})= \int\limits_{AdS_{p+1}} d^{p+1} \hat{z}_{2} \ G_{\Delta}(z_{1},\hat{z}_{2}). 
\end{align}

\noindent The explicit expression for $G_{\Delta}(x_{1},\hat{x}_{2})$ is 
\begin{align}
    G_{\Delta}(x_{1},\hat{x}_{2})
    &= 2^{\Delta}C_{\Delta} \zeta^{\Delta} \ _{2}F_{1}\left(\frac{\Delta}{2},\frac{\Delta+1}{2},\Delta-\frac{d}{2}+1,\zeta^{2}\right)
\end{align}

\noindent where
\begin{align}
    \zeta=\frac{2z_{1}^{0}\hat{z}_{2}^{0}}{(z_{1}^{0})^{2}+(\hat{z}_{2}^{0})^{2}+(z_{1,\perp})^{2}+(z_{1}^{a}-\hat{z}_{2}^{a})^{2}}, \quad C_{\Delta} = \frac{\Gamma(\Delta)\Gamma\left(\Delta-\frac{d-1}{2}\right)}{(4\pi)^{\frac{d+1}{2}}\Gamma(2\Delta-d+1)}. 
\end{align}

\noindent Now to evaluate the $AdS_{p+1}$ integral, we use the series representation of the hypergeometric function to get,
\begin{align}
     \mathcal{I}(z_{1})&=2^{2\Delta}C_{\Delta} (z_{1}^{0})^{\Delta}\sum_{k=0}^{\infty}\frac{2^{2k}\left(\frac{\Delta}{2}\right)_{k}\left(\frac{\Delta+1}{2}\right)_{k}}{k!\left(\Delta-\frac{d}{2}+1\right)_{k}} \ (z_{1}^{0})^{2k} \times \nonumber\\
     &\times \int\limits_{AdS_{p+1}} d^{p+1} \hat{z}_{2} \ \frac{(\hat{z}_{2}^{0})^{\Delta+2k}}{\left[z_{1}^{0})^{2}+(\hat{z}_{2}^{0})^{2}+(z_{1,\perp})^{2}+(z_{1}^{a}-\hat{z}_{2}^{a})^{2}\right]^{\Delta+2k}}
\end{align}

\noindent The integral over $AdS_{p+1}$ can now be performed easily by introducing Schwinger parameters, and we get
\begin{align}
  & \int\limits_{AdS_{p+1}} d^{p+1} \hat{z}_{2} \ \frac{(\hat{z}_{2}^{0})^{\Delta+2k}}{\left[z_{1}^{0})^{2}+(\hat{z}_{2}^{0})^{2}+(z_{1,\perp})^{2}+(z_{1}^{a}-\hat{z}_{2}^{a})^{2}\right]^{\Delta+2k}}  \nonumber\\
  &= \frac{\pi^{\frac{p}{2}}\Gamma\left(\frac{\Delta-p}{2}+k\right)\Gamma\left(\frac{\Delta}{2}+k\right)}{2\Gamma(\Delta+2k)} \ \frac{1}{\left[(z_{1}^{0})^{2}+(z_{1,\perp})^{2}\right]^{\frac{\Delta}{2}+k}}
\end{align}

\noindent The integral in Eq.~\eqref{bulk1ptfnc} is then given by
\begin{align}
\label{bulk1ptfnc1}
     \mathcal{I}(z_{1})&=2^{2\Delta-1}\pi^{\frac{p}{2}}C_{\Delta} \sum_{k=0}^{\infty}\frac{2^{2k}\left(\frac{\Delta}{2}\right)_{k}\left(\frac{\Delta+1}{2}\right)_{k}\Gamma\left(\frac{\Delta-p}{2}+k\right)\Gamma\left(\frac{\Delta}{2}+k\right)}{k!\left(\Delta-\frac{d}{2}+1\right)_{k}\Gamma(\Delta+2k)} \ \left(\frac{(z_{1}^{0})^{2}}{(z_{1}^{0})^{2}+(z_{1,\perp})^{2}}\right)^{\frac{\Delta}{2}+k} 
\end{align}

\noindent The above infinite sum can be easily performed, and we get
\begin{align}
\label{bulk1ptinteg}
     \mathcal{I}(z_{1})
     & = 2^{\Delta}C_{\Delta}\pi^{\frac{p+1}{2}}\frac{\Gamma\left(\frac{\Delta-p}{2}\right)}{\Gamma\left(\frac{\Delta+1}{2}\right)} \left(\frac{(x_{1}^{0})^{2}}{(x_{1}^{0})^{2}+(x_{1}^{i})^{2}}\right)^{\frac{\Delta}{2}} \ _{2}F_{1}\left(\frac{\Delta}{2},\frac{\Delta-p}{2},\Delta-\frac{d}{2}+1, \frac{(z_{1}^{0})^{2}}{(z_{1}^{0})^{2}+(z_{1,\perp})^{2}}\right). 
\end{align}

\section{Computation of loop-corrected $1$-point functions}
\label{app:1ptloop}

In this Appendix, we compute the one-loop corrected one-point function coefficient\footnote{The results for the one-point function coefficients presented here potentially suffer from bulk UV divergences in AdS. We postpone a systematic understanding of how to regulate and renormalize these divergences to future work.} which appears in subsection \ref{subsec:loop1}. The corresponding one-point diagram is shown in Fig.~\ref{fig:1ptloop1}(a). We also consider the one-loop in Fig.~\ref{fig:1ptloop1}(c), and show that the one-point function coefficient in this case is given in terms of the defect-to-bulk crossing kernel.

\subsection*{Diagram $9$(a)}
\begin{figure}[htp]
    \centering
    \includegraphics[width=7.5cm]{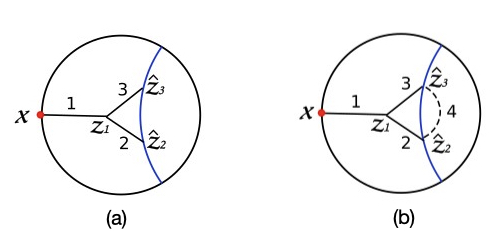}
    \caption{$1$-loop corrections to one-point function of a bulk CFT operator in the presence of a defect.}
    \label{fig:1ptloop1}
\end{figure}

Consider the $1$-point Witten diagram in Fig.~\ref{fig:1ptloop1}(a). We take the external operator as well as the exchanges in this diagram to be scalars. 

\noindent The integral expression for this diagram is
\begin{align}
\label{1ptloopa}
    I_{\Delta_{1},\Delta_{2},\Delta_{3}}(x)= \int\limits_{AdS_{d+1}} d^{d+1}z_{1} \hspace{-0.3cm}\int\limits_{AdS_{p+1}} d^{p+1}\hat{z}_{2}d^{p+1}\hat{z}_{3} \ K_{\Delta_{1}}(x,z_{1}) G_{\Delta_{2}}(z_{1},\hat{z}_{2}) G_{\Delta_{3}}(z_{1},\hat{z}_{3}).  
\end{align}

\noindent The AdS$_{p+1}$ integrals involving the bulk-to-brane propagators in  Eq.~\eqref{1ptloopa} can be done using the identity derived in Eq.~\eqref{bulk1ptinteg}. We then get
\begin{align}
\label{1ptloopb}
    I_{\Delta_{1},\Delta_{2},\Delta_{3}}(x)& = \widetilde{C}_{\Delta_{2}}\widetilde{C}_{\Delta_{3}}\hspace{-0.2cm}\int\limits_{AdS_{d+1}} d^{d+1}z_{1} \ K_{\Delta_{1}}(x,z_{1}) \ \chi^{\frac{\Delta_{2}+\Delta_{3}}{2}} \ _{2}F_{1}\left(\frac{\Delta_{2}}{2},\frac{\Delta_{2}-p}{2},\Delta_{2}-\frac{d}{2}+1, \chi\right) \times \nonumber\\
    & \hspace{3.0cm}  \ _{2}F_{1}\left(\frac{\Delta_{3}}{2},\frac{\Delta_{3}-p}{2},\Delta_{3}-\frac{d}{2}+1, \chi\right)
\end{align}

\noindent where
\begin{align}
\label{ctilde}
    \widetilde{C}_{\Delta}= \frac{2^{\Delta-d-1} \Gamma(\Delta)\Gamma\left(\frac{\Delta-p}{2}\right)\Gamma\left(\Delta-\frac{d-1}{2}\right)}{\pi^{\frac{d-p}{2}}\Gamma\left(\frac{\Delta+1}{2}\right)\Gamma(2\Delta-d+1)}, \quad \chi = \frac{(z_{1}^{0})^{2}}{(z_{1}^{0})^{2}+(z_{1,\perp})^{2}} .
\end{align}

\noindent Now let us expand the product of hypergeometric functions in Eq.~\eqref{1ptloopb} as
\begin{align}
    \prod_{i=2}^{3}\ _{2}F_{1}\left(\frac{\Delta_{i}}{2},\frac{\Delta_{i}-p}{2},\Delta_{i}-\frac{d}{2}+1, \chi\right) = \sum_{n=0}^{\infty} \alpha_{n} \ \chi^{n}, 
\end{align}

\noindent where the coefficients $\alpha_{n}$ are given by
\begin{align}
    &\alpha_{n}= \frac{\left(\frac{\Delta_{2}}{2}\right)_{n}\left(\frac{\Delta_{2}-p}{2}\right)_{n}}{n!\left(\Delta_{2}+1-\frac{d}{2}\right)_{n}}  \ {}_4F_3\left(
\begin{array}{c}
-n,\, \frac{d}{2}-n-\Delta_{2},\, \frac{\Delta_{3}-p}{2}, \, \frac{\Delta_{3}}{2} \\
1-n-\frac{\Delta_{2}}{2},\, 1-n+\frac{p-\Delta_{2}}{2}, 1-n+\frac{p-\Delta_{2}}{2},\Delta_{3}-\frac{d}{2}+1 \, 
\end{array}
; 1\right) .
\end{align}

\noindent Then, interchanging the order of the sum and integration, we can express $I_{123}(x)$ as
\begin{align}
   I_{\Delta_{1},\Delta_{2},\Delta_{3}}(x)& = \widetilde{C}_{\Delta_{2}}\widetilde{C}_{\Delta_{3}} \sum_{n=0}^{\infty} \alpha_{n}  \mathcal{I}_{n}(x), 
\end{align}

\noindent where $ \mathcal{I}_{n}(x)$ is the following integral
\begin{align}
    \mathcal{I}_{n}(x) = \int_{AdS_{d+1}} d^{d+1}z_{1} \ \frac{(z_{1}^{0})^{\Sigma+2n}}{\left[(z_{1}^{0})^{2}+(x^{i}_{\perp}-z^{i}_{\perp})^{2}+(x^{a}-z^{a})^{2}\right]^{\Delta_{1}}\left[(z_{1}^{0})^{2}+(z_{1,\perp})^{2}\right]^{\frac{\Sigma-\Delta_{1}}{2}+n}},  
\end{align}

\noindent and we have defined $\Sigma=\Delta_{1}+\Delta_{2}+\Delta_{3}$. This integral can be straightforwardly evaluated using the Schwinger parameterization. The result is as follows.
\begin{align}
    \mathcal{I}_{n}(x) 
    &= \frac{\pi^{\frac{d}{2}}\Gamma\left(\frac{\Delta_{1}}{2}\right)\Gamma\left(\frac{\Delta_{1}-p}{2}\right)\Gamma\left(\frac{\Sigma+2n-d}{2}\right)\Gamma\left(\frac{\Sigma-2\Delta_{1}+2n}{2}\right)}{2\Gamma\left(\Delta\right)\Gamma\left(\frac{\Delta_{2}+\Delta_{3}}{2}+n\right)\Gamma\left(\frac{\Delta_{2}+\Delta_{3}+2n-p}{2}\right)}  \ \frac{1}{|x_{\perp}|^{\Delta_{1}}}
\end{align}

\noindent The result for the integral in Eq.~\eqref{1ptloopa} is then 
\begin{align}
\label{1ptblk1loppafinal}
    I_{\Delta_{1},\Delta_{2},\Delta_{3}}(x)& = \frac{a^{1-\mathrm{loop}}(\Delta_{1},\Delta_{2},\Delta_{3})}{|x_{\perp}|^{\Delta_{1}}},  
\end{align}

\noindent where the $1$-loop corrected $1$-point function coefficient $a^{1-\mathrm{loop}}(\Delta_{1},\Delta_{2},\Delta_{3})$ is given by
\begin{align}
\label{1pt1loopacoeff}
a^{1-\mathrm{loop}}(\Delta_{1},\Delta_{2},\Delta_{3}) = \frac{\Gamma\left(\frac{\Delta_{1}}{2}\right)\Gamma\left(\frac{\Delta_{1}-p}{2}\right)\Gamma\left(\frac{\Delta_{2}+\Delta_{3}-\Delta_{1}}{2}\right)\Gamma\left(\frac{\Delta_{1}+\Delta_{2}+\Delta_{3}-d}{2}\right)}{\Gamma\left(\Delta_{1}\right)} \ \widetilde{a}(\Delta_{1},\Delta_{2},\Delta_{3}), 
\end{align}

\noindent and $\widetilde{a}(\Delta_{1},\Delta_{2},\Delta_{3})$ is given by
\begin{align}
\label{1pt1loopacoeff1}
&\widetilde{a}(\Delta_{1},\Delta_{2},\Delta_{3}) = \frac{2^{\Delta_{2}+\Delta_{3}-2d-1}}{\pi^{\frac{d}{2}-p}} \left(\prod_{i=2}^{3}\frac{ \Gamma(\Delta_{i})\Gamma\left(\frac{\Delta_{i}-p}{2}\right)\Gamma\left(\Delta_{i}-\frac{d-1}{2}\right)}{\Gamma\left(\frac{\Delta_{i}+1}{2}\right)\Gamma(2\Delta_{i}-d+1)} \right)\times \nonumber\\
    & \sum_{n=0}^{\infty} \bigg[ \frac{\left(\frac{\Delta_{2}+\Delta_{3}-\Delta_{1}}{2}\right)_{n}\left(\frac{\Delta_{1}+\Delta_{2}+\Delta_{3}-d}{2}\right)_{n}\left(\frac{\Delta_{2}}{2}\right)_{n}\left(\frac{\Delta_{2}-p}{2}\right)_{n}}{n!\left(\Delta_{2}+1-\frac{d}{2}\right)_{n}\Gamma\left(\frac{\Delta_{2}+\Delta_{3}}{2}+n\right)\Gamma\left(\frac{\Delta_{2}+\Delta_{3}+2n-p}{2}\right)} \times \nonumber\\ 
    &  \ _{4}F_{3}\left(-n,\frac{d}{2}-n-\Delta_{2},\frac{\Delta_{3}-p}{2}, \frac{\Delta_{3}}{2}; 1-n-\frac{\Delta_{2}}{2}, 1-n+\frac{p-\Delta_{2}}{2},\Delta_{3}-\frac{d}{2}+1;1\right)\bigg].
\end{align}

\noindent From the above expression we observe that $a^{1-\mathrm{loop}}$ has poles at $\Delta_{1}= \Delta_{2}+\Delta_{3}+2m$, and $d-\Delta_{1}= \Delta_{2}+\Delta_{3}+2m$, where $m\in\mathbb{Z}_{\ge 0}$.

\subsection*{Diagram $9$(b)}
\label{app:1ptloop6j}

We will now consider the diagram in Fig.~\ref{fig:1ptloop1}(b). This is given by the integral
\begin{align}
    \label{1pt6j}
    I(x) = \int\limits_{AdS_{d+1}} d^{d+1}z_{1} \hspace{-0.3cm}\int\limits_{AdS_{p+1}} d^{p+1}\hat{z}_{2}d^{p+1}\hat{z}_{3} \ K_{\Delta_{1}}(x,z_{1}) G_{\Delta_{2}}(z_{1},\hat{z}_{2}) \widehat{G}_{\hat{\Delta}_{4}}(\hat{z}_{2},\hat{z}_{3})G_{\Delta_{3}}(z_{1},\hat{z}_{3}).
\end{align}

\noindent Due to $SO(p+1,1)\times SO(d-p)$ symmetry, the result of this integral takes the form
\begin{align}
\label{1pt6ja}
    I(x) = \frac{a(\Delta_{1},\Delta_{2},\Delta_{3},\hat{\Delta}_{4})}{|x_{\perp}|^{\Delta_{1}}}.
\end{align}

\noindent We will now demonstrate how the coefficient in Eq.~\eqref{1pt6ja} is determined by the defect-to-bulk channel crossing kernel. For this purpose, let us apply the split representation of the propagators $2,3$. This yields
\begin{align}
    \label{1pt6j1}
    I(x) = \int_{-\infty}^{\infty} \left(\prod_{i=2}^{3}d\nu_{i} \ P(\nu_{i},\Delta_{i})\right) \int\limits_{\partial AdS_{d+1}} d^{d}x_{2} d^{d}x_{3} \ \mathcal{A}_{1\underline{2}\hspace{0.02cm}\underline{3}}(x_{1},x_{2},x_{3})  \mathbf{A}^{\underline{\tilde{2}}\hspace{0.02cm}\underline{\tilde{3}}}_{\hat{\Delta}_{4},\mathrm{exch}}(x_{2},x_{3}), 
\end{align}

\noindent where $\mathcal{A}_{1\underline{2}\hspace{0.02cm}\underline{3}}$ is a tree-level $3$-point function in a CFT without defects and $\mathbf{A}^{\underline{\tilde{2}}\hspace{0.02cm}\underline{\tilde{3}}}_{\hat{\Delta}_{4},\mathrm{exch}}$ is a tree-level $2$-point defect-exchange diagram. Now, let us decompose the exchange diagram in terms of bulk channel partial waves using the defect-to-bulk channel crossing kernel. This enables us to write Eq.~\eqref{1pt6j1} as
\begin{align}
    \label{1pt6j2}
     &I(x) = \int_{-\infty}^{\infty} d\nu_{2}d\nu_{3}d\nu_{4} \ P(\nu_{2},\Delta_{2})P(\nu_{3},\Delta_{3}) \widehat{P}(\nu_{4},\hat{\Delta}_{4})\lambda_{1\underline{2}\hspace{0.02cm}\underline{3}} b_{\underline{\tilde{2}}\hspace{0.02cm}\underline{\hat{4}}} b_{\underline{\tilde{3}}\hspace{0.02cm}\underline{\tilde{\hat{4}}}} \times \nonumber\\
     & \sum_{\ell=0}^{\infty} \int_{\frac{d}{2}}^{\frac{d}{2}+i\infty} \frac{d\Delta}{2\pi i} \ \mathcal{J}_{d,p}(\underline{\hat{\Delta}_{4}},0;\Delta,\ell|\underline{\widetilde{\Delta}_{2}},\underline{\widetilde{\Delta}_{3}})\hspace{-0.3cm}\int\limits_{\partial AdS_{d+1}} \hspace{-0.3cm}d^{d}x_{2} d^{d}x_{3} \ \langle\mathcal{O}_{\Delta_{1}}(x)\underline{\mathcal{O}_{2}}(x_{2})\underline{\mathcal{O}_{3}}(x_{3})\rangle  \mathcal{F}^{\underline{\tilde{2}}\hspace{0.02cm}\underline{\tilde{3}}}_{\Delta,\ell}(x_{2},x_{3}) .
\end{align}

\noindent Now we need to show that the conformal integral in Eq.~\eqref{1pt6j2} yields the expected position dependence of a one-point function. To this end, we use the integral representation of the bulk channel partial wave. The conformal integral in the above equation then becomes
\begin{align}
    \label{1pt6j3}
     \int\limits_{\partial AdS_{d+1}} d^{d}x_{2} d^{d}x_{3}d^{d}y \ \langle\mathcal{O}_{\Delta_{1}}(x)\underline{\mathcal{O}_{2}}(x_{2})\underline{\mathcal{O}_{3}}(x_{3})\rangle \langle\underline{\widetilde{\mathcal{O}}_{2}}(x_{2})\underline{\widetilde{\mathcal{O}}_{3}}(x_{3})\mathcal{O}_{\Delta,\ell}(y)\rangle  \langle\hspace{-0.09cm}\langle\mathcal{O}_{d-\Delta,\ell}(y)\rangle\hspace{-0.09cm}\rangle.
\end{align}

\noindent Here we have suppressed the Lorentz indices of the spinning operators for simplicity. Now, the $x_{2},x_{3}$ integrals can be done using the CFT bubble identity as follows
\begin{align}
    \label{bubbleid2}
     & \int\limits_{\partial AdS_{d+1}} \hspace{-0.2cm}d^{d}x_{2} d^{d}x_{3} \ \langle\mathcal{O}_{\Delta_{1}}(x)\underline{\mathcal{O}_{2}}(x_{2})\underline{\mathcal{O}_{3}}(x_{3})\rangle \langle\underline{\widetilde{\mathcal{O}}_{2}}(x_{2})\underline{\widetilde{\mathcal{O}}_{3}}(x_{3})\mathcal{O}_{\Delta,\ell}(y)\rangle  = B_{\Delta_{1}} \delta_{\ell,0} \delta^{d}(x-y) \delta(\nu+\nu'), 
\end{align}

\noindent where $\Delta=\frac{d}{2}+i\nu'$ and we have also taken $\Delta_{1}=\frac{d}{2}+i\nu$ in order to apply the bubble identity. Applying the above identity in Eq.~\eqref{1pt6j3} and performing the integral over $x$ using the delta function $\delta(x-y)$, we get 
\begin{align}
    \label{1pt6j3}
     & \int\limits_{\partial AdS_{d+1}} d^{d}x_{2} d^{d}x_{3}d^{d}y \ \langle\mathcal{O}_{\Delta_{1}}(x)\underline{\mathcal{O}_{2}}(x_{2})\underline{\mathcal{O}_{3}}(x_{3})\rangle \langle\underline{\widetilde{\mathcal{O}}_{2}}(x_{2})\underline{\widetilde{\mathcal{O}}_{3}}(x_{3})\mathcal{O}_{\Delta,\ell}(y)\rangle  \langle\hspace{-0.09cm}\langle\mathcal{O}_{d-\Delta,\ell}(y)\rangle\hspace{-0.09cm}\rangle\nonumber\\
     &= B_{\Delta_{1}} \delta_{\ell,0} \ \delta(\nu+\nu') \ \langle\hspace{-0.09cm}\langle\mathcal{O}_{\frac{d}{2}+i\nu,0}(x)\rangle\hspace{-0.09cm}\rangle
\end{align}

\noindent Using the delta function $\delta(\nu+\nu')$, we can further carry out the spectral integral over $\Delta$ in Eq.~\eqref{1pt6j2}. Due to the $\delta_{\ell,0}$ factor coming from Eq.~\eqref{bubbleid2}, only the $\ell=0$ term survives in the sum over spins in Eq.~\eqref{1pt6j2}. Consequently, the integral $I(x)$ takes the expected form in Eq.\eqref{1pt6ja}. We then get the following spectral representation of the $1$-loop coefficient, involving the bulk-to-defect crossing kernel
\begin{align}
    \label{1pt6j4}
    & a(\Delta_{1},\Delta_{2},\Delta_{3},\hat{\Delta}_{4}) \nonumber\\
    &= \int_{-\infty}^{\infty} \left(\prod_{i=2}^{3}d\nu_{i}  P(\nu_{i},\Delta_{i})\right)d\nu_{4} \widehat{P}(\nu_{4},\hat{\Delta}_{4})\lambda_{1\underline{2}\hspace{0.02cm}\underline{3}} b_{\underline{\tilde{2}}\hspace{0.02cm}\underline{\hat{4}}} b_{\underline{\tilde{3}}\hspace{0.02cm}\underline{\tilde{\hat{4}}}} B_{\Delta_{1}} \ \mathcal{J}_{d,p}(\underline{\hat{\Delta}_{4}},0;\Delta_{1},0|\underline{\widetilde{\Delta}_{2}},\underline{\widetilde{\Delta}_{3}}) . 
\end{align} 


\bibliographystyle{utphys}
\bibliography{draft.bib}

\end{document}